\documentclass[AMA,STIX1COL]{WileyNJD-v2}
\usepackage{moreverb}
\usepackage{latexsym,amsmath,amssymb,amsbsy,amstext,amscd,amsfonts}
\usepackage{float}
\DeclareMathOperator{\arctanh}{arctanh}

\newcommand\BibTeX{{\rmfamily B\kern-.05em \textsc{i\kern-.025em b}\kern-.08em
T\kern-.1667em\lower.7ex\hbox{E}\kern-.125emX}}

\articletype{Research article}%

\received{<day> <Month>, <year>}
\revised{<day> <Month>, <year>}
\accepted{<day> <Month>, <year>}

\begin{document}

\title{Numerical simulation of hydraulic fracturing: a hybrid FEM based algorithm\protect\thanks{Preprint submitted to \textit{International Journal for Numerical and Analytical Methods in Geomechanics}.}}

\author[1]{Michal Wrobel*}

\author[1]{Panos Papanastasiou}

\author[2]{Daniel Peck}

\authormark{WROBEL \textsc{et al}}

\address[1]{\orgdiv{Department of Civil and Environmental Engineering}, \orgname{University of Cyprus}, \orgaddress{\state{Nicosia}, \country{Cyprus}}}

\address[2]{\orgdiv{Department of Mathematics}, \orgname{Aberystwyth University}, \orgaddress{\state{Aberystwyth}, \country{UK}}}

\corres{*M. Wrobel, Department of Civil and Environmental Engineering, University of Cyprus. \email{wrobel.michal@ucy.ac.cy}}

\abstract[Abstract]{In this paper a problem of numerical simulation of hydraulic fractures is considered. An efficient algorithm of solution is proposed for the plain strain model of hydraulic fracturing. The algorithm utilizes a FEM based subroutine to compute deformation of the fractured material. The flow of generalized Newtonian fluid in the fracture is modelled in the framework of lubrication theory. In this way, the architecture of the computational scheme is relatively simple and enables one to deal with  advanced cases of the fractured material properties and configurations as well as various rheological models of fluid. In particular, the problems of poroelasticity, plasticity and spatially varying properties of the fractured material can be analyzed. The accuracy and efficiency of the proposed algorithm are verified against analytical benchmark solutions.  The algorithm capabilities are demonstrated using the example of the hydraulic fracture propagating in complex geological settings. }


\maketitle

\section{Introduction}\label{Introduction}

Hydraulic fracturing refers to the phenomenon of hydraulically induced fractures propagating in a solid material. It can be observed in multiple physical processes, such as subglacial drainage of water or extension of magma intrusions in the Earth's crust. The most prominent technological application of hydraulic fracture (HF) is fracking, a technique used to stimulate low permeability hydrocarbon reservoirs. In the last few decades this technology has revolutionized the exploitation of unconventional oil and gas resources. On the other hand, hydraulic fracturing can constitute an undesired and dangerous accompanying effect of the main technological process, for example in  CO$_2$ sequestration \cite{Papanastasiou_CO2}. Similarly, the stability of hydraulic structures, such as dams, can be jeopardized by propagation of hydraulic fractures \cite{Roth_2020}. Credible prediction and control of the HF process is therefore of great importance in designing  fracking treatments as well as in preventing  catastrophic failures in various technological installations. Towards this end, mathematical modelling of the underlying physics can be effectively employed.

Mathematical simulation of hydraulic fracturing constitutes a formidable task. It results from a very complex nature of interaction between the respective physical fields that comprise the HF process. In terms of mathematical description the main difficulties are caused by: i) strong non-linearities originating from the non-local interaction between the solid and the fluid phases, as well as from their respective non-linear properties, ii) singularities of the component physical fields, iii) moving boundaries, iv) degeneration of the governing equations in the singular points of the domain, v) multiscale effects, vi) possible plastic deformation, and many others. Clearly, it is impossible to tackle all these challenges at once. Instead,  simplified HF models have been studied to improve our understanding of the underlying physics and to develop computational methods. Even though an immense progress has been achieved since the time when the pioneering works were done, still the simple 1D models such as PKN \cite{Nordgren}, KGD (plane strain) \cite{Khristianovic,Geertsma} and penny shaped \cite{Sneddon} are employed towards this effort. Despite their geometrical simplicity, these models reflect properly the basic physical mechanisms that govern the HF process. Thus, they enabled e. g. to define the regimes of crack propagation  \cite{Detournay_2004,Garagash_2009,Bao_2014}, estimate the influence of the non-Newtonian rheology of fracturing fluid on the HF process \cite{Adachi_Detournay,Perkowska_2016,Pereira_2021}, analyze the phenomenon of subcritical fracture growth \cite{Lu_2017} or investigate the effect of hydraulically induced tangential tractions exerted on the crack faces \cite{Wrobel_2017,Wrobel_2018,Papanastasiou_2018,Piccolroaz_2021}.
Moreover, solutions obtained for the simplifed models can be used for benchmarking purposes in more complex cases \cite{Peirce_2008,Dontsov_2017}.

Nevertheless, in many instances the geometrical and/or material simplifications accepted when deriving the classical HF models cannot be justified. Such situations may involve a complex  configuration of the fractured material with spatially varying properties, poroelastic or plastic deformation, presence of stress contrast and many others. Then, the general system of the governing time dependent equations in 2D or 3D  form has to to be used. Needless to say, the complication of the resulting mathematical and numerical models becomes significant, with the computational cost increased accordingly (or becoming prohibitive). A representative example of such a problem is the case of hydraulic fracture propagating in an elasto-plastic material. In this case,  simulations are typically performed using the Finite Element Method (FEM) in various settings \cite{Papanastasiou_1997,Papanastasiou_algorithm,Papanastasiou_1999a,
Papanastasiou_2000,Sarris_2012,Wang_2015,Zeng_2020}. Even though the FEM constitutes a very flexible and versatile numerical tool, its application entails numerous problems originating for example from the character of the fluid-solid interaction, a need to account for the localized effects or the necessity of employing dedicated stabilization techniques \cite{Liu_2017}. Thus, any development of the mathematical model that can reduce the complication of the problem or improve the efficiency of the computations is of prime importance.

In the paper of Wrobel $\&$ Mishuris \cite{Wrobel_2015} a flexible computational algorithm for the problem of hydraulic fracture propagating in elastic material was proposed, utilizing an approach later referred to as `universal algorithm'. Its performance was demonstrated for the classical PKN and KGD models, with the respective extension to the penny shaped model provided later by Peck et al.  \cite{Peck_2018_1,Peck_2018_2}. Under this scheme, computations are performed for two basic dependent variables: the crack opening and the fluid velocity, while the fracture front tracing mechanism is based on the Stefan type condition \cite{Kemp}.
The algorithm has a modular structure with the basic blocks pertaining to the respective dependent variables. As a result of this construction,  the scheme can be easily modified to account for different fluid flow and solid deformation models.  The algorithm proved to be a very efficient and flexible tool to simulate the HF problem. It was modified by Perkowska et al. \cite{Perkowska_2016} to account for the power-law rheology of the fracturing fluid. In the papers by Wrobel et al. \cite{Wrobel_2017,Wrobel_2018} the effect of tangential traction exerted by the fluid on the crack faces was included. Further development of the scheme\cite{Wrobel_2020,Wrobel_2021} involved application of the rheological model of the generalized Newtonian fluid \cite{Bird_1987}. All of the above variants of the algorithm were designed assuming elastic deformation of the solid (even though different elasticity operators were used) with the boundary equation of elasticity defining the relation between the net fluid pressure and the crack opening. Thus, the inelastic models of the fractured material and/or complex material setting of the computational domain cannot be analyzed with the original version of the scheme. 

In this paper we propose a modification of the aforementioned algorithm in which the module for computing the crack opening, based on a boundary operator,  is substituted by a dedicated FEM block. Consequently, the computational scheme retains the relative simplicity of its original version and simultaneously enables one to deal with more advanced descriptions of the fractured material properties and configuration. The approach to solving the fluid mechanics equations remains the same as in the original  scheme, with the respective fluid flow model based on lubrication theory. This combination contributes to efficiency and stability of computations. The presented modified algorithm has already been used to analyze the problem of elasto-plastic HF in the study by Wrobel et al. \cite{Wrobel_plasticity} where the shielding effect introduced by the plastic deformation was investigated. Just as in a number of our previous publications, the considered crack geometry is that of the KGD model. According to the above discussion, even such a simplified geometrical configuration of the problem allows one to analyze the interactions between the underlying physical mechanisms of hydraulic fracture. Moreover, some of the modelling concepts can be extended to the more general cases of fracture geometry.

The paper is structured as follows. In Section \ref{formulation} we introduce a general form of the analyzed mathematical formulation of the HF problem. Section \ref{alg_desc} includes presentation of the algorithm (Subsection \ref{gen_alg}) with a detailed description of the newly introduced FEM-based module (Subsection \ref{FEM_module}). The numerical analysis is conducted in Section \ref{num_an}. Here, the accuracy tests of the FEM module (Subsection \ref{ver_FEM}) are followed by the investigation of performance of the complete algorithm (Subsection \ref{ver_alg}). A computational example, concerning the hydraulic fracture crossing an interface between two dissimilar layers of rock, is described in Subsection \ref{num_ex}. The final conclusions are listed in Section \ref{conc}. The supplementary material includes: i) the self-similar formulation of the HF problem used to construct a benchmark example (Appendix \ref{ap_A}), ii) the analytical benchmark example (Appendix \ref{ap_B}).

\section{Mathematical formulation of the HF problem}
\label{formulation}

\begin{figure}[htb!]
\begin{center}
\includegraphics[scale=0.55]{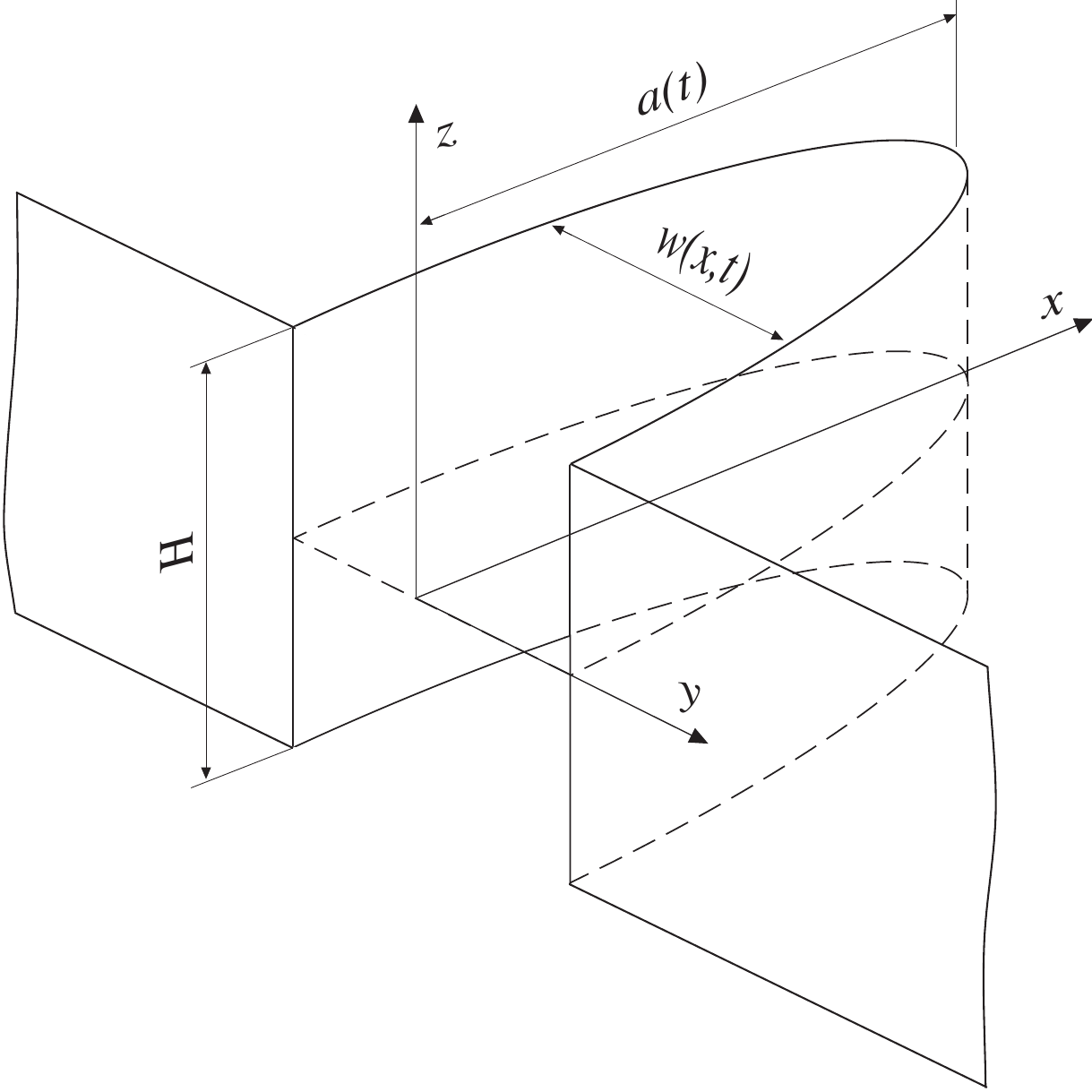}
\caption{The KGD fracture geometry.}
\label{KGD_geom}
\end{center}
\end{figure}

Let us consider the problem of a plane strain hydraulic fracture described by the standard KGD geometry. Due to the problem symmetry we analyze only one of the crack wings, as schematically shown in Figure \ref{KGD_geom}. The time-dependent fracture geometry is described by: i) the crack length $a(t)$ [m], ii) the crack opening $w(x,t)$ [m], iii) and the crack height $H$ [m]. In the following we specify the governing system of equations for the analyzed problem.

\subsection{Fluid flow equations}

The local mass balance inside the fracture is described by the continuity equation:
\begin{equation}
\label{cont}
\frac{\partial w}{\partial t}+\frac{\partial q}{\partial x}+q_\text{L}=0, \quad x\in[0,a(t)],
\end{equation}
where $t$ [s] is time,  $q(x,t)$ $\left[\frac{\text{m}^2}{\text{s}}\right]$ denotes the average normalized (by the fracture height $H$) fluid flow rate through the fracture while $q_\text{L}$ $\left[\frac{\text{m}}{\text{s}}\right]$ stands for the leak-off function. 

The fracture propagation is propelled by injection of the generalized Newtonian fluid. The shear rate dependent apparent viscosity is denoted as $\eta_\text{a}(\dot \gamma)$ [Pa$\cdot$s], where $\dot \gamma$ [s$^{-1}$] is the fluid shear rate. For the apparent viscosity we accept the following general assumption concerning its limiting values:
\begin{equation}
\label{TP_def}
\eta_\text{a}=
  \begin{cases}
		\eta_0       & \quad \text{for } \quad |\dot \gamma |<|\dot \gamma_1|,\\	
    \eta_\infty  & \quad \text{for } \quad |\dot \gamma|>|\dot \gamma_2|,
  \end{cases}
\end{equation}
for some predefined magnitudes of $\dot \gamma_1$ and $\dot \gamma_2$. From the condition \eqref{TP_def} it follows that fluid viscosity reaches plateaus below and above some values of $\dot \gamma$. Condition \eqref{TP_def} does not detract from the generality of our analysis as, in those cases  where only one or no plateaus are present, one can set $|\dot \gamma_1|=0$ or/and $|\dot \gamma_2|=\infty$. The interim behaviour of $\eta_\text{a}(\dot \gamma)$ can be arbitrary, provided that a unique  relation (continuous or discrete) between $\eta_\text{a}$ and $\dot \gamma$ is defined.
In the paper by Wrobel \cite{Wrobel_Arxiv} an efficient algorithm of solution was proposed for the flow of the generalized Newtonian fluid in a flat channel. In the framework of the proposed scheme the fluid flow rate inside the fracture can be described by the following Poiseulle-like equation:
\begin{equation}
\label{Poiseulle}
q(x,t)=-\frac{1}{12 \eta_\infty}w^3\frac{\partial p}{\partial x} F(x,t),
\end{equation}
where $p(x,t)$ [Pa] is the fluid pressure and $F(x,t)$ is a dimensionless function defined as:
\begin{equation}
\label{F_def}
F(x,t)=-\frac{24\eta_\infty}{w^3}\left(\frac{\partial p}{\partial x} \right)^{-1}\int_0^{w/2}V(x,y,t)\text{d}y.
\end{equation}
In the above formula $V(x,y,t)$ $\left[\frac{\text{m}^2}{\text{s}}\right]$ describes the fluid velocity profile inside the fracture. Note that the velocity profile depends on the pressure gradient, the crack opening and the specific rheological properties of fracturing fluid. Thus, equation \eqref{F_def} is just a formal definition of $F(x,t)$, not a computational formula. As the full computational relation derived by Wrobel \cite{Wrobel_Arxiv} is rather complicated, we refer the reader to the cited paper and subsequent publications of the author \cite{Wrobel_2020,Wrobel_2021} for its complete form and instructions on numerical implementation.

 The Poiseulle-type structure of relation \eqref{Poiseulle} facilitates its implementation in the framework of the proposed scheme (see subsection \ref{gen_alg} for further details). Furthermore, in its physical interpretation the function $F(x,t)$ informs to what degree the solution in a certain spatial and temporal location deviates from the high shear rate Newtonian regime of flow. In particular:
\begin{equation}
\label{F_est}
F \to 1, \quad x \to a(t),
\end{equation}
which means that at the fracture tip the fluid behaves like a Newtonian fluid. Moreover, in the case of Newtonian fluid $F$ turns identically to unity.  
When applied in the PKN \cite{Wrobel_2020} and KGD \cite{Wrobel_2021} models, the algorithm based on equation \eqref{Poiseulle} proved to be an efficient and flexible tool to investigate the impact of fluid rheology on hydraulic fracture evolution.

For the fluid flow rate the following boundary conditions hold:
\begin{itemize}
\item{the tip boundary condition:
\begin{equation}
\label{BCs_tip}
\quad q(a,t)=0,
\end{equation}}
\item{the influx boundary condition:
\begin{equation}
\label{BC_q}
q(0,t)=q_0(t).
\end{equation}}
\end{itemize}

In computations we also employ a function of the average fluid velocity, $v(x,t)$ $\left[\frac{\text{m}}{\text{s}}\right]$, defined as:
\begin{equation}
\label{v_def}
v(x,t)=\frac{q}{w}.
\end{equation}
The advantages of using the fluid velocity as a dependent variable in numerical simulation of the HF problem are thoroughly demonstrated in the papers of Wrobel $\&$ Mishuris\cite{Wrobel_2015} and  Kusmierczyk et al.\cite{Kusmierczyk}. 

The hydraulically induced tangential traction exerted by fluid on the fracture walls, $\tau(x,t)$, is determined according to the following definition \cite{Wrobel_2017}:
\begin{equation}
\label{tau_def}
\tau(x,t)=-\frac{w}{2}\frac{\partial p}{\partial x}.
\end{equation}
When combining \eqref{tau_def} with \eqref{Poiseulle} and \eqref{v_def} one arrives at an alternative definition of $\tau$ which will be used later  in computations:
\begin{equation}
\label{tau_1}
\tau(x,t)=\frac{6 \eta_\infty v}{w F}.
\end{equation}

The so-called no lag assumption is adopted in our study which means that the crack tip coincides with the fluid front. This implies that the following Stefan-type condition has to be satisfied:
\begin{equation}
\label{SE}
v(a,t)=\frac{\text{d}a}{\text{d}t}.
\end{equation}
The above condition holds provided that the leak-off function is bounded at the crack tip. Otherwise (e.g. when the Carter leak-off model is used), equation \eqref{SE} should be modified to account for an additional leak-off dependent term.
In the framework of the employed algorithm formula \eqref{SE} is used to trace the fracture front. Description of the respective mechanism of the fracture front tracing can be found in the publication by Wrobel $\&$ Mishuris \cite{Wrobel_2015}.

\subsection{Solid mechanics equations}
\label{sol_mech}

In the classical formulation of the hydraulic fracture problem this group of equations describes the deformation of the fractured material under the applied hydraulic pressure. In numerous publications \cite{Wrobel_2017,Wrobel_2018,Papanastasiou_2018,Perkowska_2017,Wrobel_redirection} it was shown that also the tangential traction exerted by the fluid on the crack faces can be relevant to the HF process. In our analysis we take into account both types of the hydraulic loading.

The universal algorithm of computations first introduced by Wrobel $\&$ Mishuris  \cite{Wrobel_2015} has been used so far to analyse various aspects of the HF problem\cite{Wrobel_2017,Wrobel_2018,Wrobel_2021,Wrobel_redirection} for linear elastic model of the fractured material. The rock  deformation under the applied hydraulic loading was described by the boundary integral equation of elasticity. However, in many situations of interest for the modelling of HF, the problem of interaction between solid and fluid cannot be reduced to the boundary integral equation or such a reduction can be very problematic and/or computationally prohibitive. This may occur in the cases of e.g.  inelastic behaviour or non-uniform  properties of the fractured material, history dependent physical fields, directional or non-uniform confining stress. Moreover, in many applications a knowledge on the full 2D stress and strain fields is required. For such cases we employ in our algorithm the FEM based module for the solid mechanics equations constructed in the ABAQUS FEA software. The below description of the solid mechanics equations is general in its form as we do not restrict ourselves to any particular constitutive model from those available in the ABAQUS FEA library.

The general system of equations describing the deformation of the fractured material under the applied load should include the momentum balance equation, the constitutive relations that describe the material behaviour under applied load and the respective boundary conditions. The linear momentum balance equation reads:
\begin{equation}
\label{sigma_bal}
\nabla \cdot \mbox{$\bf \sigma$}+{\textbf{f}}=\bf{0},
\end{equation}
where $\bf{\sigma}$ is the stress tensor and $\bf{f}$ is the body force vector. In the case of porous elastic medium, the stress tensor in \eqref{sigma_bal} refers to the total stress.

A broad range of constitutive models for different materials are available in ABAQUS FEA \cite{ABAQUS}. These  models often consider elastic and inelastic response. The latter is most commonly described by various plasticity theories. In those models where inelastic deformation effects are taken into account, the elastic and inelastic responses are distinguished by separating the deformation into recoverable (elastic) and nonrecoverable (inelastic) parts. This separation is based on the assumption that there is an additive relationship between strain rates: 
\begin{equation}
\label{ep_dot}
\dot \varepsilon=\dot \varepsilon^\text{el}+\dot \varepsilon^\text{pl},
\end{equation}
where $\dot \varepsilon$ is the total strain rate, $\dot \varepsilon^\text{el}$ denotes the rate of change of the elastic strain, while $\dot \varepsilon^\text{pl}$ stands for the rate of change of inelastic strain. Equation \eqref{ep_dot} , with the rate of deformation employed to define the total strain rate, is used in all the ABAQUS plasticity models.

The elastic behaviour is usually described by the linear elasticity:
\begin{equation}
\label{lin_el}
\mbox{$\bf \sigma$} = \textbf{D}^\text{el} :\varepsilon^\text{el},
\end{equation}
where the elasticity matrix $\bf D^\text{el}$ may be temperature dependent but it does not depend on the deformation (unless such a dependence is introduced in the damage model). This elasticity model is intended to be used for small-strain
problems or to model the elasticity in an elastic-plastic model in which the elastic strains are always small.

The purely elastic response of material is limited to the region in which the yield function, $f$, has negative values:
\begin{equation}
\label{yield_f}
f(\mbox{$\bf \sigma$},T,H_\alpha)<0.
\end{equation}
In the above relation $T$ stands for temperature while $H_\alpha$ denotes the set of hardening parameters for a particular plasticity model. The post-yield behaviour (i. e. when $f=0$) is governed by the flow rule which in the general form can be expressed as:
\begin{equation}
\label{flow_rule}
\text{d} \varepsilon^\text{pl}=\text{d} \lambda \frac{\partial g}{\partial \mbox{$\bf \sigma$}},
\end{equation}
where $g(\mbox{$\bf \sigma$},T,H_\alpha)$ is the flow potential and $\text{d} \lambda$ is the so-called plastic multiplier determined from the consistency condition ($f=0$) .

The exact forms of respective constitutive relations and requirements concerning their application depend on the particular model chosen. A comprehensive description of various models can be found in the software documentation of ABAQUS FEA \cite{ABAQUS}. The respective boundary conditions imposed for the rock deformation component problem are detailed with the reference to the problem geometry in subsection \ref{FEM_module}.

The above system of the solid mechanics equations (in a particular form pertaining to the selected constitutive model) is solved by the ABAQUS package. The numerical implementation, in the framework of the proposed approach, amounts to specifying in the ABAQUS input file the solid deformation model and the material constants together with the respective boundary conditions and external loading (the fluid pressure, $p$, and tangential traction, $\tau$). Thus, we do not construct any in-house scheme for discretization and solution of the equations. The example of application of this technique can be found  in  the paper by Wrobel et al.\cite{Wrobel_plasticity} where the problem of hydraulic fracture propagating in elasto-plastic pressure sensitive material was analyzed for the the Mohr-Coulomb plastic deformation model.

\subsection{Crack propagation condition}

The flexibility of computational scheme introduced by Wrobel $\&$ Mishuris \cite{Wrobel_2015} facilitates its combination with different crack propagation conditions. So far it has been employed with the standard LEFM crack propagation condition \cite{Perkowska_2016,Wrobel_2015,Peck_2018_1,Peck_2018_2}, the crack propagation condition that accounts for the hydraulically induced tangential traction \cite{Wrobel_2017,Wrobel_2018,Wrobel_2021} and the elasto-plastic crack propagation condition \cite{Wrobel_plasticity}. Following these results let us introduce here a general form of the crack propagation condition:
\begin{equation}
\label{crack_gen}
K_I^2+\zeta K_I K_\text{f}=\left(\frac{K_{I\text{c}}}{\alpha}\right)^2,
\end{equation}
where $\zeta=4(1-\nu)$. In the above formula $K_I$ denotes the standard mode I stress intensity factor (SIF), $K_\text{f}$ is the so-called shear stress intensity factor related to the hydraulically induced tangential traction \cite{Wrobel_2017}, $K_{I\text{c}}$ stands for the material fracture toughness and $\alpha\leq1$ is a plasticity-dependent toughness magnification coefficient \cite{Wrobel_plasticity}. Particular cases can be derived from the condition \eqref{crack_gen} by: i) setting $\zeta=0$ for elimination of the tangential traction on the crack faces, ii) setting $\alpha=1$ for elimination of the plastic deformation effect \footnote{The plasticity dependent variant of the crack propagation condition is based on the concept of an effective fracture toughness  defined as $K_{I\text{c}}^\text{eff}=K_{I\text{c}}/\alpha$. The toughness magnification coefficient $\alpha<1$ is an element of solution which depends on the plastic deformation model, material constants and loading magnitude. For a detailed description of the condition and its implementation see the publication by Wrobel et al.\cite{Wrobel_plasticity}}. Note that if both of the above assumptions are made, condition \eqref{crack_gen} converts to the standard LEFM crack propagation condition ($K_I=K_{I\text{c}}$). For the details of derivation of condition \eqref{crack_gen} the reader is referred to the previous papers of the authors \cite{Wrobel_2017,Piccolroaz_2021,Wrobel_plasticity}.

\section{Computational algorithm}
\label{alg_desc}

The computational algorithm presented in this paper is based on the universal scheme for simulation of hydraulic fractures first introduced by Wrobel $\&$ Mishuris \cite{Wrobel_2015}. The algorithm originally proposed for the classic PKN and KGD models was later developed to account for non-Newtonian fluid rheologies \cite{Perkowska_2016,Wrobel_2020,Wrobel_2021}, hydraulically induced tangential traction \cite{Wrobel_2017,Wrobel_2018} and elasto-plastic model of the fractured material \cite{Wrobel_plasticity}. Moreover, its extension to the case of a penny shape fracture was done by Peck et al.\cite{Peck_2018_1,Peck_2018_2}. This iterative scheme of computations comprises two basic modules: i) the module to compute the fluid velocity from the continuity equation \eqref{cont}, ii) the module for computing the crack opening from the boundary integral equation of elasticity. An additional subroutine, based on equation \eqref{SE}, is employed for the fracture front tracing. The algorithm utilizes rigorous application of the solution tip asymptotics. For a detailed description of the techniques employed in the original scheme the  reader is directed to the above cited papers. Below we present a modification of the universal algorithm in which the FEM based module, instead of the one utilizing the boundary integral equation of elasticity,  is employed for the solid deformation. 

\subsection{The general scheme}
\label{gen_alg}

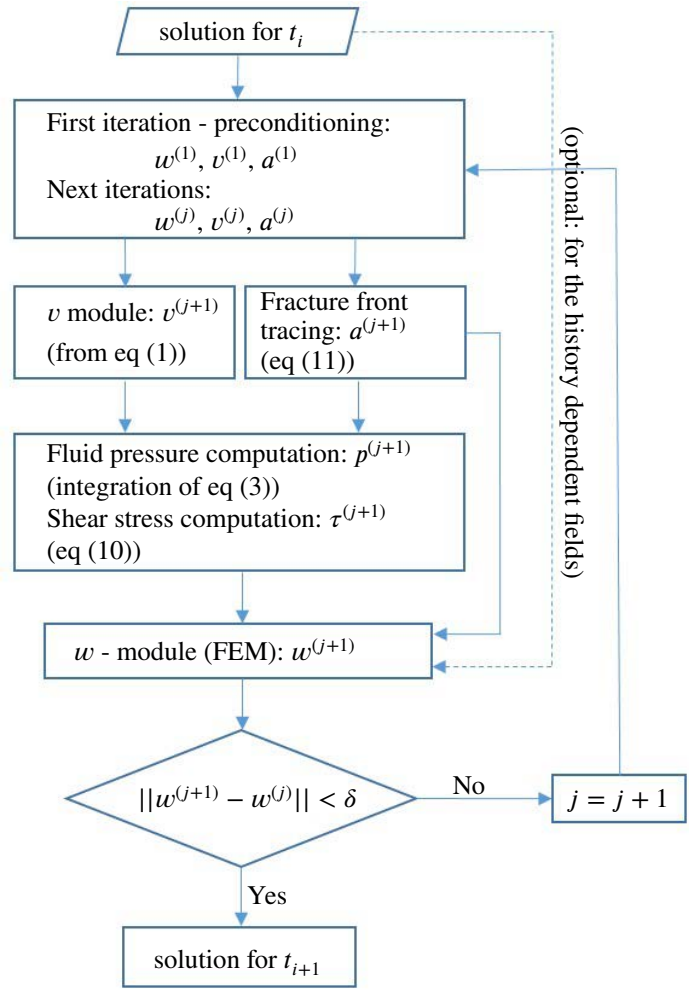
\begin{figure}[htb!]
\begin{center}
\includegraphics[scale=0.60]{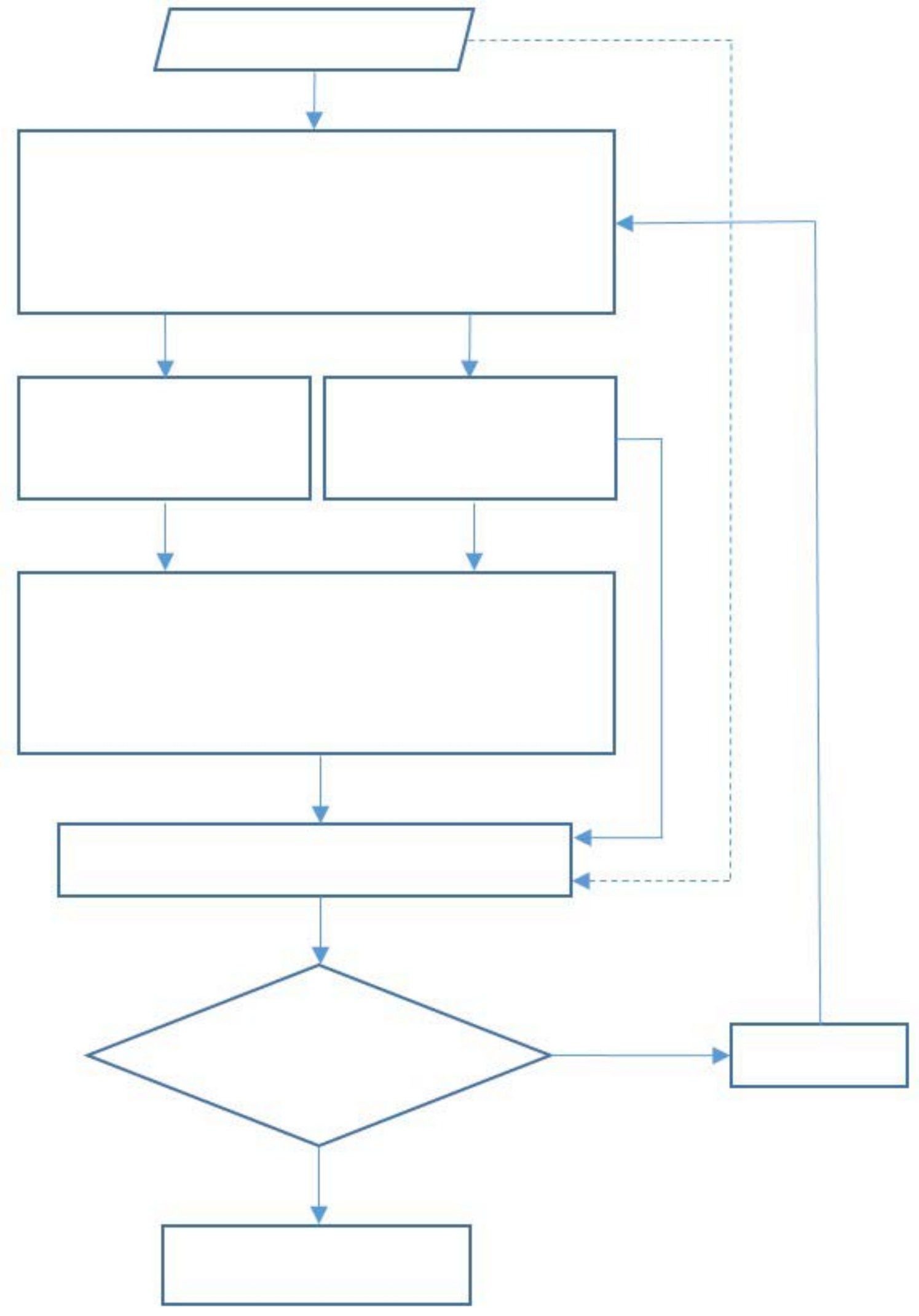}
\put(-200,357){solution for $t_i$}
\put(-242,323){First iteration - preconditioning:}
\put(-202,309){$w^{(1)}$, $v^{(1)}$, $a^{(1)}$}
\put(-242,298){Next iterations:}
\put(-202,286){$w^{(j)}$, $v^{(j)}$, $a^{(j)}$}
\put(-242,252){$v$ module: $v^{(j+1)}$}
\put(-242,237){(from eq \eqref{cont})}
\put(-162,256){Fracture front}
\put(-162,245){tracing: $a^{(j+1)}$}
\put(-162,234){(eq \eqref{SE})}
\put(-242,199){Fluid pressure computation: $p^{(j+1)}$}
\put(-242,187){(integration of eq \eqref{Poiseulle})}
\put(-242,175){Shear stress computation: $\tau^{(j+1)}$}
\put(-242,163){(eq \eqref{tau_1})}
\put(-232,125){$w$ - module (FEM): $w^{(j+1)}$}
\put(-208,70){$|| w^{(j+1)}-w^{(j)}||<\delta$}
\put(-167,34){Yes}
\put(-202,10){solution for $t_{i+1}$}
\put(-90,75){No}
\put(-47,70){$j=j+1$}
\put(-50,325){\rotatebox{-90}{(optional: for the history dependent fields)}}
\caption{The flow chart of the algorithm of computations for a single time step $\Delta t=t_{i+1}-t_i$. The iteration number is denoted by the superscript $j$.}
\label{algorytm}
\end{center}
\end{figure}

A flow chart for the general algorithm of solution for a single time step is depicted in Figure \ref{algorytm}. The computations are carried out in Matlab environment, with ABAQUS subroutine employed for the solid mechanics equations. The iterative computational scheme is constructed with application of two basic modules:
\begin{itemize}
\item{`$v$ module' - in this block the fluid velocity is computed from the continuity equation \eqref{cont}. The temporal derivative of the crack opening is approximated by an enhanced second-order scheme (comments on its advantage in numerical simulations can be found in the paper by Wrobel $\&$ Mishuris  \cite{solver_calkowy}):  
\[
\frac{\partial w}{\partial t}\Big |_{t_{i+1}}=2\frac{w(x,t_{i+1})-w(x,t_{i})}{t_{i+1}-t_i}-\frac{\partial w}{\partial t}\Big |_{t_i}.
\]
Note that the above representation is equivalent to the classical Crank-Nicolson method, however its numerical implementation differs from the latter. The leading singular asymptotic terms of the governing equation are canceled out analytically and the equation is reduced to an integral relation. A detailed description of the respective transformations and the final form of the integral operator to compute $v$ can be found in the original publication by Wrobel $\&$ Mishuris\cite{Wrobel_2015}.  }
\item{`$w$ module' - in this block of the algorithm the FEM subroutine constructed in the ABAQUS package is used to compute the stress and displacement fields. The crack opening, $w$, is extracted from the 2D displacement field. A comprehensive description of the $w$ module is provided in the next subsection.}
\end{itemize}

In general, the spatial discretization over the fracture surface is different in the respective blocks. The mesh employed in the $v$ module is selected in a way to be optimal for the continuity equation \eqref{cont}.  Mesh density is increased at the inlet and at the tip of the crack according to the scheme proposed by Wrobel $\&$ Mishuris \cite{solver_calkowy}. On the other hand, the spatial discretization of the fracture surface in the $w$ module results directly from the employed mesh of finite elements (see Subsection \ref{FEM_module}). Thus, the computed values of $w(x,t_{i+1})$ need to be interpolated in the nodes of the former mesh before being reintroduced to the $v$ module. In the interpolation process the leading asymptotic terms of $w$ are singled out analytically so that the numerical mapping is used only for the remaining part of the crack opening. In this way, high quality of solution in the near-tip zone can be retained.

Additional auxiliary subroutines employed in the algorithm include the fracture front tracing module and the block for computing the fluid pressure. The former utilizes the solution tip asymptotics together with equation \eqref{SE}. Technical implementation of this concept and discussion on its advantages can be found in the paper by Wrobel $\&$ Mishuris \cite{Wrobel_2015}. The fluid pressure is obtained by integration of the pressure derivative resulting from the transformed form of equation \eqref{Poiseulle}. Also at this stage, the leading singular terms of pressure derivative are singled out and integrated analytically, which facilitates accuracy and stability of computations. The hydraulically induced tangential traction on the fracture walls is obtained from the formula \eqref{tau_1}.
Function $F(x,t)$ is computed numerically according to the algorithm proposed by Wrobel \cite{Wrobel_Arxiv}. In each iteration, the entries of the subroutine for $F(x,t)$ are taken from the previous iterative step.   Moreover, when the problem of history dependent physical fields is considered (e.g. plastic deformation of the fractured material is accounted for) the final solution from the time step $t_i$ is mapped to the initial state in the time instant $t_{i+1}$, as shown in the flow chart. In the case where the pore fluid pressure in the rock is taken into account, the pore pressure field produced by the FEM subroutine is used to compute the leak-off function, $q_\text{L}$. The values of $q_\text{L}$ are then used in the $v$-module. The overall iterative process continues as long as the relative difference between two consecutive approximations of the crack opening, $|| w^{(j+1)}-w^{(j)}||$, is greater than some predefined threshold value $\delta$.

\subsection{FEM module}\label{FEM_module}

In the following we explain construction of the `$w$ module' employed to solve the solid mechanics equations.

\begin{figure}[htb!]
\begin{center}
\includegraphics[scale=0.37]{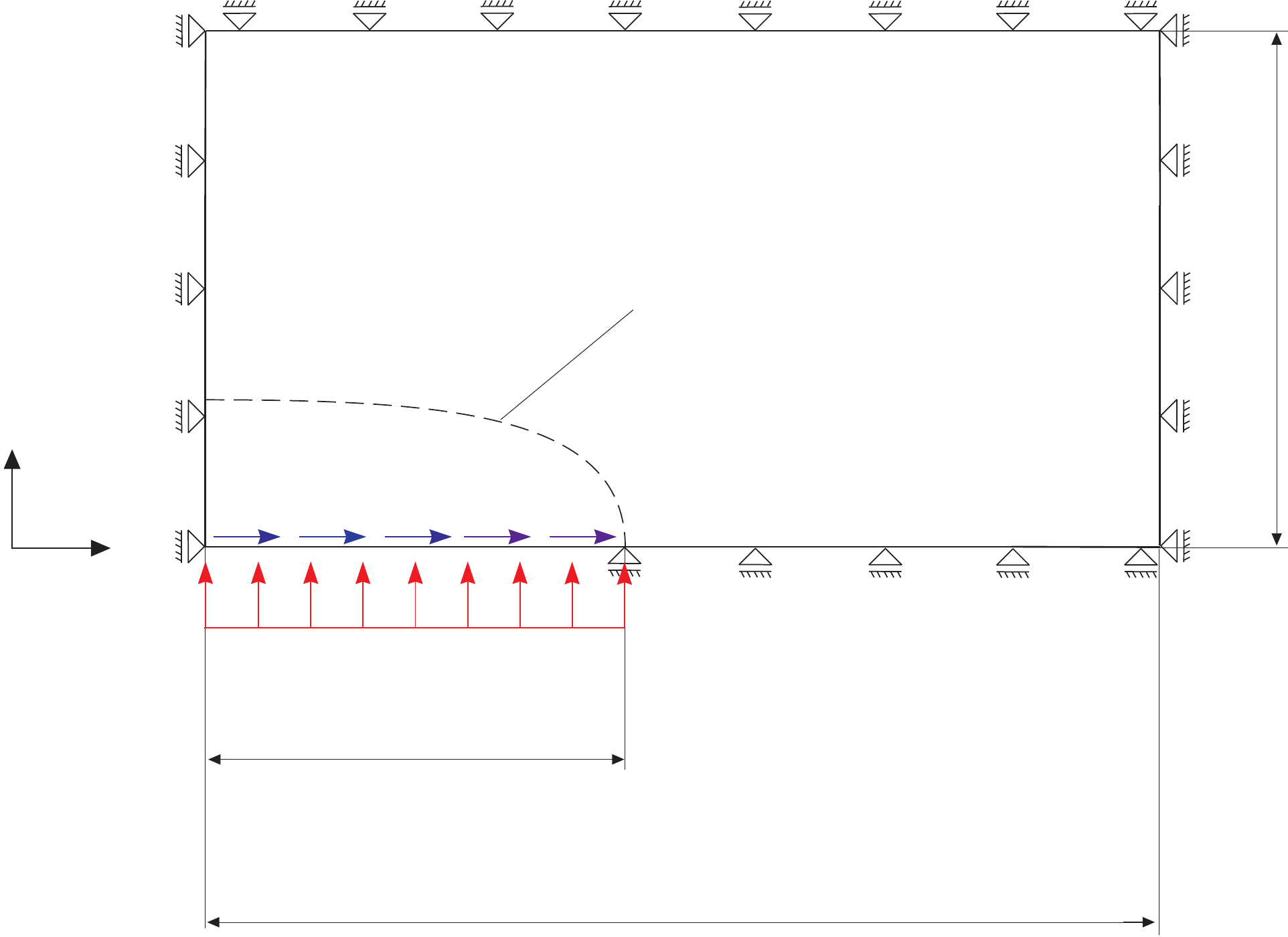}
\put(-193,65){$\tilde x$}
\put(-201,72){$\tilde y$}
\put(-10,97){\rotatebox{90}{100}}
\put(-155,40){$p(\tilde x,t)$}
\put(-155,68){$\tau(\tilde x,t)$}
\put(-142,29){1}
\put(-105,5){$101$}
\put(-120,105){$ \frac{1}{2}w(\tilde x,t)$}
\hspace{0mm}
\includegraphics[scale=0.37]{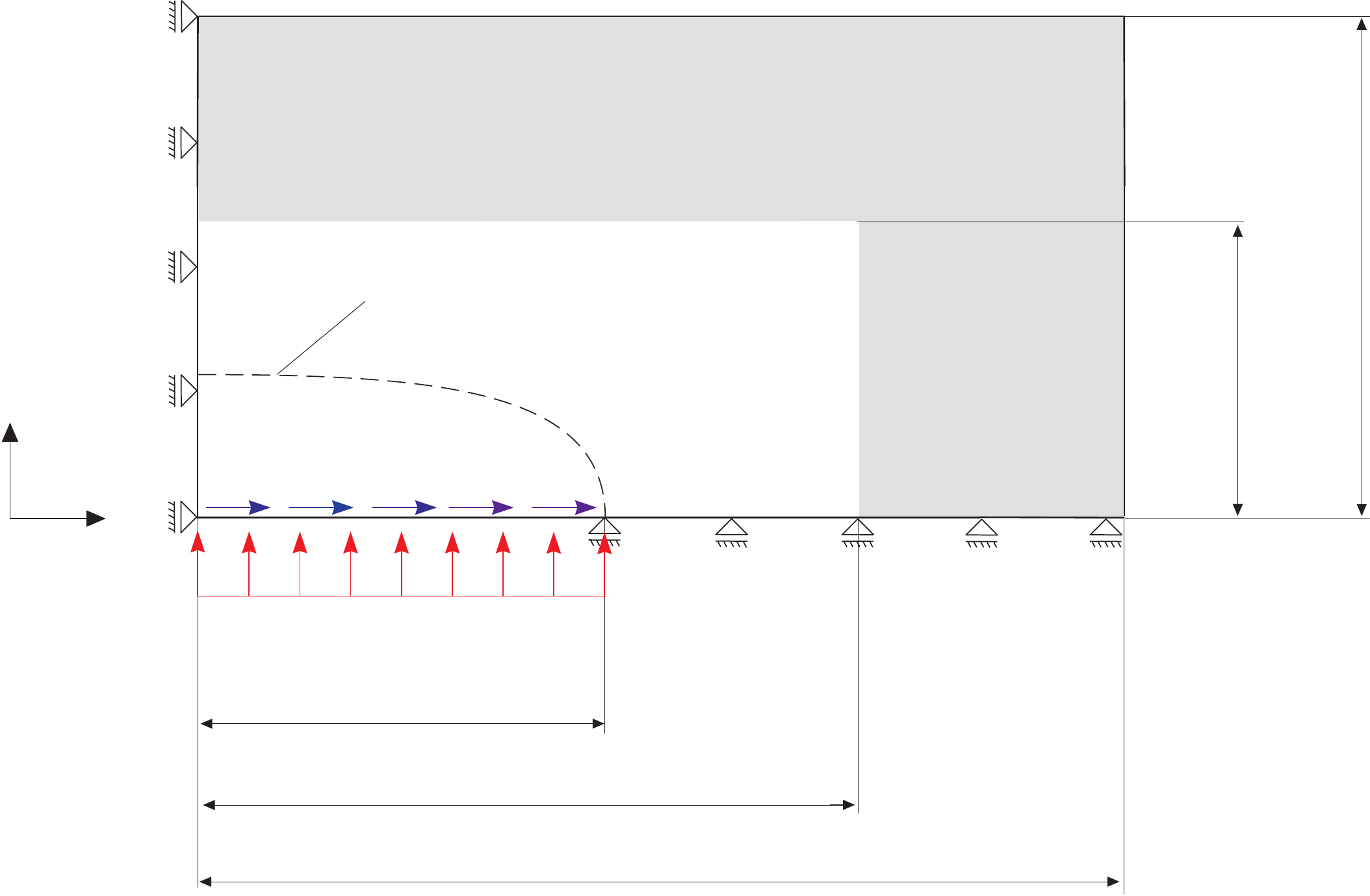}
\put(-213,65){$\tilde x$}
\put(-221,72){$\tilde y$}
\put(-120,5){$21$}
\put(-145,17){$11$}
\put(-160,29){$1$}
\put(-173,40){$p(\tilde x,t)$}
\put(-173,68){$\tau(\tilde x,t)$}
\put(-178,100){$ \frac{1}{2}w(\tilde x,t)$}
\put(-10,97){\rotatebox{90}{20}}
\put(-30,80){\rotatebox{90}{10}}
\put(-440,155){$\textbf{a)}$}
\put(-225,155){$\textbf{b)}$}
\caption{A sketch of the pattern geometry of the FEM problem in the normalized variables, $\tilde x=x/a$, $\tilde y=y/a$, for: a) the case where only the plane strain elements CPE8R are used, b) the case where the plane strain elements CPE8R are combined with infinite plane strain elements CINPE4 (the shaded area).}
\label{FEM_geometry}
\end{center}
\end{figure}
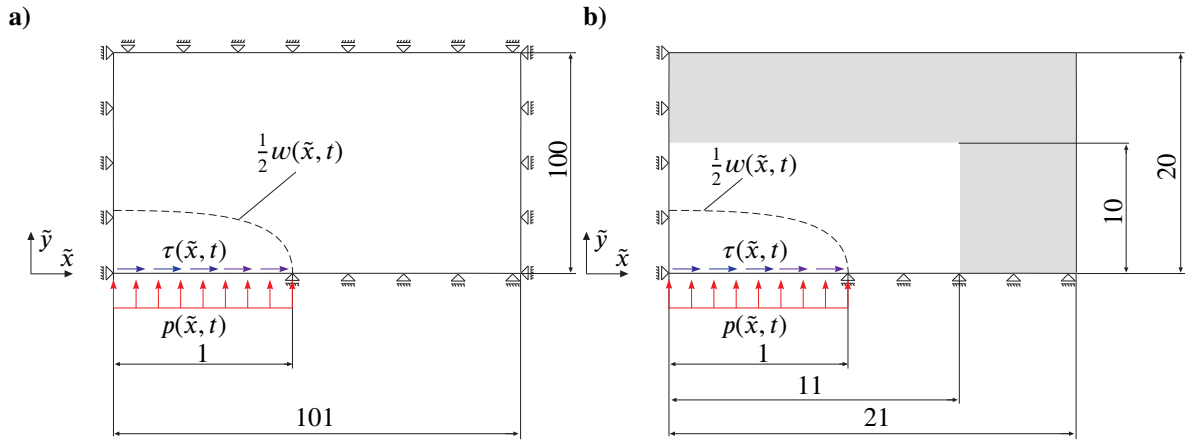

Let us consider the problem of a stationary crack of length $a$ under plane strain conditions schematically depicted in Figure \ref{FEM_geometry}. The crack is loaded by the normal pressure, $p$, and the hydraulically induced tangential traction, $\tau$. Due to the problem symmetry we consider only one quadrant of the original domain. This problem is solved with FEM in the framework of ABAQUS FEA software. The interaction with Matlab environment is provided by the \textit{Abaqus2Matlab} interface \cite{Abaqus2Matlab}. The full 2D solution in the plane $(x,y)$ is used to retrieve the crack opening profile, $w$, which is employed in the main computational scheme. 

In the analysis we will use two arrangements of the basic FEM problem.  In the first one, shown in Figure \ref{FEM_geometry}a), no infinite elements are employed. Consequently, the proportions of the external domain dimensions to the crack length have to be sufficiently large to mimic the infinite size of the original (physical) domain. We set this ratios to 1/101 and 1/100 in the $x$ and $y$ directions, respectively. In every time step the pattern FEM geometry (as it is shown in Figure \ref{FEM_geometry}) is rescaled by the value of crack length $a(t)$. In computational implementation this rescaling amounts to a simple multiplication of the nodal coordinates in the ABAQUS input file by a constant. Thus, the whole process is efficient as there is no need to use the mesh generator. The symmetry boundary conditions are specified along the edges $x=0$ and $y=0, \ x>1$.  On the external boundaries of the computational domain we block the translations in $x$ and $y$ directions, correspondingly, so that constant values of respective confining stress components can be imposed (i.e. the field of confining stress does not introduce initial deformation of the domain). In this variant of the problem only the eight-node bi-quadratic plane strain elements (CPE8R) are used. 

In the second analyzed configuration of the FEM problem (Figure \ref{FEM_geometry}b)) we use infinite elements. Thus, the overall size of the computational domain can be reduced with respect to the first variant. This time the ratios of the crack length to the external dimensions of the domain are: 1/21 along the $x$  axis, and 1/20 along the $y$ axis. As previously, the respective symmetry boundary conditions hold. However, there is no need to block translations at the external boundaries as the infinite elements already assume zero displacements in infinity. Again, the confining stress is imposed in a form of a predefined field. For the internal part of the domain we use CPE8R elements, while the external part is discretized with a layer of four-node plane strain infinite elements (CINPE4). 
This version of the FEM problem configuration can be employed to establish whether the span of the computational domain in the first variant is sufficient to mimic the infinite size of the physical domain. 

The ABAQUS software does not allow mapping of the solution between different meshes  if the infinite elements are in use. Thus, when applying our solver to the problem of history dependent physical fields (e.g. elasto-plastic hydraulic fractures), which requires mapping the final results from the time step $t_i$ to the initial state of the time instant $t_{i+1}$, only the first variant of the FEM module (without infinite elements) can be employed.

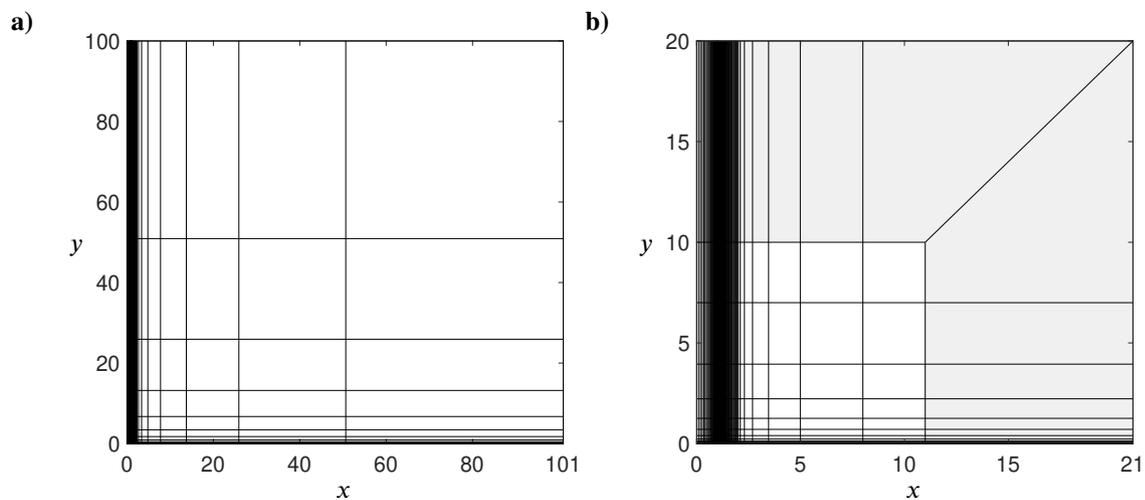
\begin{figure}[htb!]
\begin{center}
\includegraphics[scale=0.40]{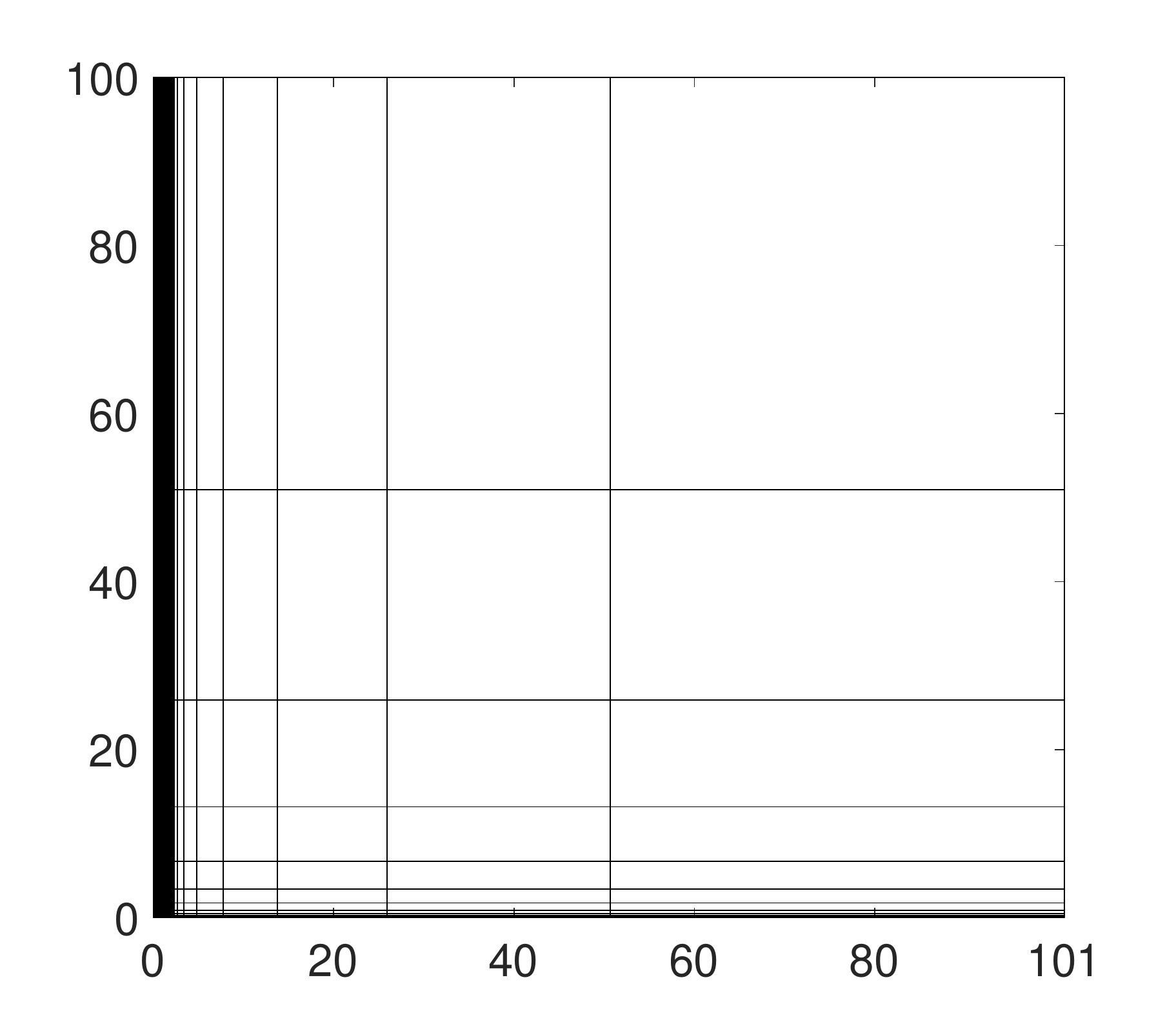}
\put(-205,92){$y$}
\put(-105,0){$x$}
\hspace{0mm}
\includegraphics[scale=0.40]{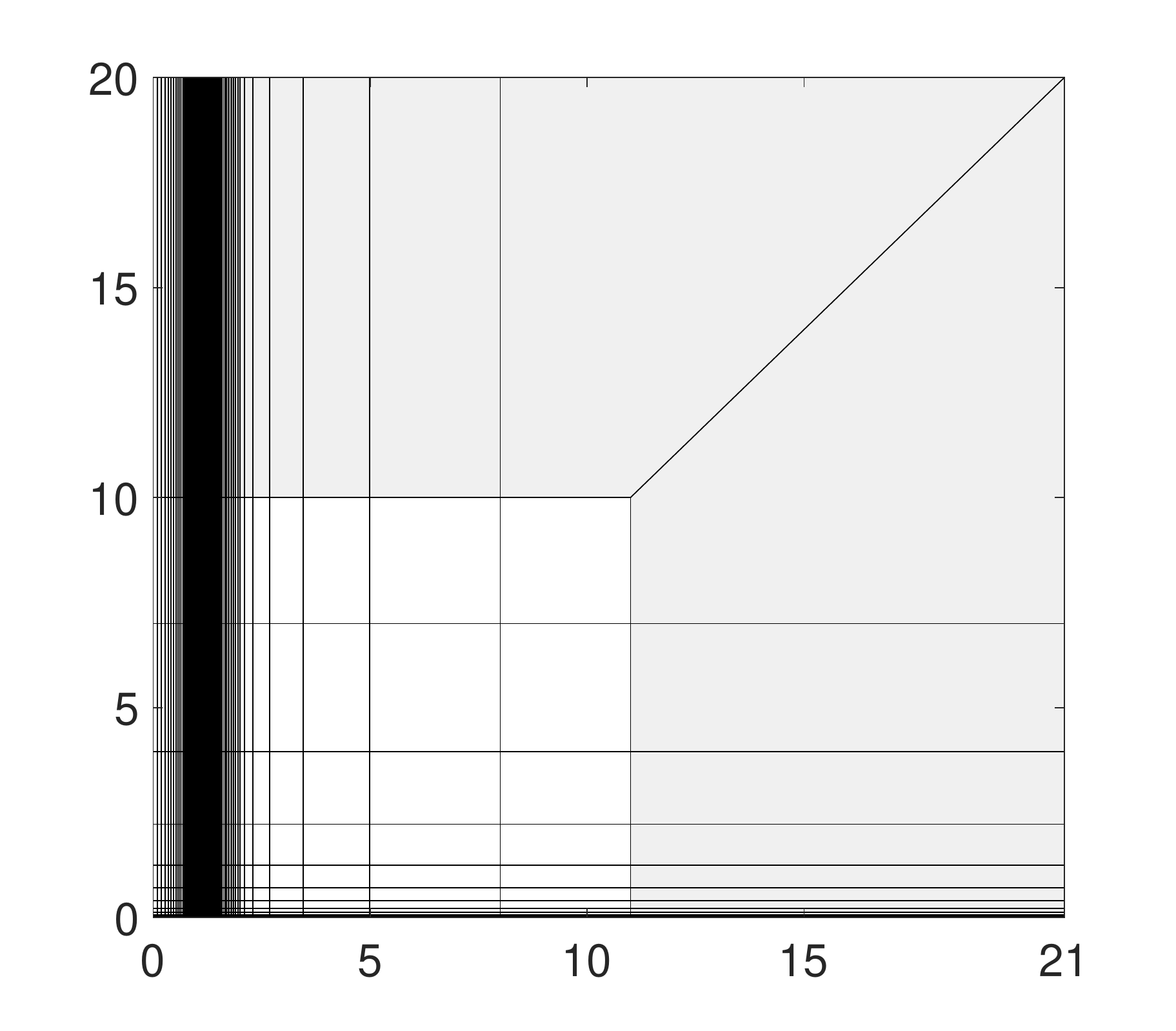}
\put(-105,0){$x$}
\put(-205,92){$y$}
\put(-440,175){$\textbf{a)}$}
\put(-225,175){$\textbf{b)}$}
\caption{A mesh of finite elements, for: a) the case where only the plane strain elements CPE8R are used, b) the case where the plane strain elements CPE8R are combined with infinite plane strain elements CINPE4 (the shaded area). In both cases the mesh density is increased (in the same way) near the fracture tip.}
\label{FEM_mesh}
\end{center}
\end{figure}

In both variants of the FEM problem configuration a regular mesh pattern  is used (see Figure \ref{FEM_mesh}). The mesh density is increased near the fracture tip. This is necessary not only to properly capture the steep gradients of respective fields but also to provide sufficiently good approximation of the fracture profile near the crack front. The importance of the latter requirement results from the fact that the universal algorithm proposed in by Wrobel $\&$ Mishuris \cite{Wrobel_2015} and adopted in our present study relies heavily on the solution  behavior (in terms of $w$ and $v$) in the near tip zone. The chosen mesh pattern  facilitates the control of mesh density over the fracture surface and interpolation of the results between the respective modules of the algorithm. Consequently, the aspect ratio of the finite elements adjacent to the plane $x = a$ becomes very high away from the fracture plane. On the other hand, the elements with high aspect ratio do not undergo large deformations.
Thus, in our algorithm the aforementioned drawback does not negatively affect the accuracy and stability of
computations, which will be shown below. Depending on the type of analyzed problem (elastic HF, elasto-plastic HF, etc.) the crack tip element is adjusted to reflect a desired singular behaviour of the stress and strain fields. In ABAQUS implementation one can choose between: i) the square root singularity, $\varepsilon \sim r^{-1/2}$, for linear elasticity, ii) the reciprocal type singularity, $\varepsilon \sim 1/r$, for perfect plasticity, and iii) the power law singularity, $\varepsilon \sim r^{-n/(n+1)}$, for the power-law hardening associative materials ($\varepsilon$ stands for the strain and $r$ is the distance from the crack tip). For non-associative materials the power law singularity is also a function of the degree of non-associativity for both, the Drucker-Prager material \cite{Papanastasiou_2001,Durban_2003} and  for the Mohr-Coulomb material \cite{Papanastasiou_2018a}.  It should be noted that the algorithm is  not particularly sensitive to the specified type of the tip element. In other words, even when the FEM solution singularity is chosen incorrectly, the computation of crack opening can still be very accurate. It results from the fact that in the proposed algorithm the FEM sub-problem is a problem of a stationary crack of a known length (the crack length is computed by an external subroutine - compare the flow chart in Figure \ref{algorytm}). Thus, the strain field in the immediate vicinity of the crack tip does not directly affect the fracture extension and stability of computations. Furthermore, fine FEM meshing near the fracture tip yields good approximation of  strain even if the tip element is selected inappropriately, provided that the respective boundary conditions are preserved (see e.g. the study by Wrobel $\&$ Mishuris \cite{Wrobel_2009} where the square root singularity was approximated by proper handling of linear finite elements).  

\section{Numerical analysis}
\label{num_an}

In this section we present a verification of the accuracy and efficiency of computations performed by the proposed algorithm. Then, a computational example concerning hydraulic fracture crossing an interface between two dissimilar rock layers is presented.

\begin{figure}[htb!]
\begin{center}
\includegraphics[scale=0.40]{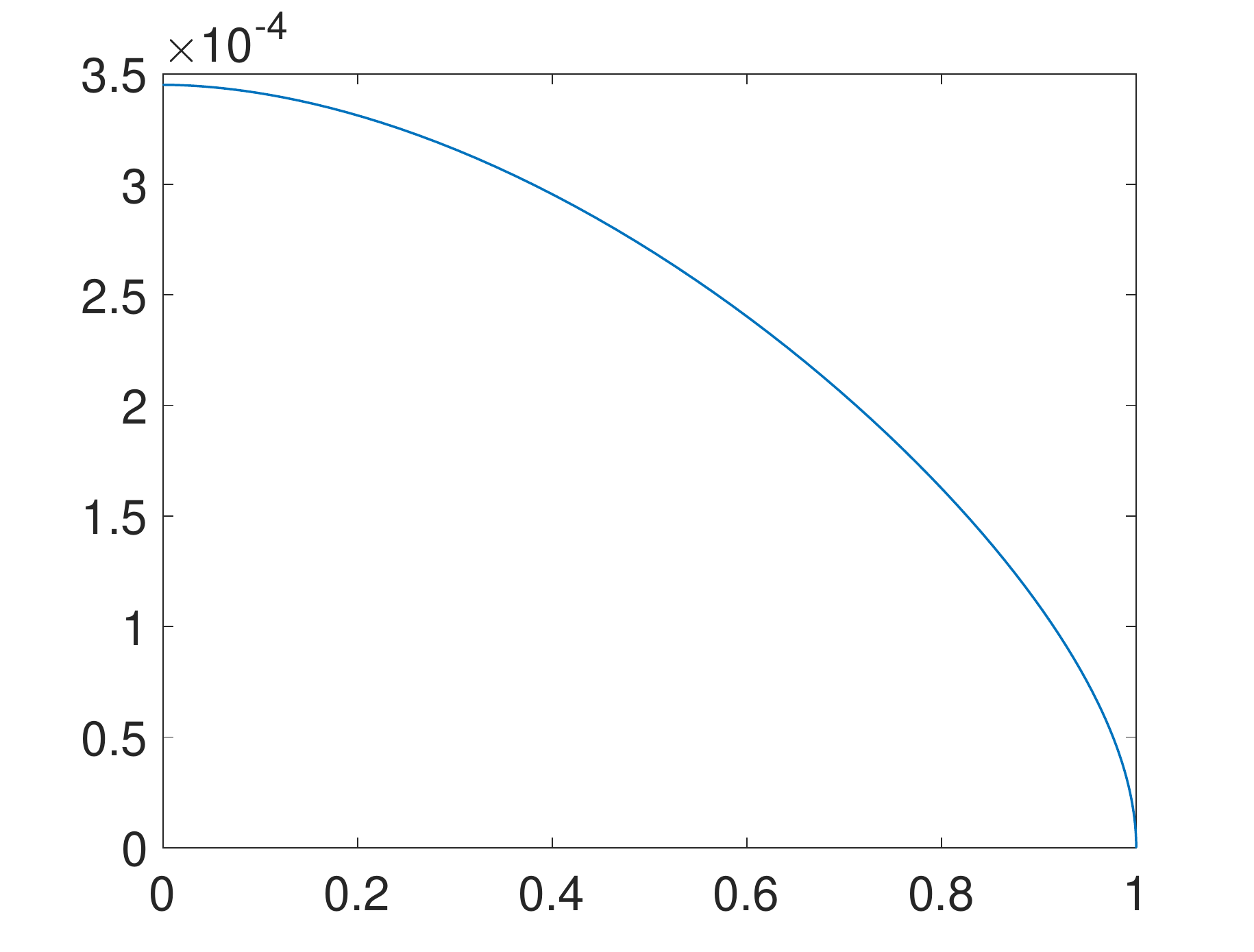}
\put(-215,80){$w_\text{b}$}
\put(-105,0){$\tilde x$}
\hspace{0mm}
\includegraphics[scale=0.40]{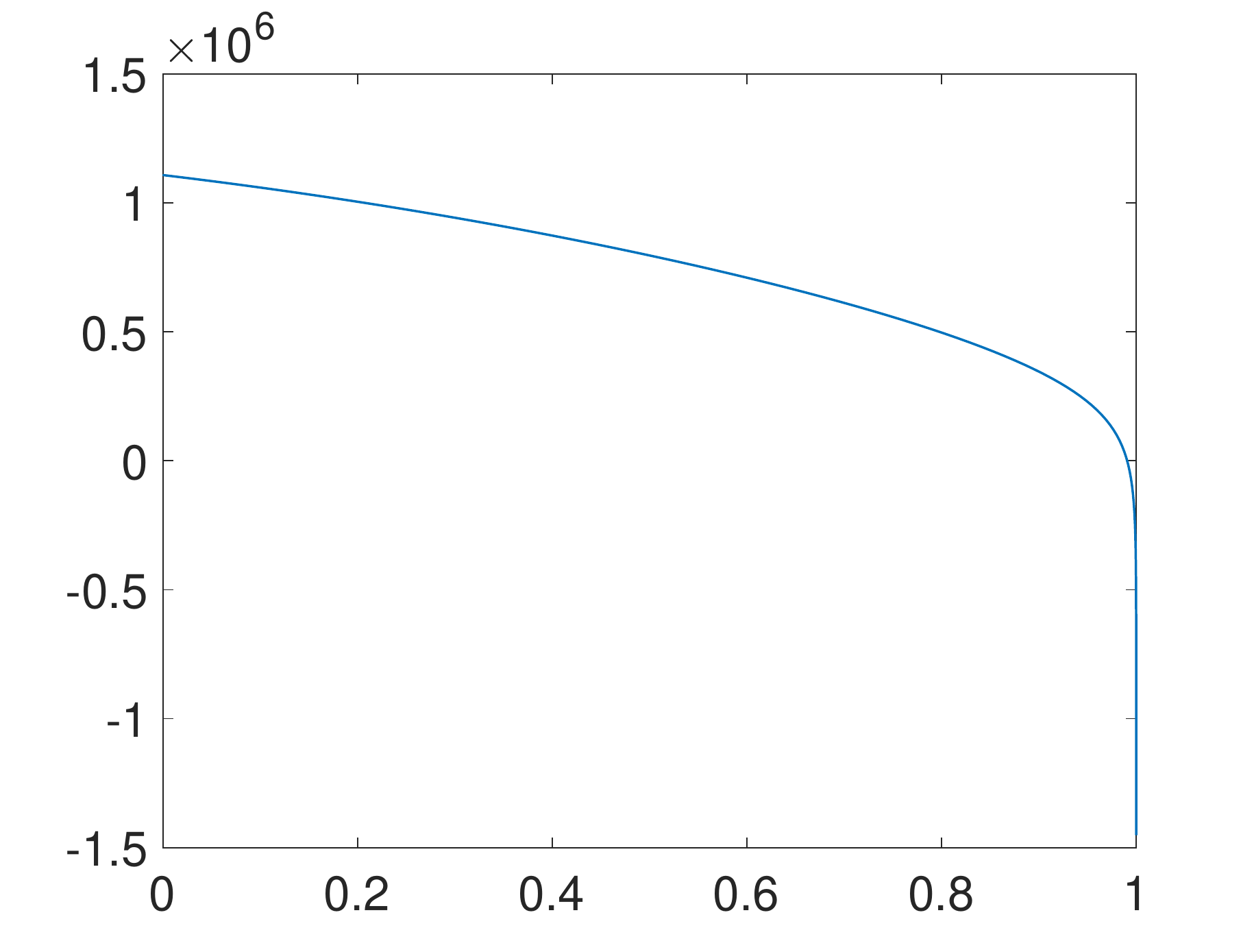}
\put(-105,0){$\tilde x$}
\put(-215,80){$p_\text{b}$}
\put(-440,160){$\textbf{a)}$}
\put(-225,160){$\textbf{b)}$}
\caption{Analytical benchmark solution at the time instant $t=0$ s: a) the crack opening, $w_\text{b}$ [m], b) the fluid pressure, $p_\text{b}$ [Pa]. The scaled spatial variable $\tilde x=x/a$ is used.}
\label{w_p_bench}
\end{center}
\end{figure}
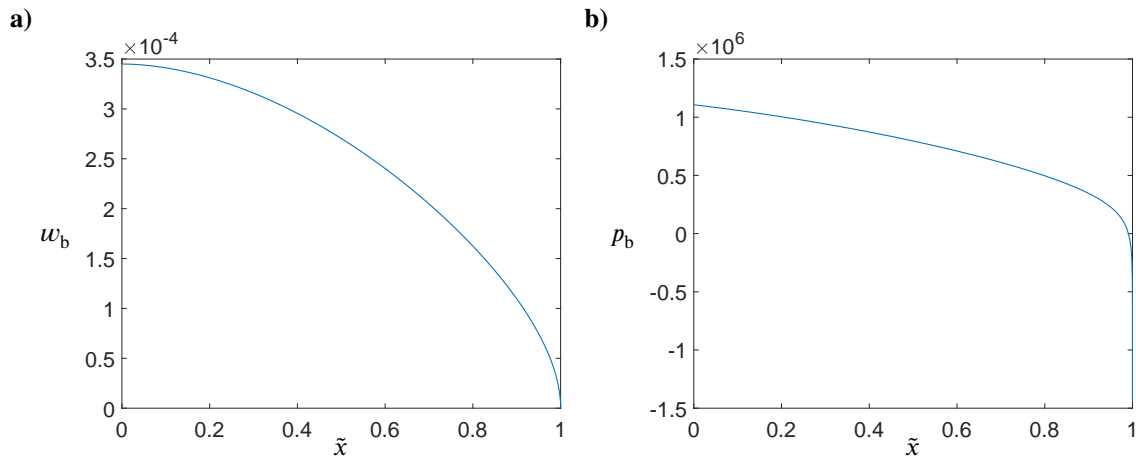

The accuracy of computations is verified against the analytical benchmark solution described in Appendix \ref{ap_B}. It involves a simplified version of the general HF formulation considered in this paper. Namely, linear elastic behaviour of the fractured material is assumed together with a constant fluid viscosity, $\eta$. Moreover, the hydraulically induced tangential traction on the fracture walls, $\tau$, is neglected. The material constants used in the computations are: $E=16.2$ GPa, $\nu=0.3$ and $\eta=10^{-3}$ Pa$\cdot$s. The values of respective multipliers in the benchmark representation of the self-similar crack opening \eqref{w_bench} are: $\hat w_0=5.67\cdot 10^{-4}$, $\hat w_1=2.05\cdot 10^{-4}$, $\hat w_2=2\cdot 10^{-5}$, $\hat w_3=7.31\cdot 10^{-4}$. This choice of coefficients means that the leak-off function is suppressed at the fracture tip. The coefficient of temporal evolution, $\beta$, from \eqref{sel_sim_scl} is set to 1/3. The spatial distributions of: i) the benchmark crack opening, $w_\text{b}$, ii) the benchmark fluid pressure, $p_\text{b}$, for $t=0$ s are depicted in Figure \ref{w_p_bench}.

\subsection{Verification of the FEM module}
\label{ver_FEM}

In this subsection we verify the accuracy of the FEM module alone. To this end we analyze a component problem of the stationary fracture, i.e. the problem shown schematically in  Figure  \ref{FEM_geometry}. As mentioned above, the tangential traction is neglected ($\tau=0$ Pa). The imposed fluid pressure complies with the benchmark example from  Appendix \ref{ap_B} for $t=0$ s (the distribution of $p$ is shown in Figure \ref{w_p_bench}b)). Consequently, the resulting benchmark crack opening is the one depicted in Figure \ref{w_p_bench}a).

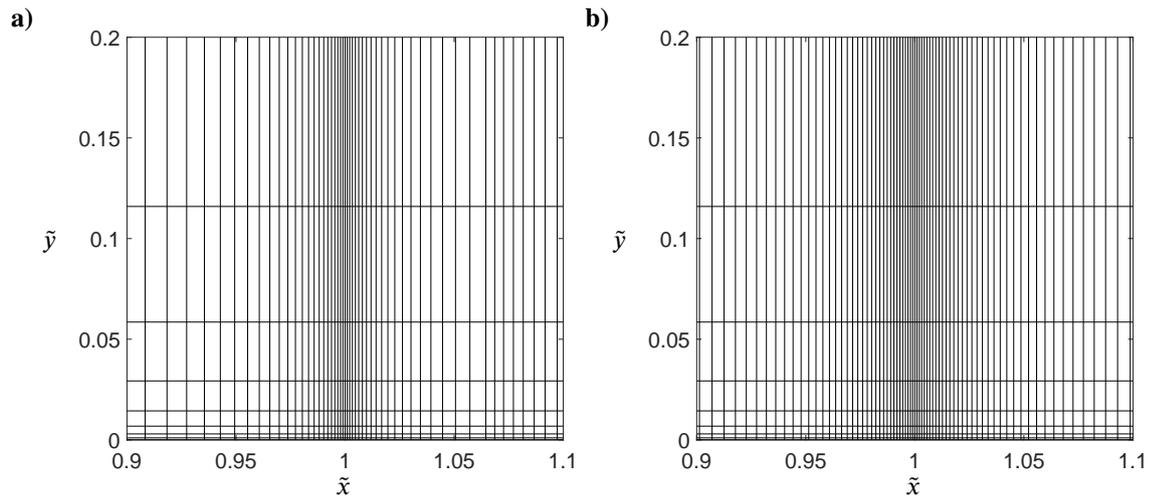
\begin{figure}[htb!]
\begin{center}
\includegraphics[scale=0.40]{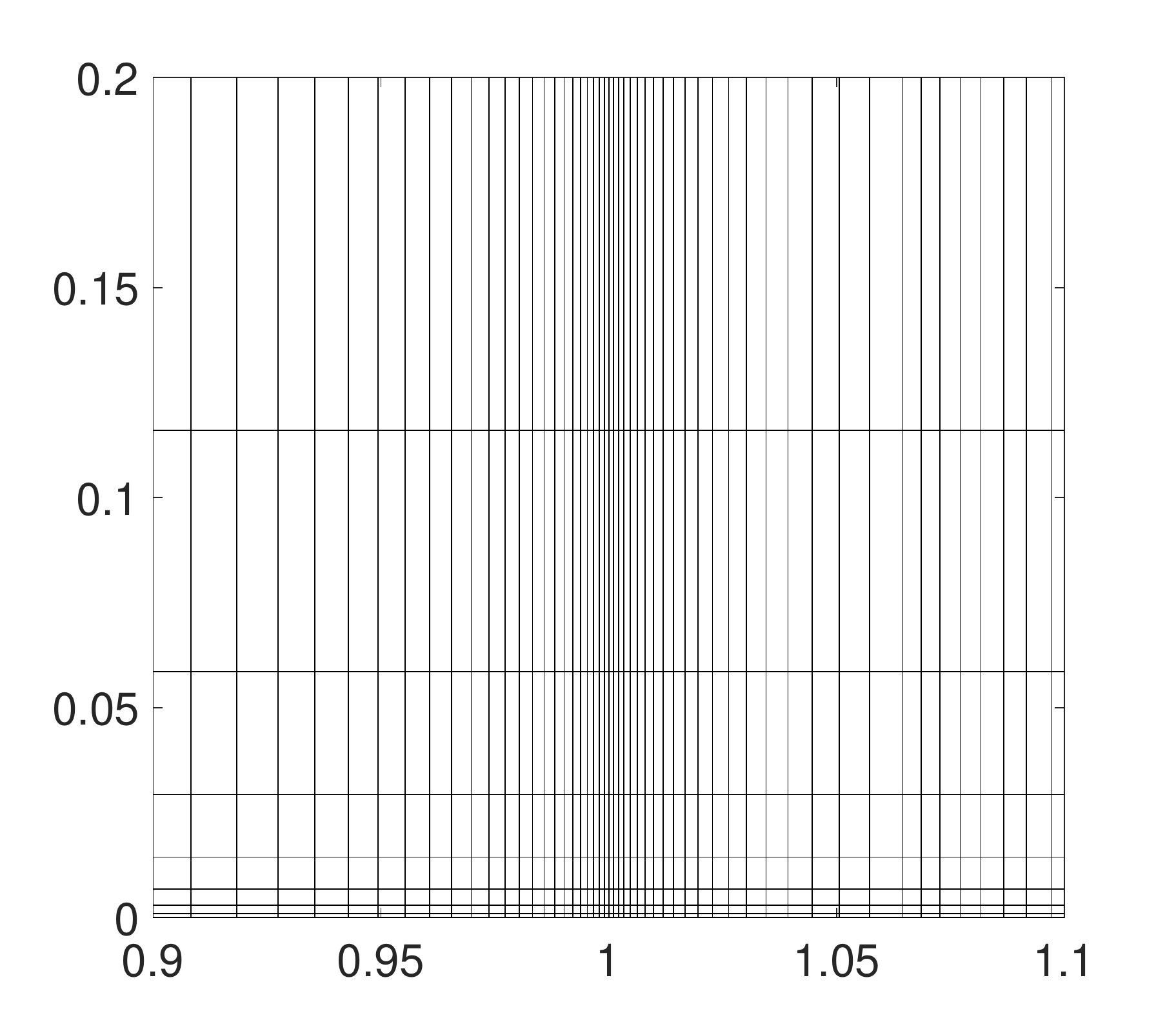}
\put(-215,92){$\tilde y$}
\put(-105,0){$\tilde x$}
\hspace{0mm}
\includegraphics[scale=0.40]{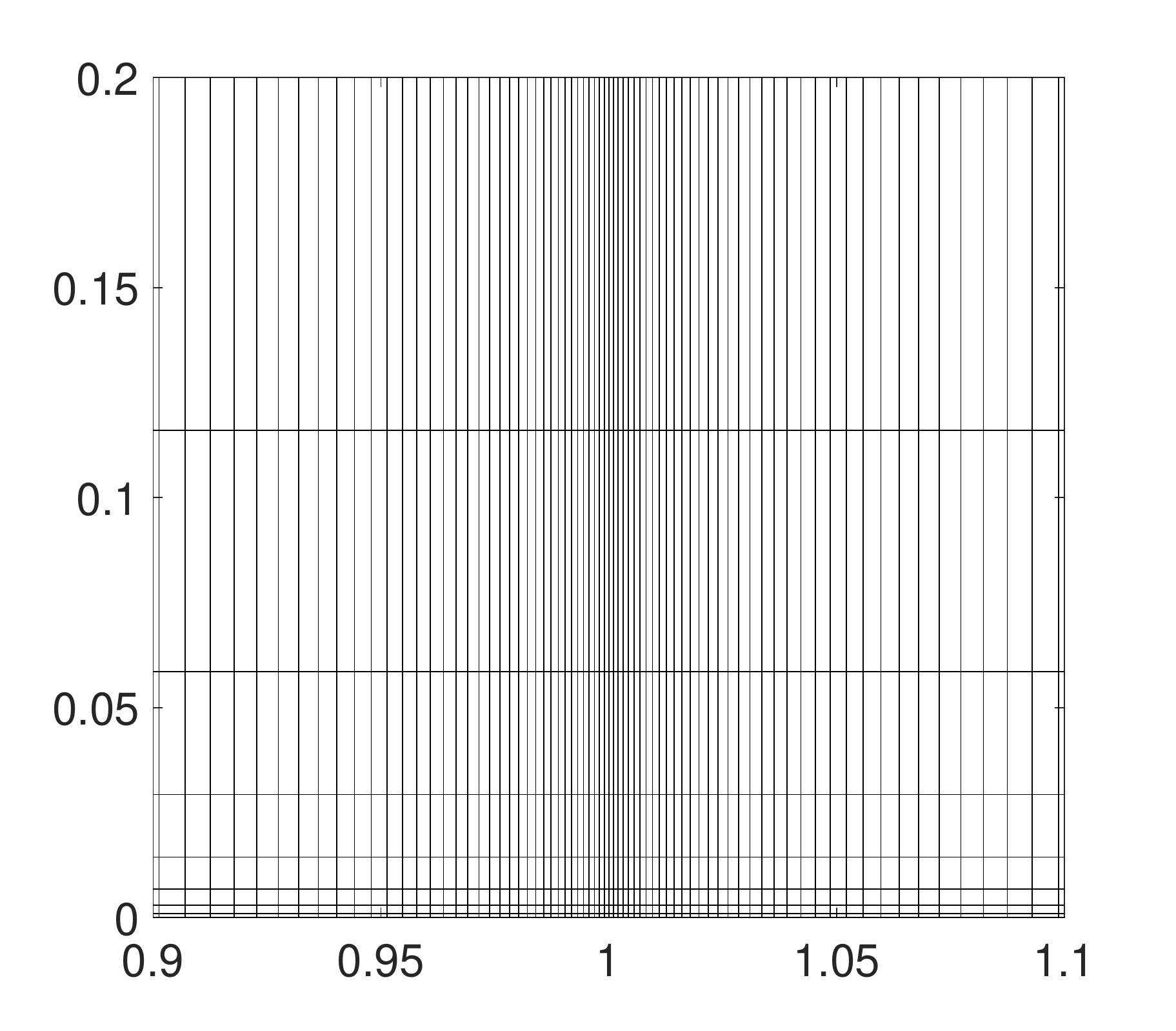}
\put(-105,0){$\tilde x$}
\put(-215,92){$\tilde y$}
\put(-440,175){$\textbf{a)}$}
\put(-225,175){$\textbf{b)}$}
\caption{Mesh of finite elements near the fracture tip: a) coarse mesh, b) dense mesh. The scaled spatial variables $\tilde x=x/a$ and $\tilde y=y/a$ are used.}
\label{mesh_tip}
\end{center}
\end{figure}

We will consider both  variants of the FEM problem configuration described in the Subsection \ref{FEM_module} with the following nomenclature in use:
\begin{itemize}
\item{configuration of the pattern geometry from  Figure \ref{FEM_geometry}a) will be named `variant 1',}
\item{ configuration from  Figure \ref{FEM_geometry}b)  (with the infinite elements) will be called `variant 2'.} 
\end{itemize}
Additionally, for each of these variants two different mesh densities will be analyzed. The mesh of lower density will be named the `coarse mesh', while the mesh with the increased density  will be called the `dense mesh'. The coarse mesh provides 91 nodes over the crack face whereas the corresponding number for the fine mesh is 159.  The near tip  discretization pattern for these two cases (symmetrical with respect to the line $\tilde x=x/a=1$) is shown in  Figure \ref{mesh_tip}. The tabular information on the employed meshes is presented in the Table \ref{mesh_tab}. The relative errors of solution are computed with respect to the benchmark solution \eqref{w_bench}.

\begin{table}[]
\begin{center}
\begin{tabular}{cc|c|c|c|}
\cline{3-5}
\multicolumn{2}{c|}{}                                          & \begin{tabular}[c]{@{}c@{}}Number\\ of elements\end{tabular} & \begin{tabular}[c]{@{}c@{}}Number\\ of nodes\end{tabular} & \begin{tabular}[c]{@{}c@{}}Number\\ of nodes\\ at the crack\\ face\end{tabular} \\ \hline
\multicolumn{1}{|c|}{\multirow{2}{*}{variant 1}} & coarse mesh & 1751                                                         & 5459                                                      & 91                                                                              \\ \cline{2-5} 
\multicolumn{1}{|c|}{}                           & dense mesh & 2907                                                         & 9098                                                      & 159                                                                             \\ \hline
\multicolumn{1}{|c|}{\multirow{2}{*}{variant2}}  & coarse mesh & 1665                                                         & 4997                                                      & 91                                                                              \\ \cline{2-5} 
\multicolumn{1}{|c|}{}                           & dense mesh & 2821                                                         & 8465                                                      & 159                                                                             \\ \hline
\end{tabular}
\caption{Considered configurations of the mesh of finite elements.}
\label{mesh_tab}
\end{center}
\end{table}

\begin{figure}[htb!]
\begin{center}
\includegraphics[scale=0.40]{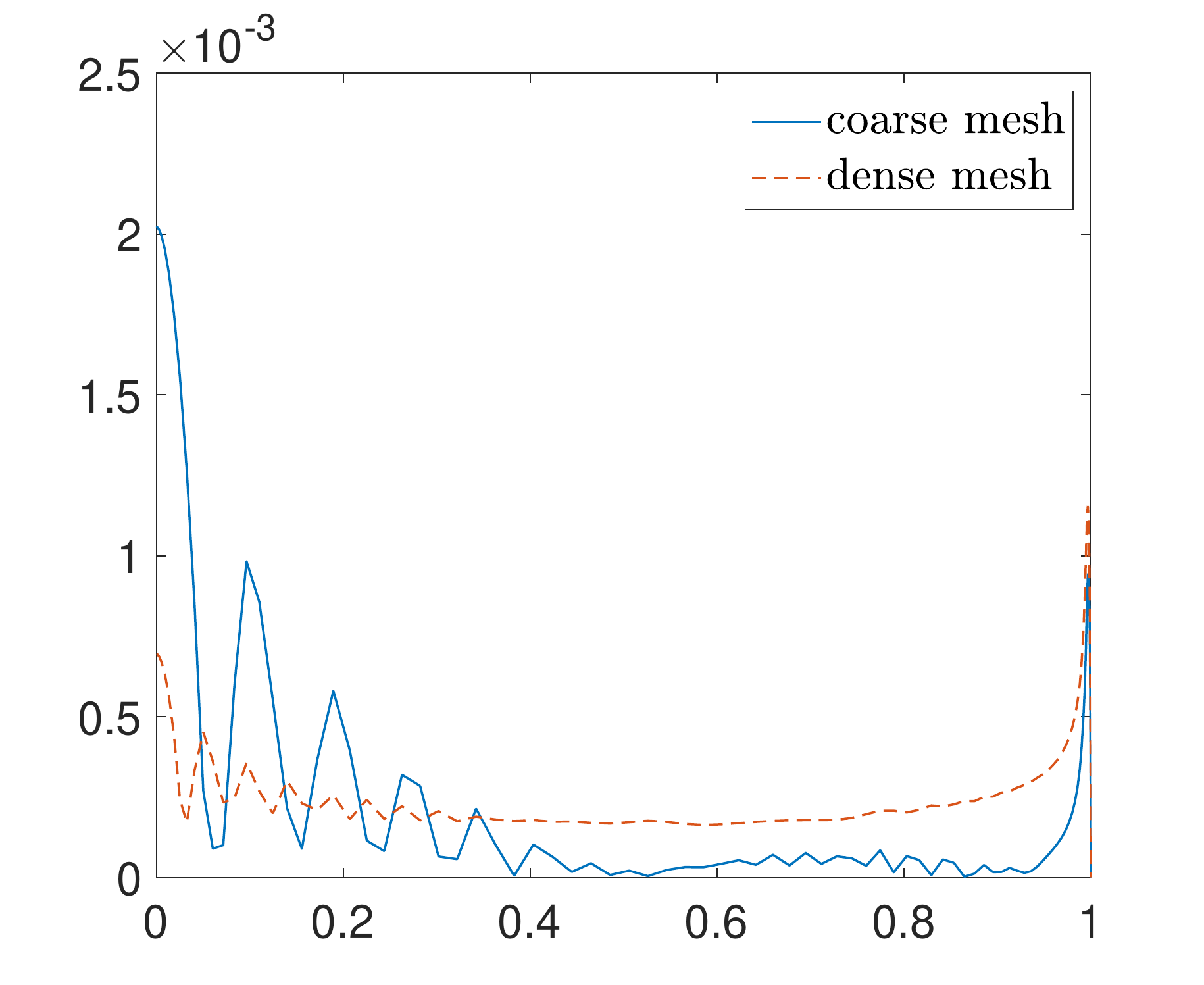}
\put(-215,92){$\delta_w$}
\put(-105,0){$\tilde x$}
\hspace{0mm}
\includegraphics[scale=0.40]{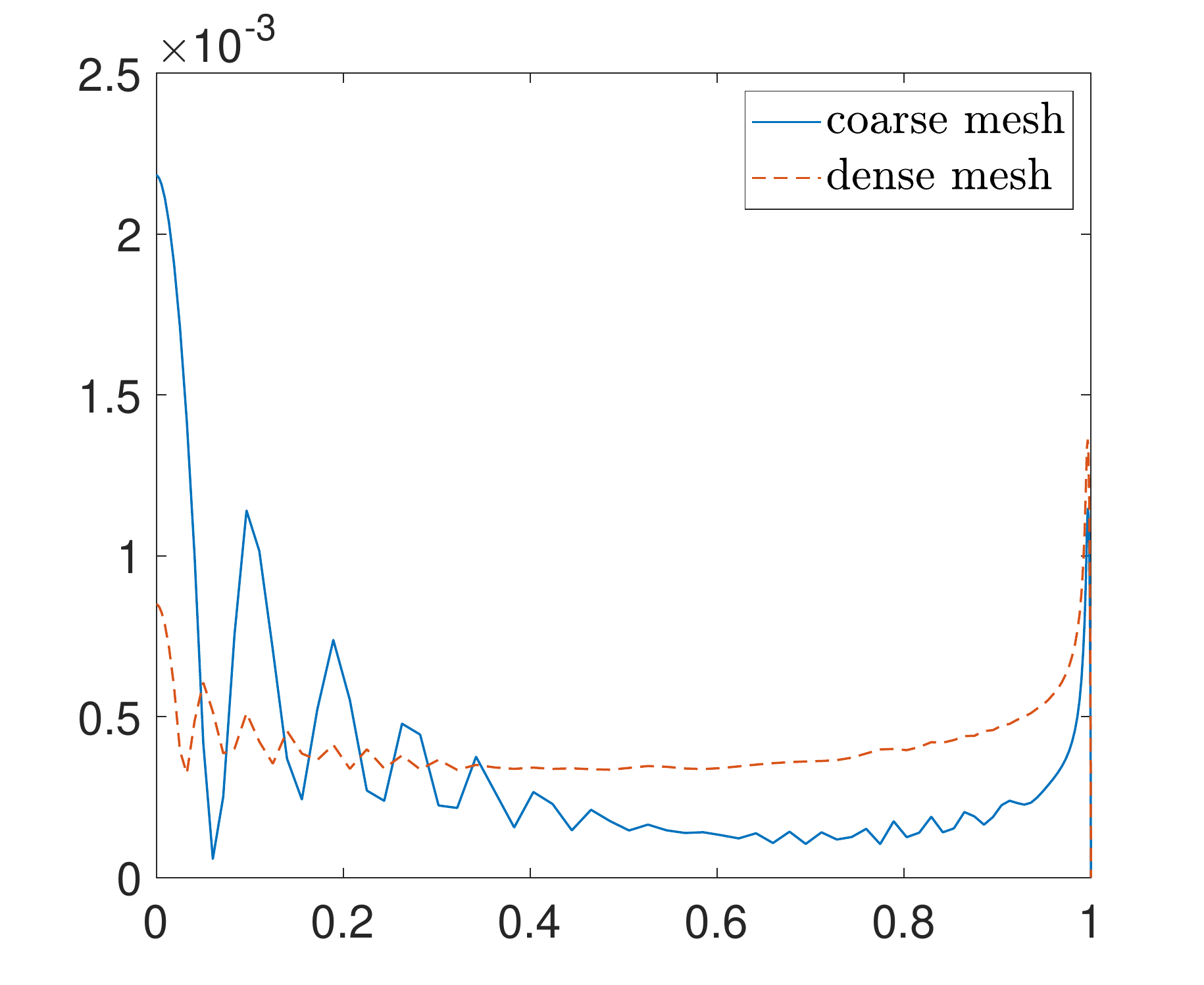}
\put(-105,0){$\tilde x$}
\put(-215,92){$\delta_w$}
\put(-440,160){$\textbf{a)}$}
\put(-225,160){$\textbf{b)}$}
\caption{The relative error of the crack opening, $\delta_w$, for: a) `variant 1' of the FEM problem configuration, b) `variant 2' of the FEM problem configuration. The scaled spatial variable $\tilde x=x/a$ is used.}
\label{delta_w_FEM}
\end{center}
\end{figure}
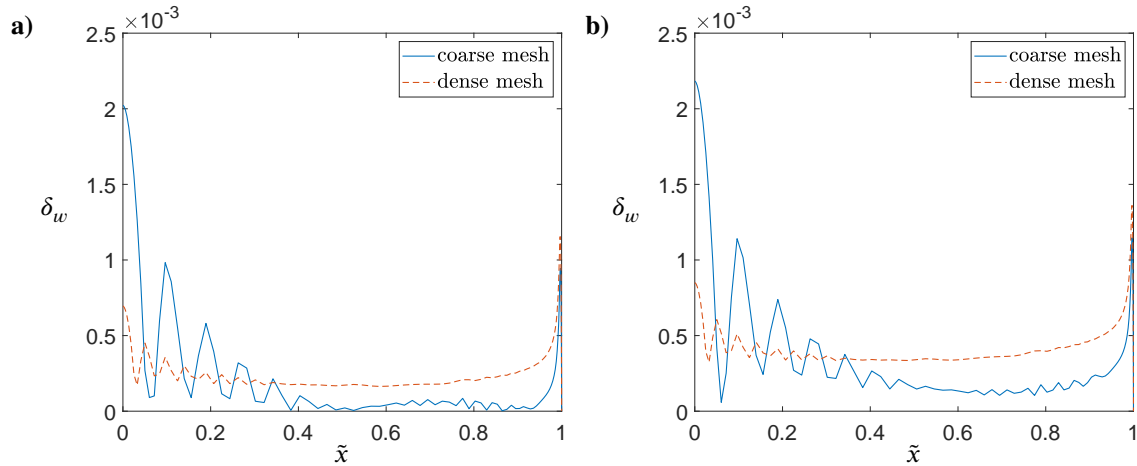

In Figure \ref{delta_w_FEM} we depict the relative error of the crack opening, $\delta_w$, obtained for the respective configurations of the FEM problem and mesh densities. It can be seen that for both variants the overall error of computations is almost the same and rarely exceeds 0.2$\%$. Moreover, under the analyzed discretization schemes the level of maximal error is quite similar for different mesh densities. Notably, the `dense mesh' provides more regular distribution of $\delta_w$ and appreciably better quality of solution near $x=0$. The values of the maximal and the average (over the crack length) errors of computations are collated in the Table \ref{bledy_tab}. Additionally, the relative error of computation of the mode I SIF, $\delta_{K_I}$, is also shown in the table. As can be seen, a stable error of little less than 1$\%$ is retained irrespective of the variant of the FEM problem configuration and mesh density being considered.

\begin{table}[]
\begin{center}
\begin{tabular}{cc|c|c|c|}
\cline{3-5}
\multicolumn{2}{c|}{}                                          & $\max(\delta_w)$      & $a^{-1}\int_0^a \delta_w \text{d}x$   & $\delta_{K_I}$    \\ \hline
\multicolumn{1}{|c|}{\multirow{2}{*}{variant 1}} & coarse mesh & $2.02 \cdot 10^{-3}$ & $2.03 \cdot 10^{-4}$                                                                             & $8.89 \cdot 10^{-3}$ \\ \cline{2-5} 
\multicolumn{1}{|c|}{}                           & dense mesh & $1.15 \cdot 10^{-3}$ & $2.33 \cdot 10^{-4}$                                                                            & $8.57 \cdot 10^{-3}$ \\ \hline
\multicolumn{1}{|c|}{\multirow{2}{*}{variant2}}  & coarse mesh & $2.18 \cdot 10^{-3}$ & $3.41 \cdot 10^{-4}$                                                                            & $8.62 \cdot 10^{-3}$ \\ \cline{2-5} 
\multicolumn{1}{|c|}{}                           & dense mesh & $1.36 \cdot 10^{-3}$ & $4.07 \cdot 10^{-4}$                                                                            & $8.30 \cdot 10^{-3}$ \\ \hline
\end{tabular}
\caption{The computational errors of the FEM module.}
\label{bledy_tab}
\end{center}
\end{table}

In order to complement this part of our analysis let us compare the distributions of principal stresses obtained for different mesh densities within respective variants of the problem. As there is no analytical solution to the considered problem in terms of $\sigma_{jj}$ ($j=1,2,3$),  we compare respective components of the stress tensor produced by the FEM module for coarse and dense meshes. The analyzed spatial domain includes the immediate vicinity of the crack surface with the limits of the normalized spatial variables being defined as: $\tilde x_\text{max}=x_\text{max}/a=2$, $\tilde y_\text{max}=y_\text{max}/a=1$. 

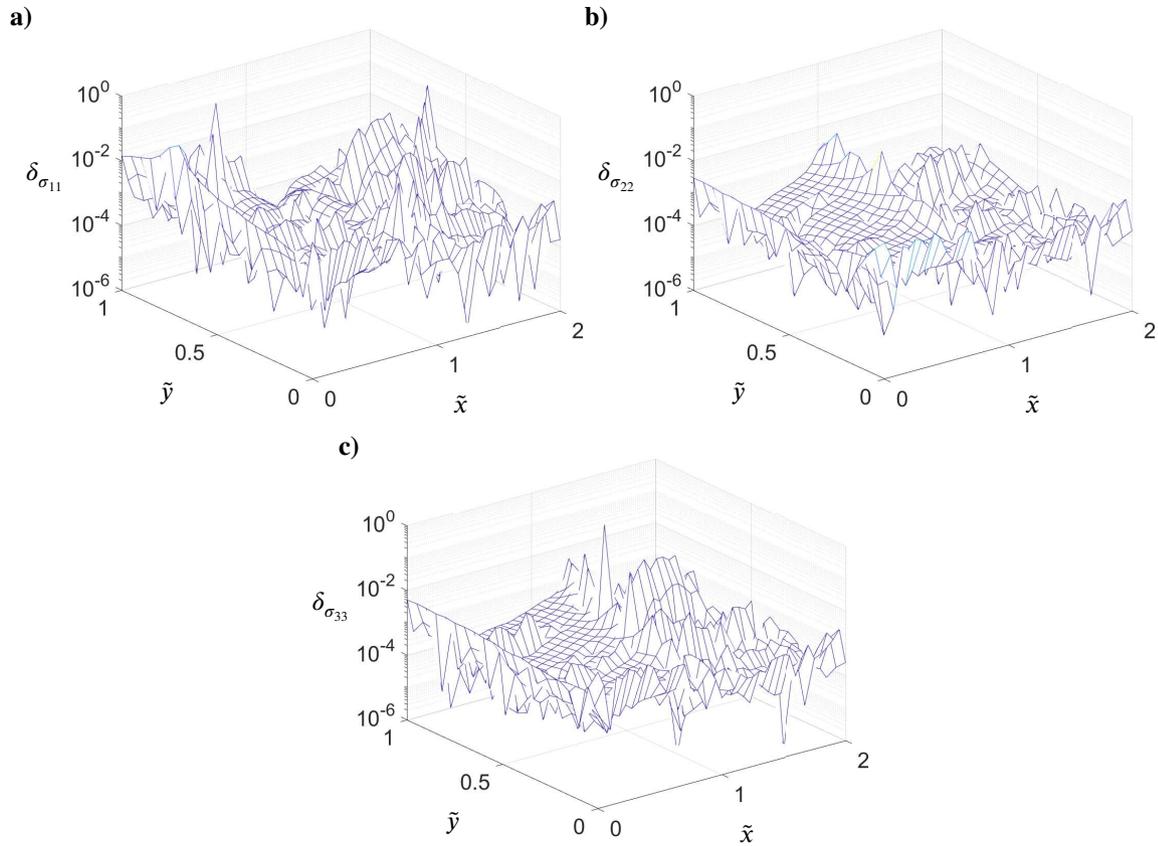
\begin{figure}[htb!]
\begin{center}
\includegraphics[scale=0.40]{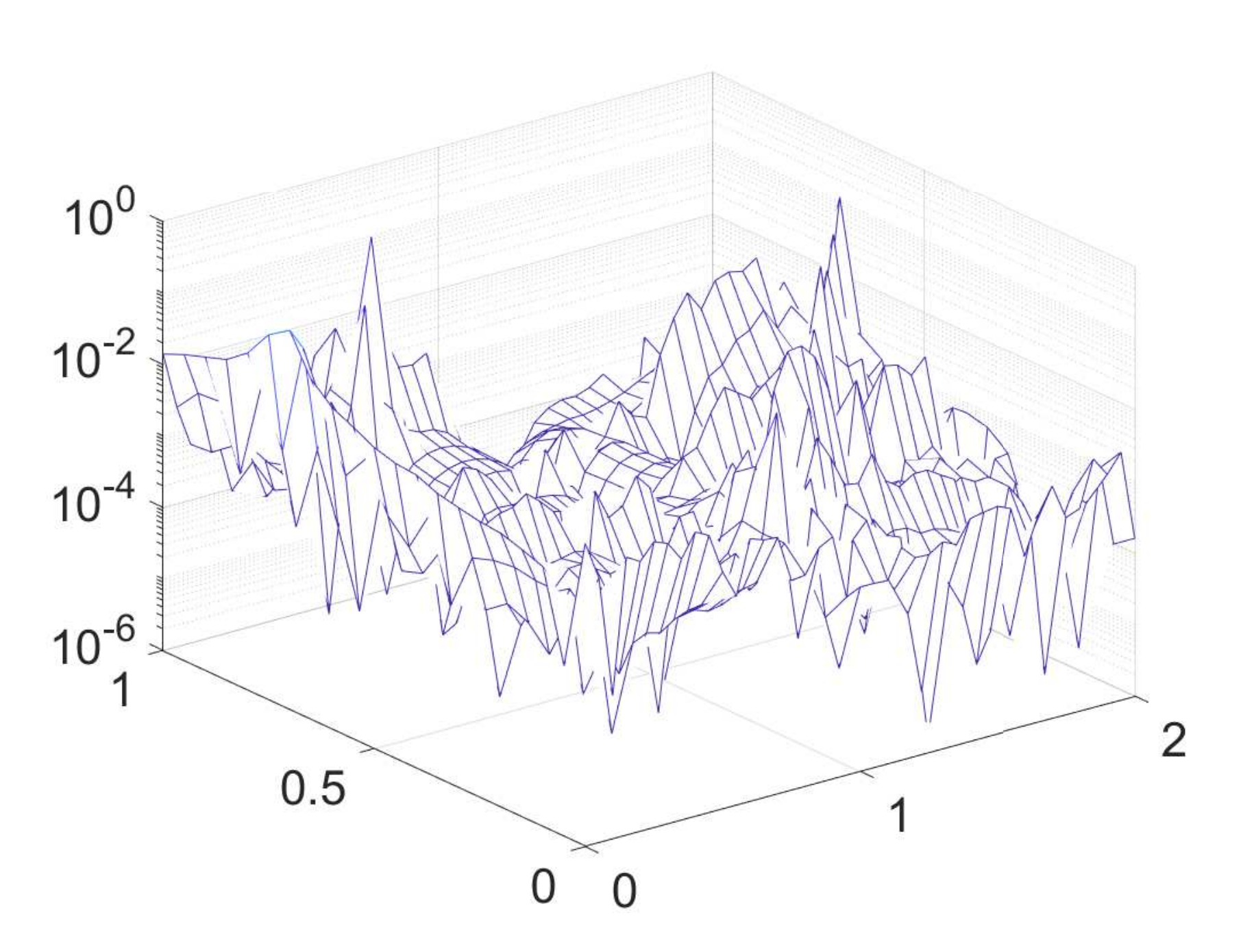}
\put(-220,92){$\delta_{\sigma_{11}}$}
\put(-60,5){$\tilde x$}
\put(-170,10){$\tilde y$}
\hspace{0mm}
\includegraphics[scale=0.40]{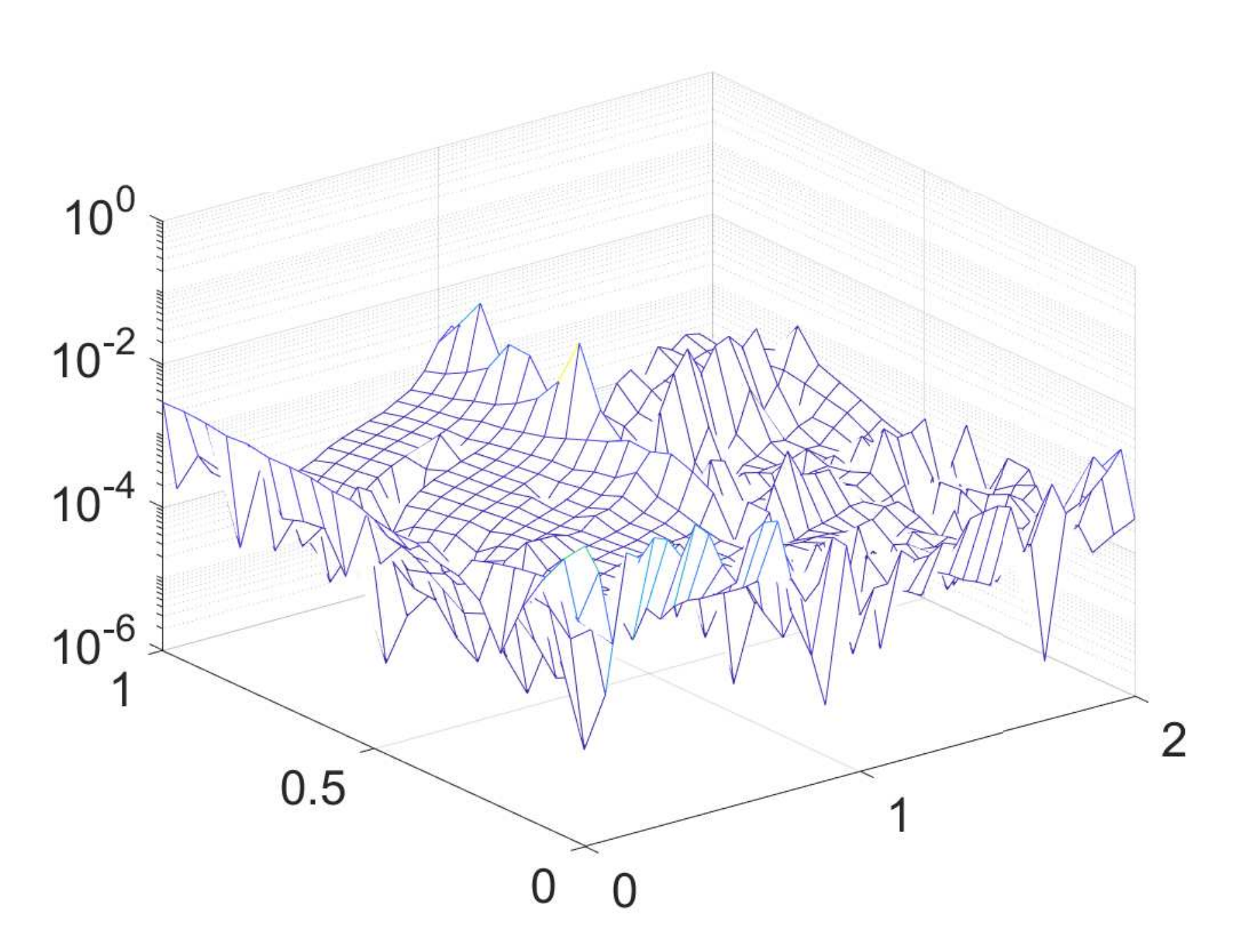}
\put(-60,5){$\tilde x$}
\put(-170,10){$\tilde y$}
\put(-220,92){$\delta_{\sigma_{22}}$}
\put(-440,150){$\textbf{a)}$}
\put(-225,150){$\textbf{b)}$}

\includegraphics[scale=0.40]{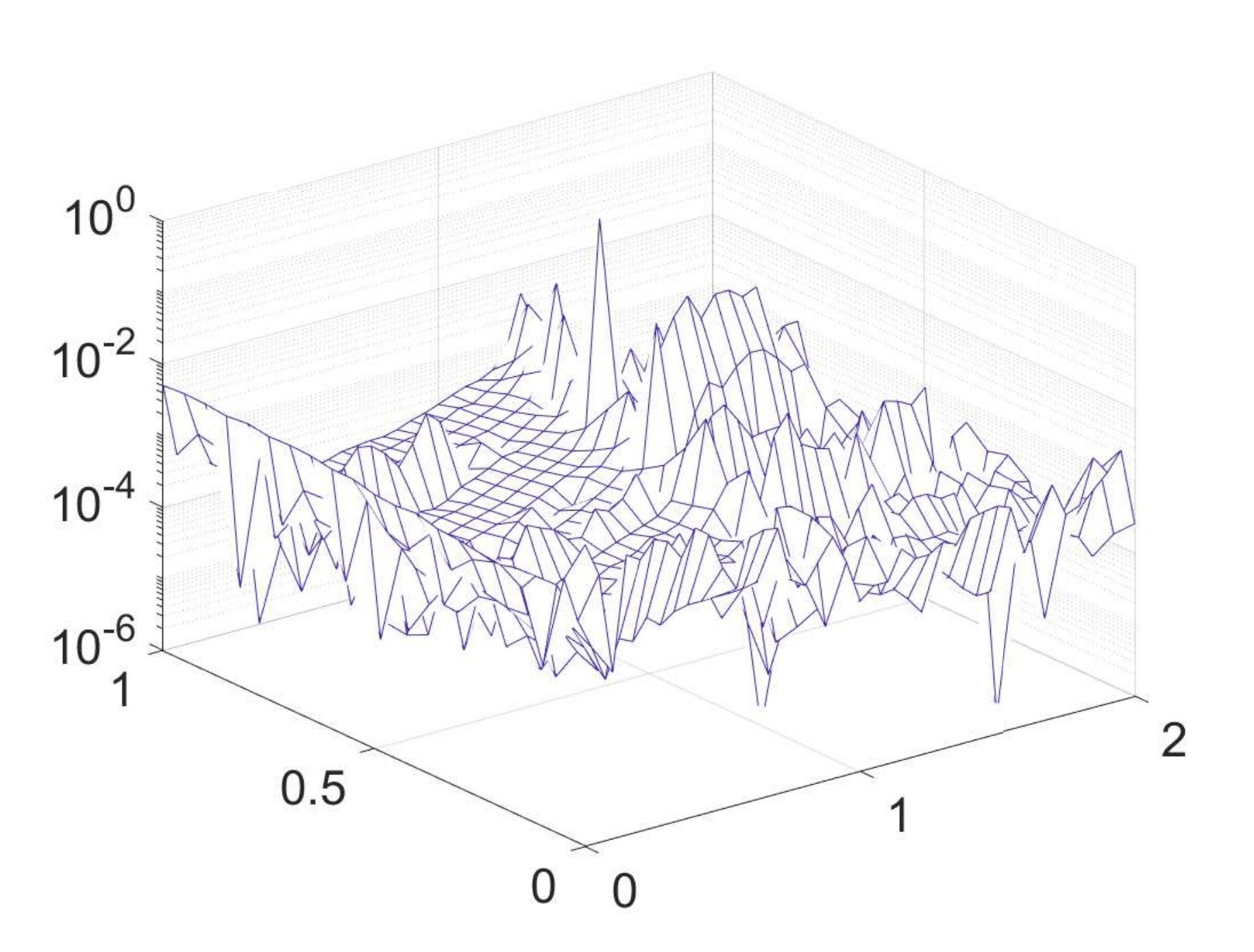}
\put(-210,150){$\textbf{c)}$}
\put(-220,92){$\delta_{\sigma_{33}}$}
\put(-60,5){$\tilde x$}
\put(-170,10){$\tilde y$}
\caption{The relative deviations, $\delta_{\sigma_{jj}}$, between the principal stresses obtained for `variant 1' with meshes of different densities (coarse and dense mesh, respectively) for: a) $\sigma_{11}$, b) $\sigma_{22}$, c) $\sigma_{33}$. The scaled spatial variables $\tilde x=x/a$ and $\tilde y=y/a$  are used.}
\label{delta_sigma_3}
\end{center}
\end{figure}

\begin{figure}[htb!]
\begin{center}
\includegraphics[scale=0.40]{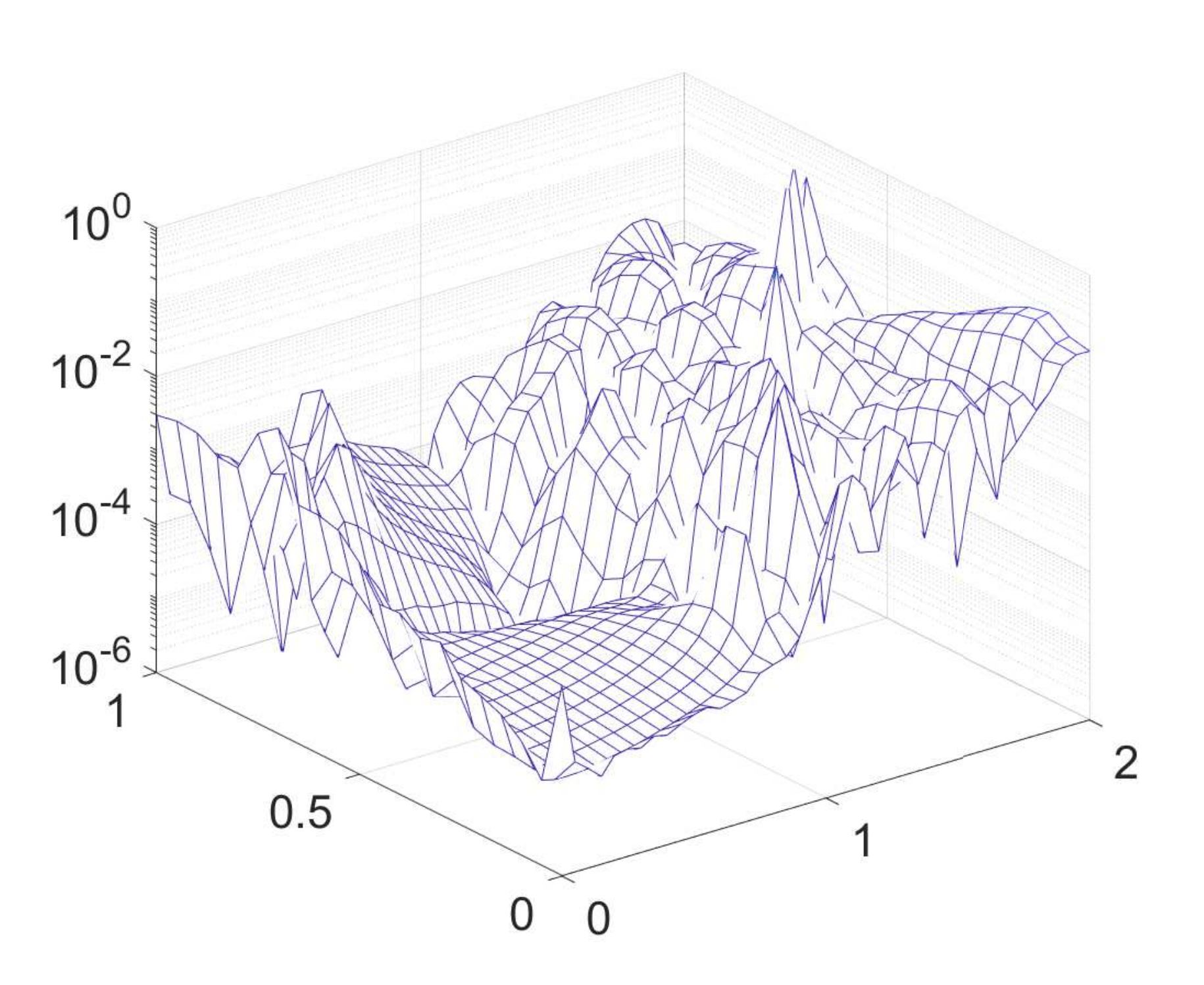}
\put(-220,92){$\delta_{\sigma_{11}}$}
\put(-60,5){$\tilde x$}
\put(-170,10){$\tilde y$}
\hspace{0mm}
\includegraphics[scale=0.40]{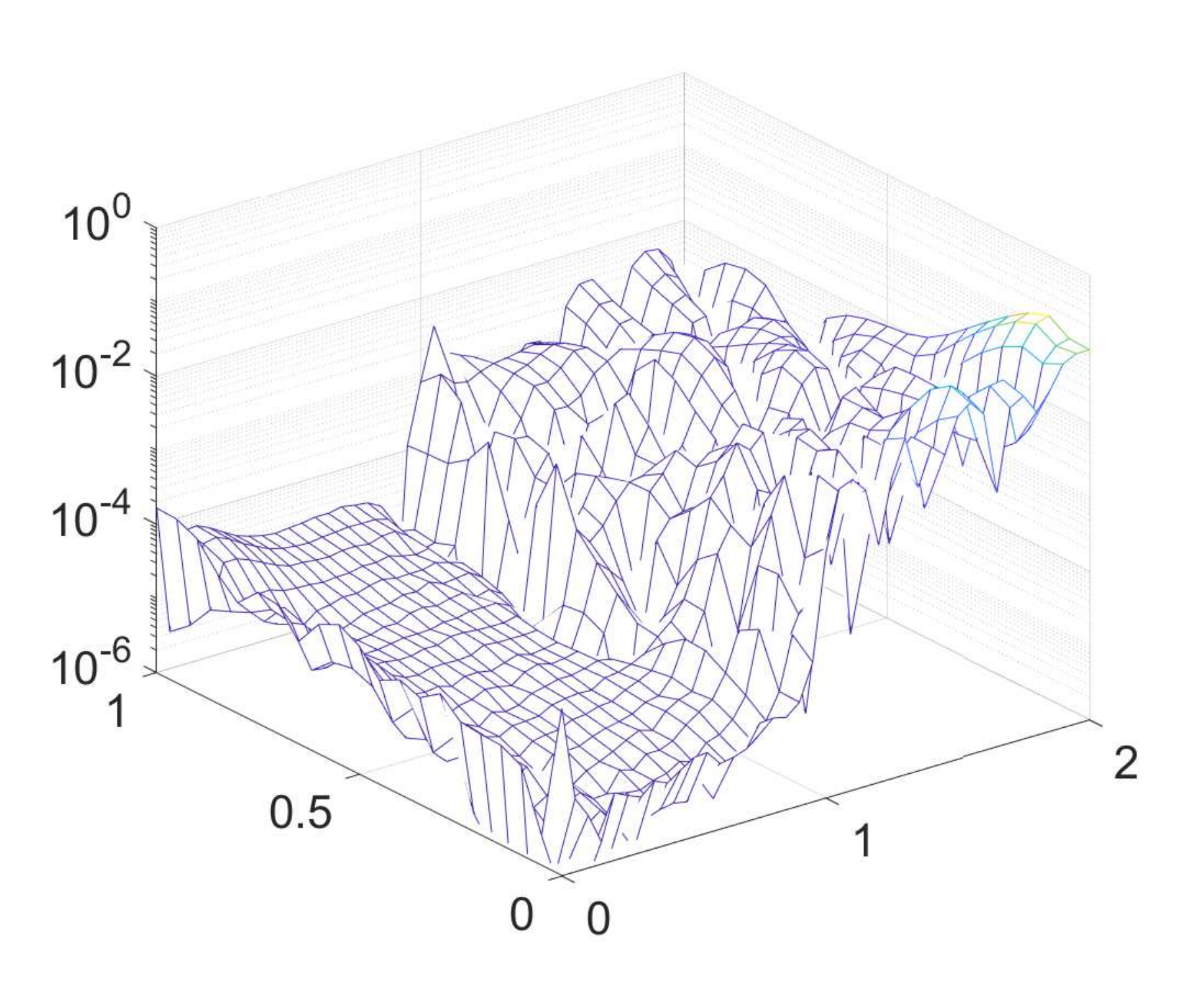}
\put(-60,5){$\tilde x$}
\put(-170,10){$\tilde y$}
\put(-220,92){$\delta_{\sigma_{22}}$}
\put(-440,150){$\textbf{a)}$}
\put(-225,150){$\textbf{b)}$}

\includegraphics[scale=0.40]{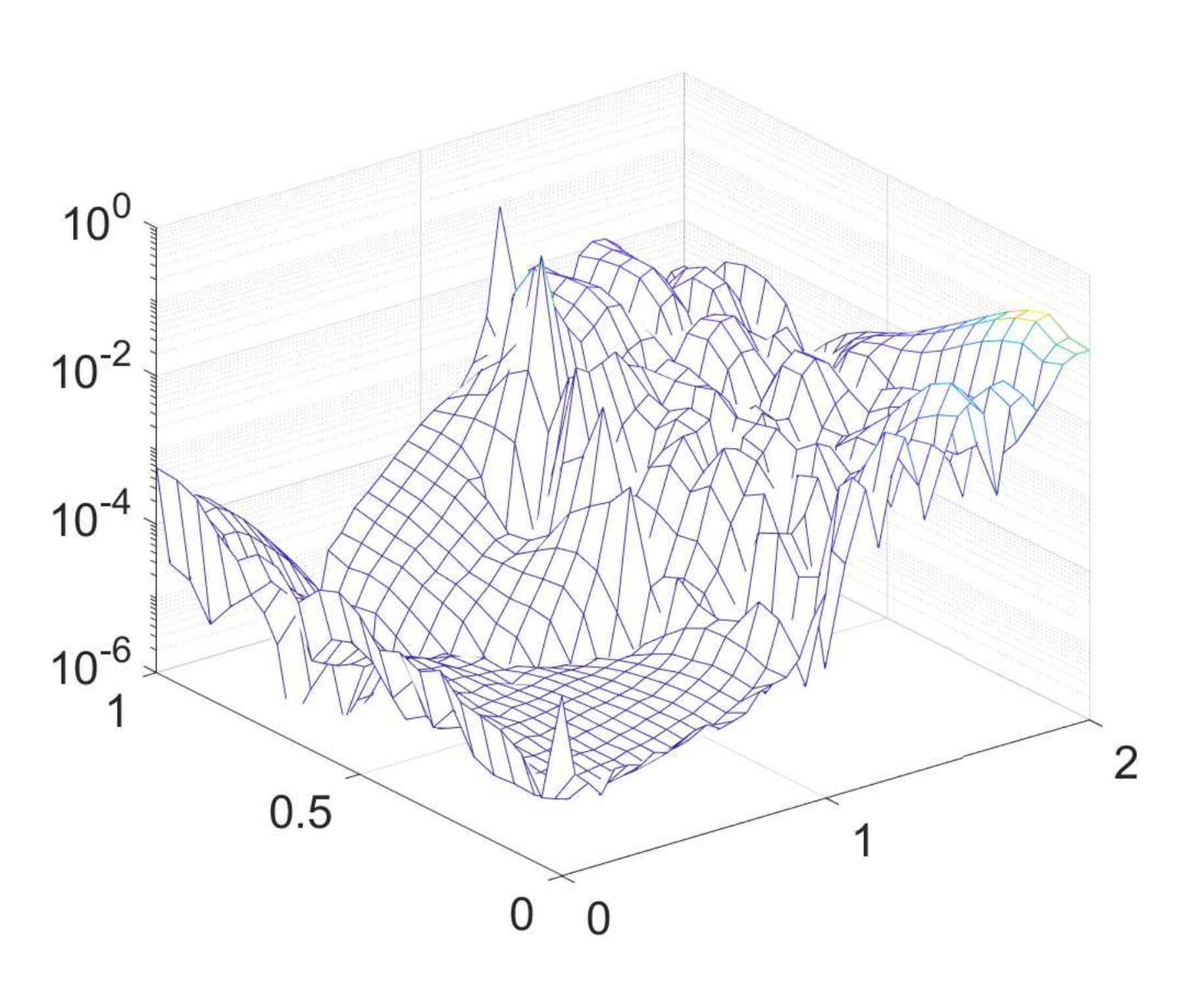}
\put(-210,150){$\textbf{c)}$}
\put(-220,92){$\delta_{\sigma_{33}}$}
\put(-60,5){$\tilde x$}
\put(-170,10){$\tilde y$}
\caption{The relative deviations, $\delta_{\sigma_{jj}}$, between the principal stresses obtained for `variant 2' with meshes of different densities (coarse and dense mesh, respectively) for: a) $\sigma_{11}$, b) $\sigma_{22}$, c) $\sigma_{33}$. The scaled spatial variables $\tilde x=x/a$ and $\tilde y=y/a$  are used.}
\label{delta_sigma_inf}
\end{center}
\end{figure}

The relative deviations of principal stresses, $\delta_{\sigma_{jj}}$, for `variant 1' of the FEM problem configuration are shown in Figure \ref{delta_sigma_3}. A very good agreement of respective results can be seen. The deviations that exceed $1\%$ are observed only in those regions where  the stress magnitude is close to zero. The deviations for `variant 2', shown in Figure \ref{delta_sigma_inf}, demonstrate that also for this configuration of the FEM problem the respective results are very close to each other. This time however, the principal stresses computed with meshes of different densities exhibit  better compatibility over this section of the spatial domain where $\tilde x<1$ than over the remaining part.

From the above analysis it is clear that the constructed FEM subroutine provides a very good accuracy of computations for the solid-mechanics equations in the HF problem. The results suggest that in the case of elastic fracture even the mesh of lower density ensures sufficient accuracy of computations  for any practical application. However, for more complicated versions of the problem (such as  elasto-plastic hydraulic fracture or complex configuration of the computational domain) the `dense mesh' seems to be the safer choice.

\subsection{Verification of the algorithm performance}
\label{ver_alg}

Having identified the level of accuracy provided by the FEM module (`$w$ - module') let us now investigate the performance of the complete computational scheme outlined in the Subsection \ref{gen_alg}. To this end we use the full form of the time-dependent analytical benchmark solution from Appendix \ref{ap_B} (note that, in the employed benchmark representation, the leak-off function $q_\text{L}$ is not an element of solution but a predefined relation). All configurations of the  FEM sub-problem analyzed in the previous subsection are employed. The spatial discretization in the $v$ module utilizes a mesh of 100 nodes with increased density  near the fracture tip in a way proposed by Wrobel $\&$ Mishuris \cite{solver_calkowy}. The implemented time stepping strategy is  the same for all the analyzed variants of the problem. It is based on an incremental increase of the fracture volume, $\Omega(t)$. Namely, for a predefined ratio of the crack volumes in the consecutive time steps, $\kappa$:
\begin{equation}
\label{kappa_def}
\kappa = \frac{\Omega_{i+1}}{\Omega_i}, \quad \Omega_k=\int_0^{a_k}w(x,t_k) \text{d}x,
\end{equation}
the continuity equation \eqref{cont} is formally integrated with respect to space and time to obtain a relation between the fracture volume change and the total influx and leak-off:
\begin{equation}
\label{time_step}
(\kappa-1)\Omega_i-\int_{t_i}^{t_{i+1}}q_0(t)\text{d}t+\int_{t_i}^{t_{i+1}} \int_0^{a_i} q_\text{L}(x,t)\text{d}x\text{d}t=0.
\end{equation}
The above relation is used to find the value of $t_{i+1}$. The scheme has been found to be quite stable, however its detailed investigation is still under way. In the conducted analysis a $5\%$ volume increase ($\kappa=1.05$) between  two consecutive time steps is imposed. The computations are carried out over the time span $t \in [0,3]$ as such an interval turned out to be sufficient for the computational error to stabilize. The accuracy of computations is described by: i) the relative error of the crack opening, $\delta_w$, ii) the relative error of the fluid velocity, $\delta_v$, iii) the relative error of the crack length, $\delta_L$, and iv) the relative error of the crack propagation speed, $\delta_{v_0}$. The initial conditions are taken in accordance with the benchmark solution.  Note that the initial crack opening and the initial fluid pressure are the same as those employed previously in the verification of the FEM module (see Figure \ref{w_p_bench}).  

\begin{figure}[htb!]
\begin{center}
\includegraphics[scale=0.40]{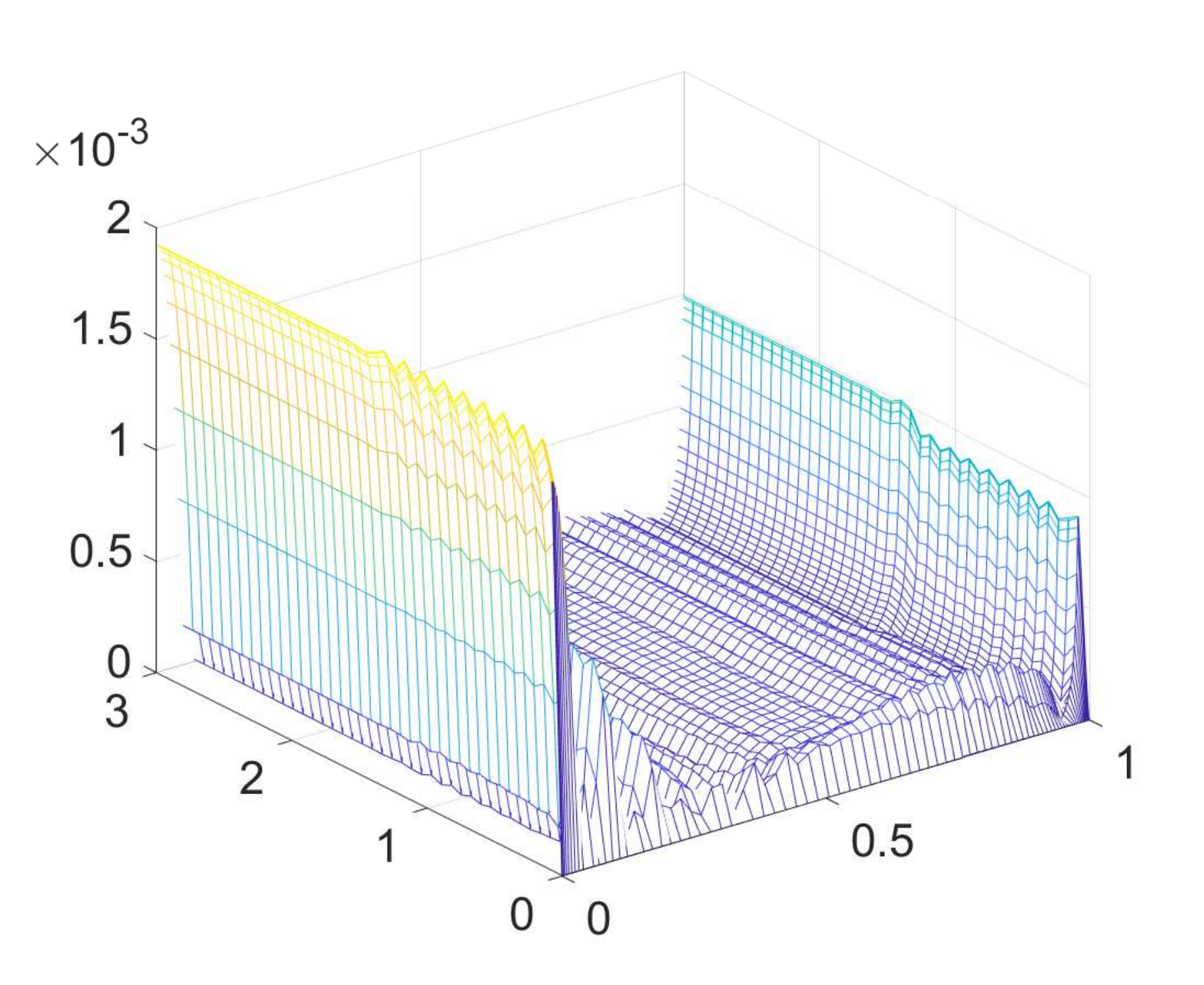}
\put(-215,92){$\delta_w$}
\put(-60,10){$\tilde x$}
\put(-170,15){$t$ [s]}
\hspace{0mm}
\includegraphics[scale=0.40]{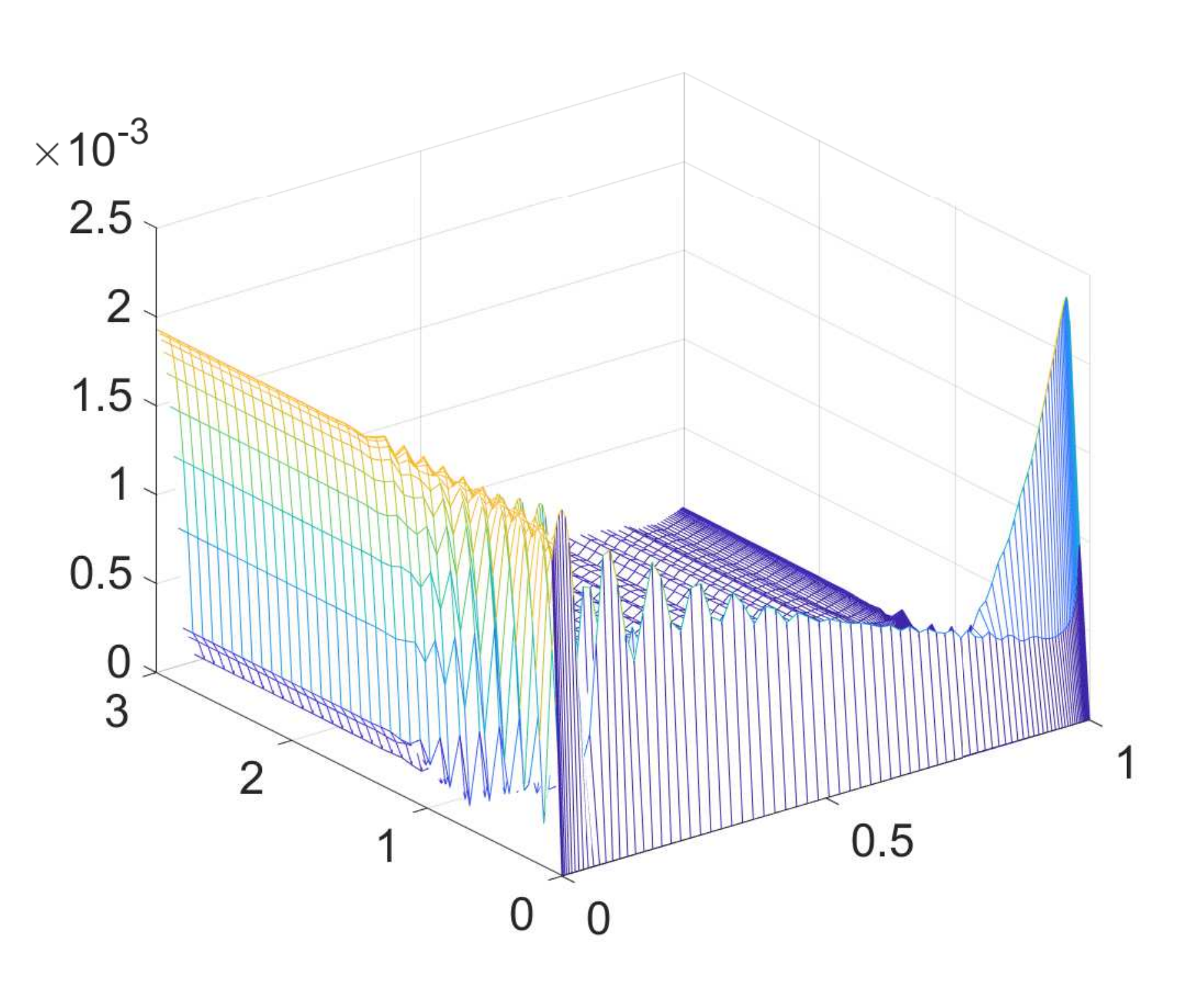}
\put(-60,10){$\tilde x$}
\put(-170,15){$t$ [s]}
\put(-215,92){$\delta_v$}
\put(-440,150){$\textbf{a)}$}
\put(-225,150){$\textbf{b)}$}
\caption{The error of solution for `variant 1' of the FEM problem configuration with coarse mesh of finite elements: a) the relative error of the crack opening, $\delta_w$, b) the relative error of the fluid velocity, $\delta_v$ . The scaled spatial variable $\tilde x=x/a$ is used.}
\label{delta_w_v_var_1_rough}
\end{center}
\end{figure}

\begin{figure}[htb!]
\begin{center}
\includegraphics[scale=0.40]{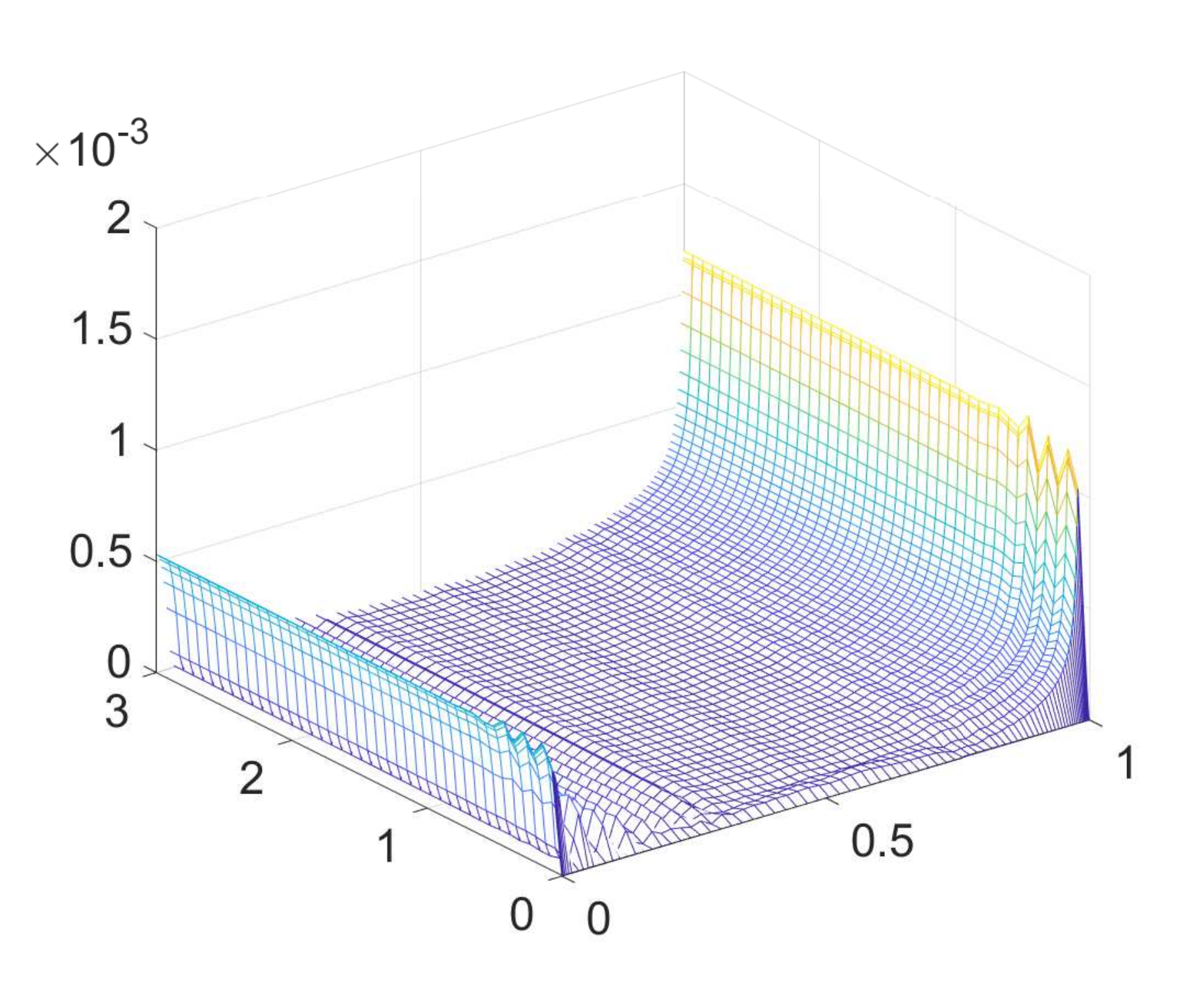}
\put(-215,92){$\delta_w$}
\put(-60,10){$\tilde x$}
\put(-170,15){$t$ [s]}
\hspace{0mm}
\includegraphics[scale=0.40]{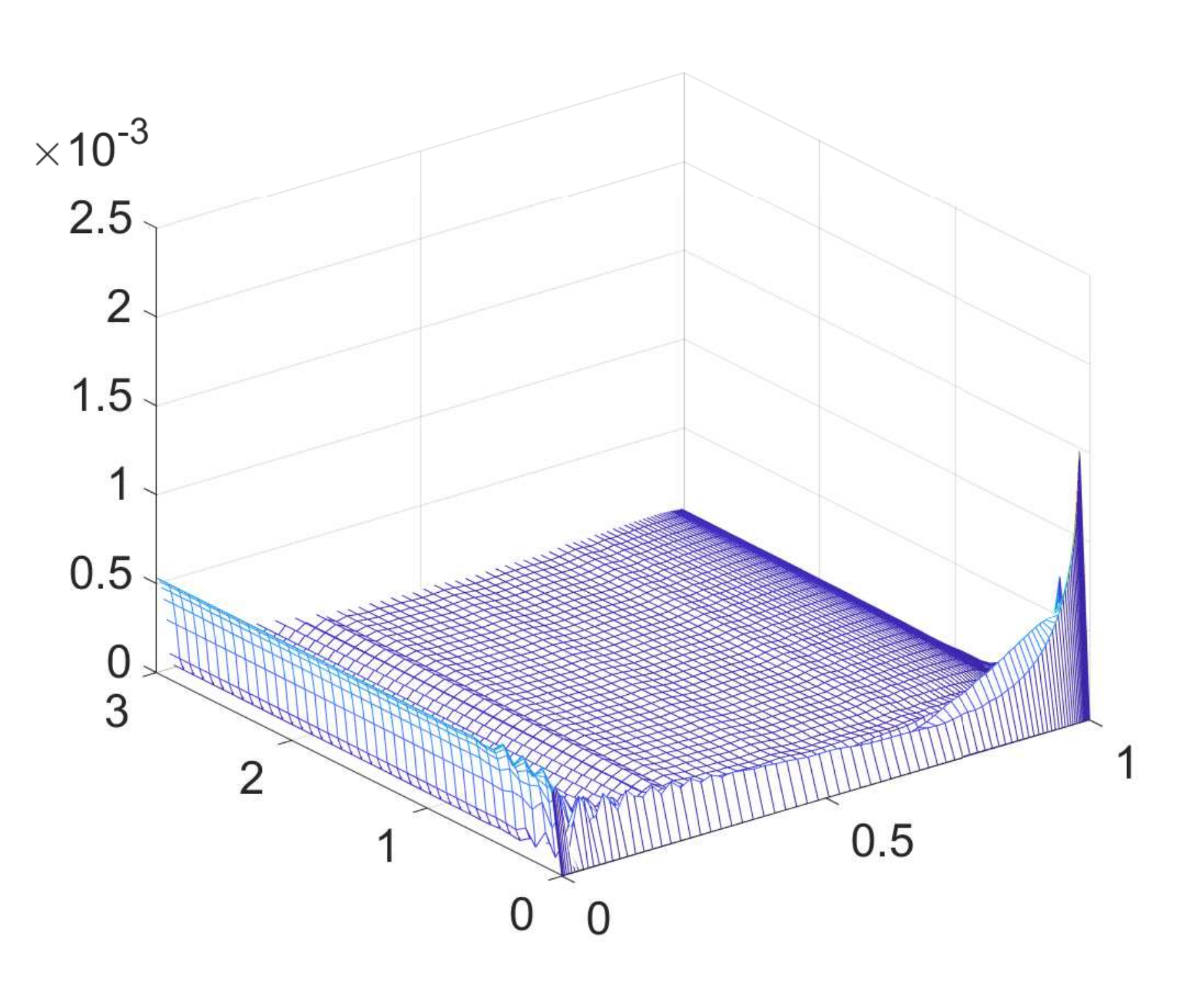}
\put(-60,10){$\tilde x$}
\put(-170,15){$t$ [s]}
\put(-215,92){$\delta_v$}
\put(-440,150){$\textbf{a)}$}
\put(-225,150){$\textbf{b)}$}
\caption{The error of solution for `variant 1' of the FEM problem configuration with dense mesh of finite elements: a) the relative error of the crack opening, $\delta_w$, b) the relative error of the fluid velocity, $\delta_v$ . The scaled spatial variable $\tilde x=x/a$ is used.}
\label{delta_w_v_var_1_dense}
\end{center}
\end{figure}

\begin{figure}[htb!]
\begin{center}
\includegraphics[scale=0.4]{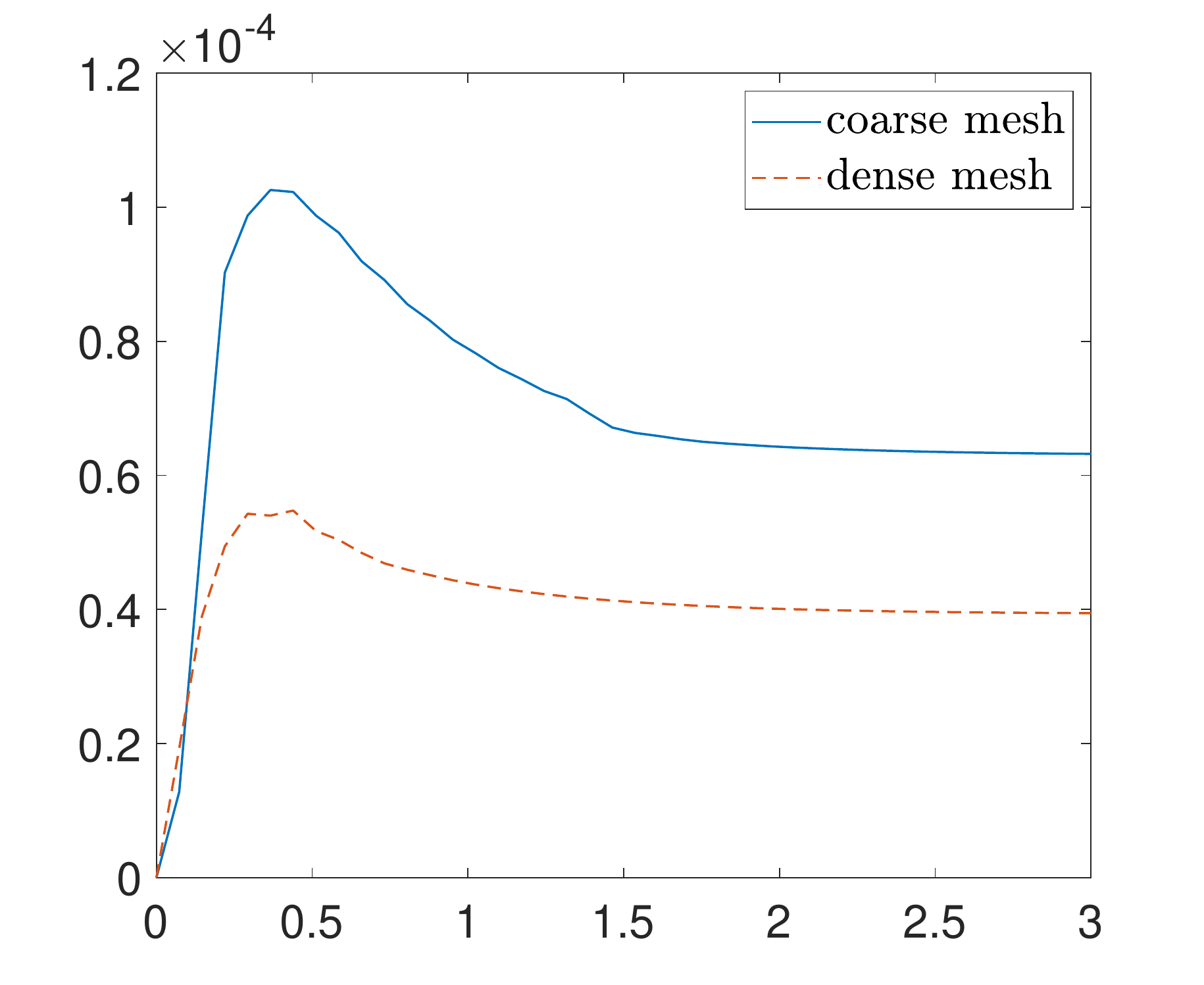}
\put(-215,92){$\delta_L$}
\put(-109,-5){$t$ [s]}
\hspace{0mm}
\includegraphics[scale=0.4]{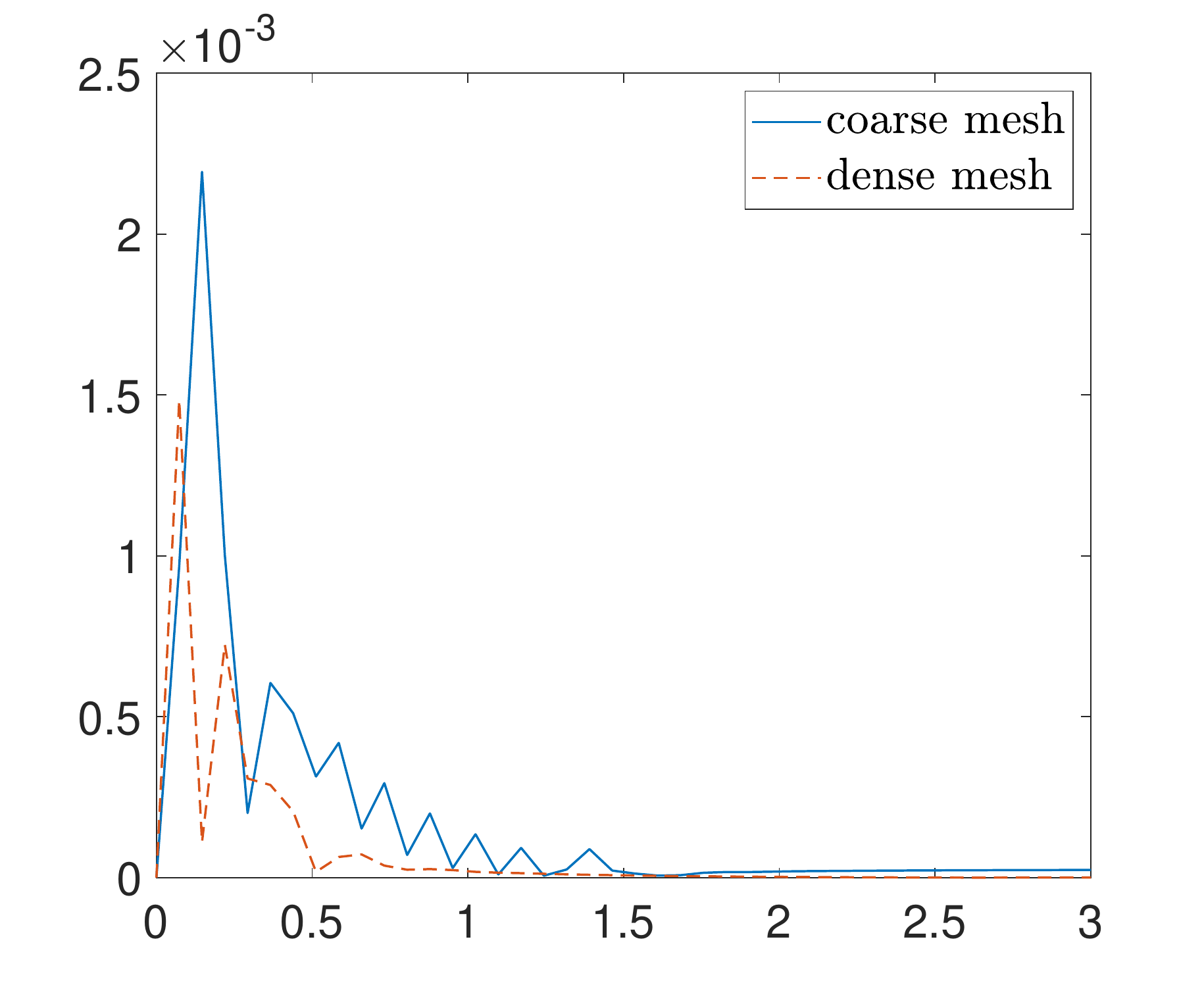}
\put(-109,-5){$t$ [s]}
\put(-215,92){$\delta_{v_0}$}
\put(-440,160){$\textbf{a)}$}
\put(-225,160){$\textbf{b)}$}
\caption{The error of solution for `variant 1' of the FEM problem configuration: a) the relative error of the crack length, $\delta_L$, b) the relative error of the crack propagation speed, $\delta_{v_0}$.}
\label{delta_L_v0_var_1}
\end{center}
\end{figure}

The results obtained with the `variant 1' of the FEM pattern geometry are depicted in Figures \ref{delta_w_v_var_1_rough} -- \ref{delta_L_v0_var_1}. It shows that, regardless of the FEM mesh density, the level of accuracy reported previously for the FEM module is retained over the whole considered time interval. When analyzing the error of the fluid velocity one can see that, after the initial increase, $\delta_v$ drops quickly and stabilizes. Once stabilized, the highest error of the fluid velocity is obtained at the fracture inlet, as a result of the corresponding magnification of $\delta_w$ at $x=0$. Irrespective of the prescribed mesh density (coarse and dense mesh, correspondingly) the maximal levels of $\delta_w$ and $\delta_v$ are very similar. A natural increase of $\delta_w$ is observed  at the fracture tip (as $w \to 0$ for $\tilde x \to 1$). The error of crack length computation, $\delta_L$, is stable over time at the level of $10^{-4}$ (see Figure \ref{delta_L_v0_var_1}a)) with noticeably better results obtained for the dense mesh. Surprisingly, for time greater than approximately 1 s, the accuracy of the crack propagation speed is better than the accuracy of the crack length (Figure \ref{delta_L_v0_var_1}b)). $\delta_{v_0}$ stabilizes at the level of $10^{-5}$ for the coarse FEM mesh and is of one order of magnitude lower for the dense mesh.

\begin{figure}[htb!]
\begin{center}
\includegraphics[scale=0.40]{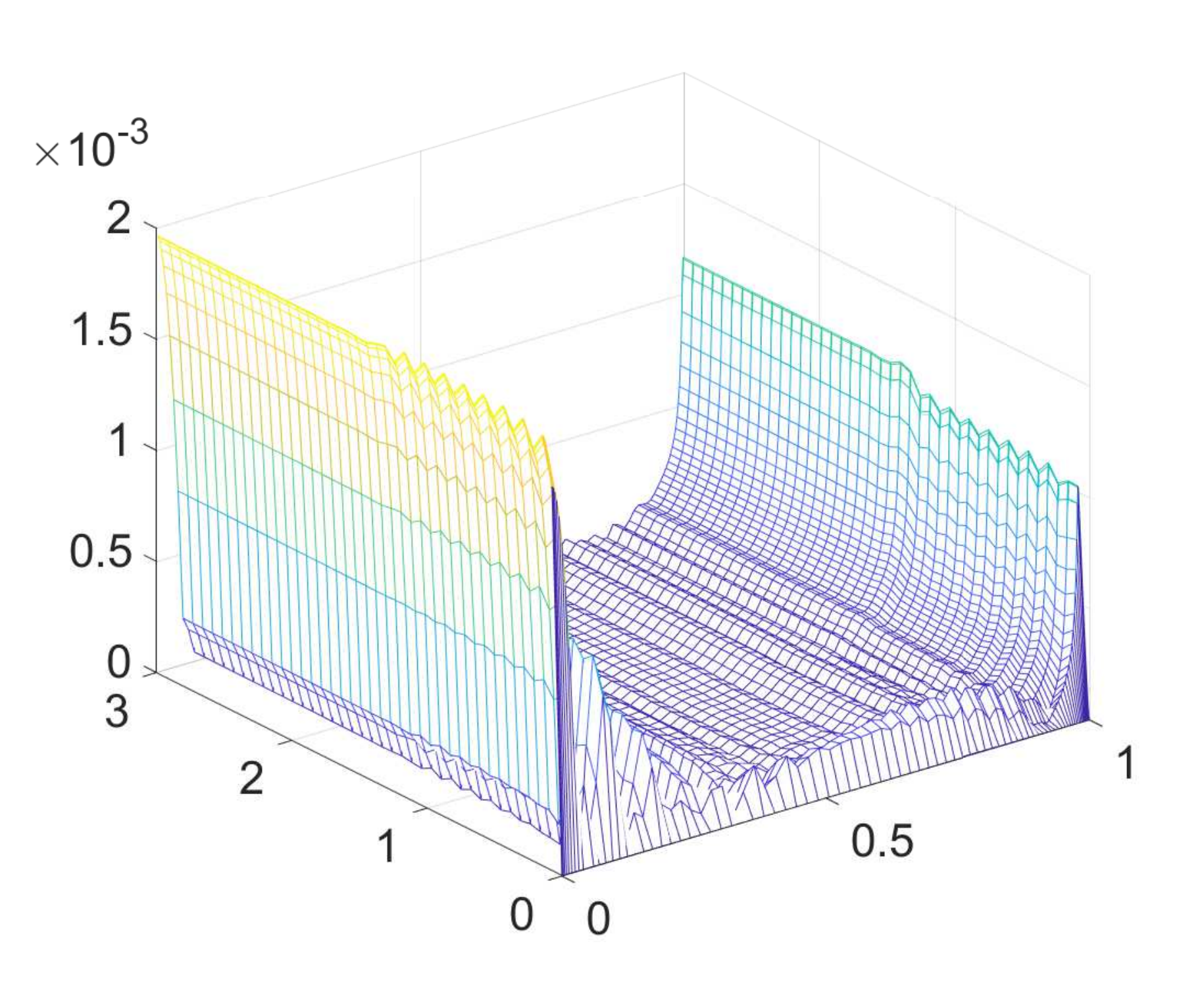}
\put(-215,92){$\delta_w$}
\put(-60,10){$\tilde x$}
\put(-170,15){$t$ [s]}
\hspace{0mm}
\includegraphics[scale=0.40]{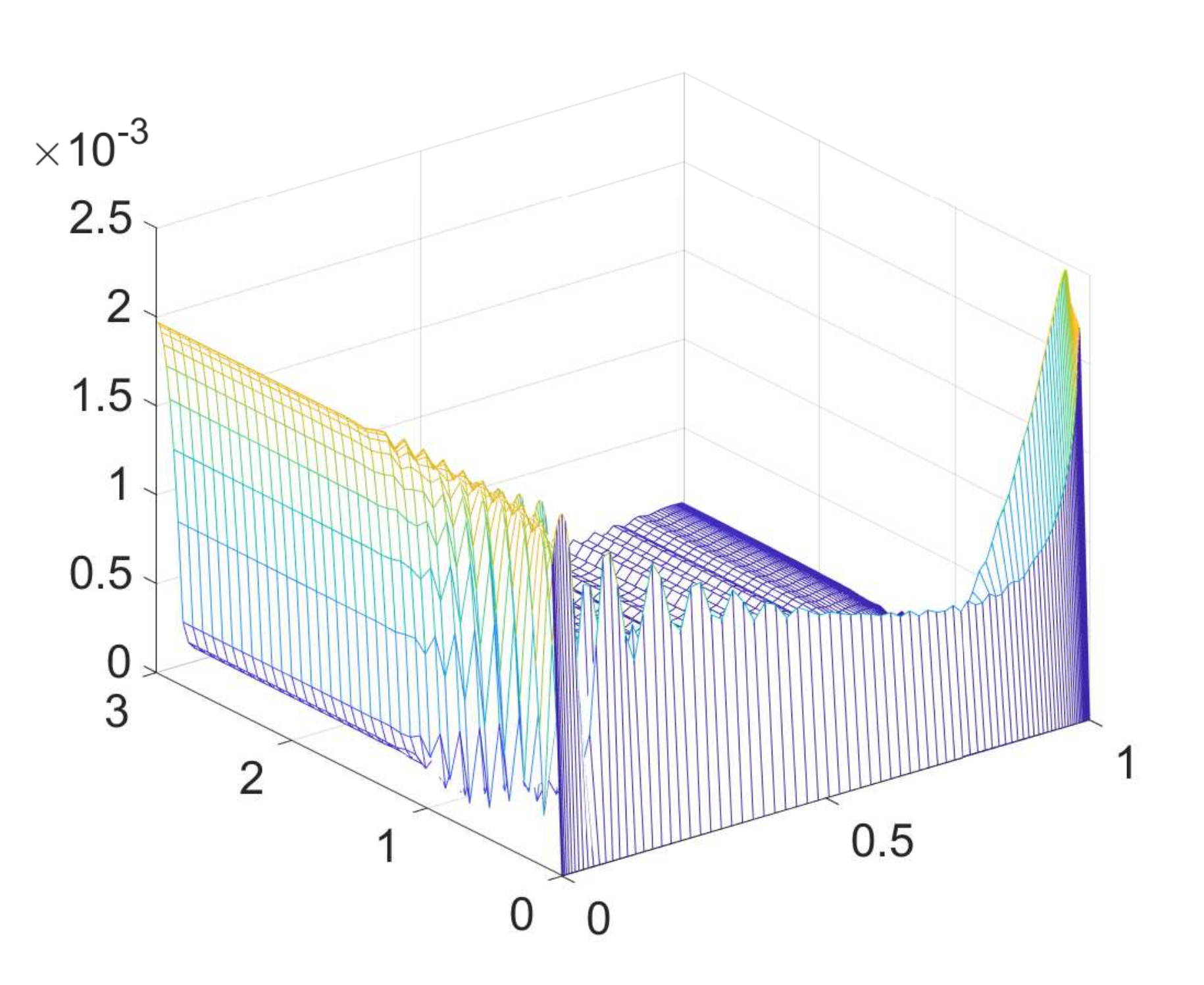}
\put(-60,10){$\tilde x$}
\put(-170,15){$t$ [s]}
\put(-215,92){$\delta_v$}
\put(-440,150){$\textbf{a)}$}
\put(-225,150){$\textbf{b)}$}
\caption{The error of solution for `variant 2' of the FEM problem configuration with coarse mesh of finite elements: a) the relative error of the crack opening, $\delta_w$, b) the relative error of the fluid velocity, $\delta_v$ . The scaled spatial variable $\tilde x=x/a$ is used.}
\label{delta_w_v_var_2_rough}
\end{center}
\end{figure}

\begin{figure}[htb!]
\begin{center}
\includegraphics[scale=0.40]{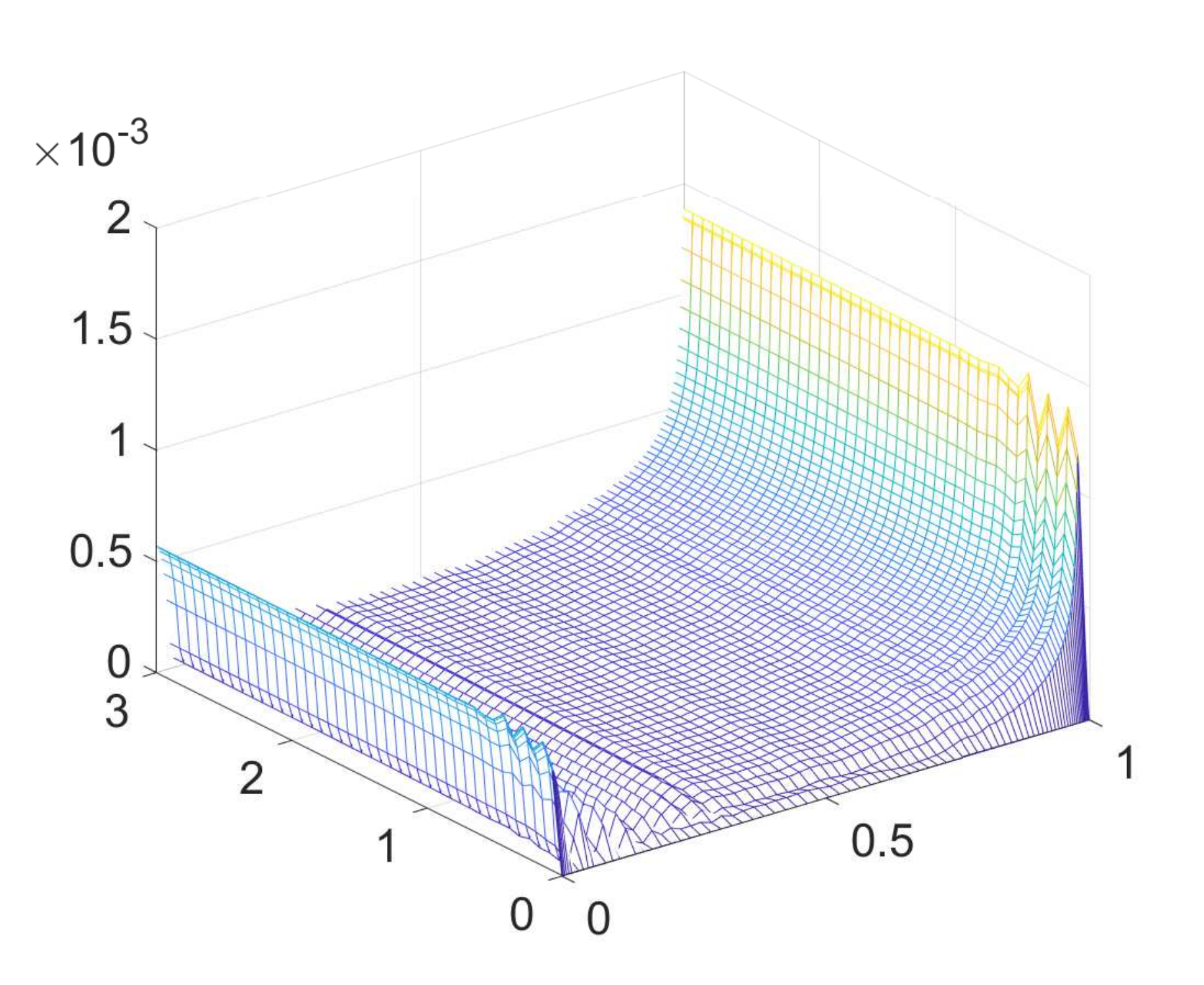}
\put(-215,92){$\delta_w$}
\put(-60,10){$\tilde x$}
\put(-170,15){$t$ [s]}
\hspace{0mm}
\includegraphics[scale=0.40]{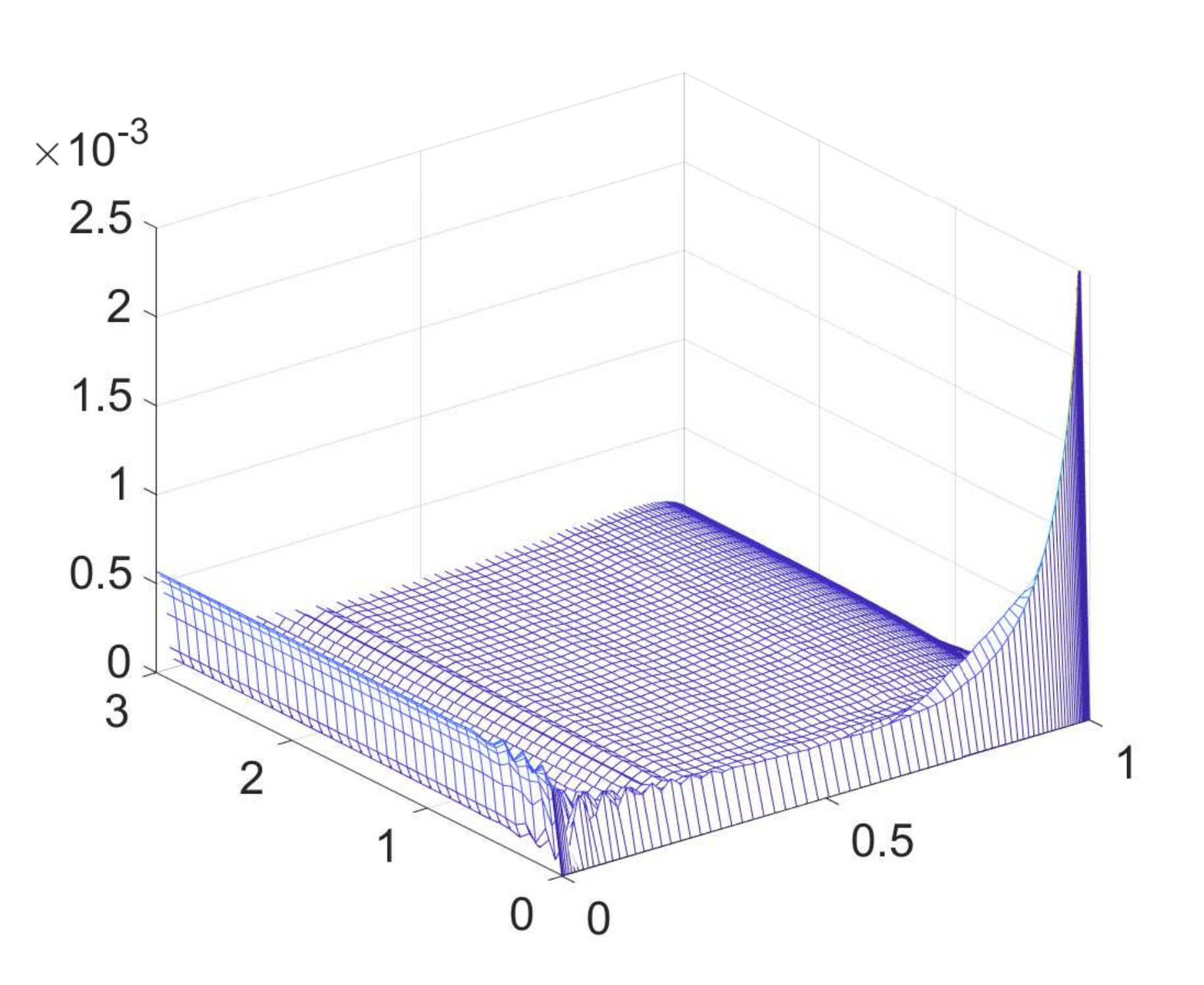}
\put(-60,10){$\tilde x$}
\put(-170,15){$t$ [s]}
\put(-215,92){$\delta_v$}
\put(-440,150){$\textbf{a)}$}
\put(-225,150){$\textbf{b)}$}
\caption{The error of solution for `variant 2' of the FEM problem configuration with dense mesh of finite elements: a) the relative error of the crack opening, $\delta_w$, b) the relative error of the fluid velocity, $\delta_v$ . The scaled spatial variable $\tilde x=x/a$ is used.}
\label{delta_w_v_var_2_dense}
\end{center}
\end{figure}

\begin{figure}[htb!]
\begin{center}
\includegraphics[scale=0.4]{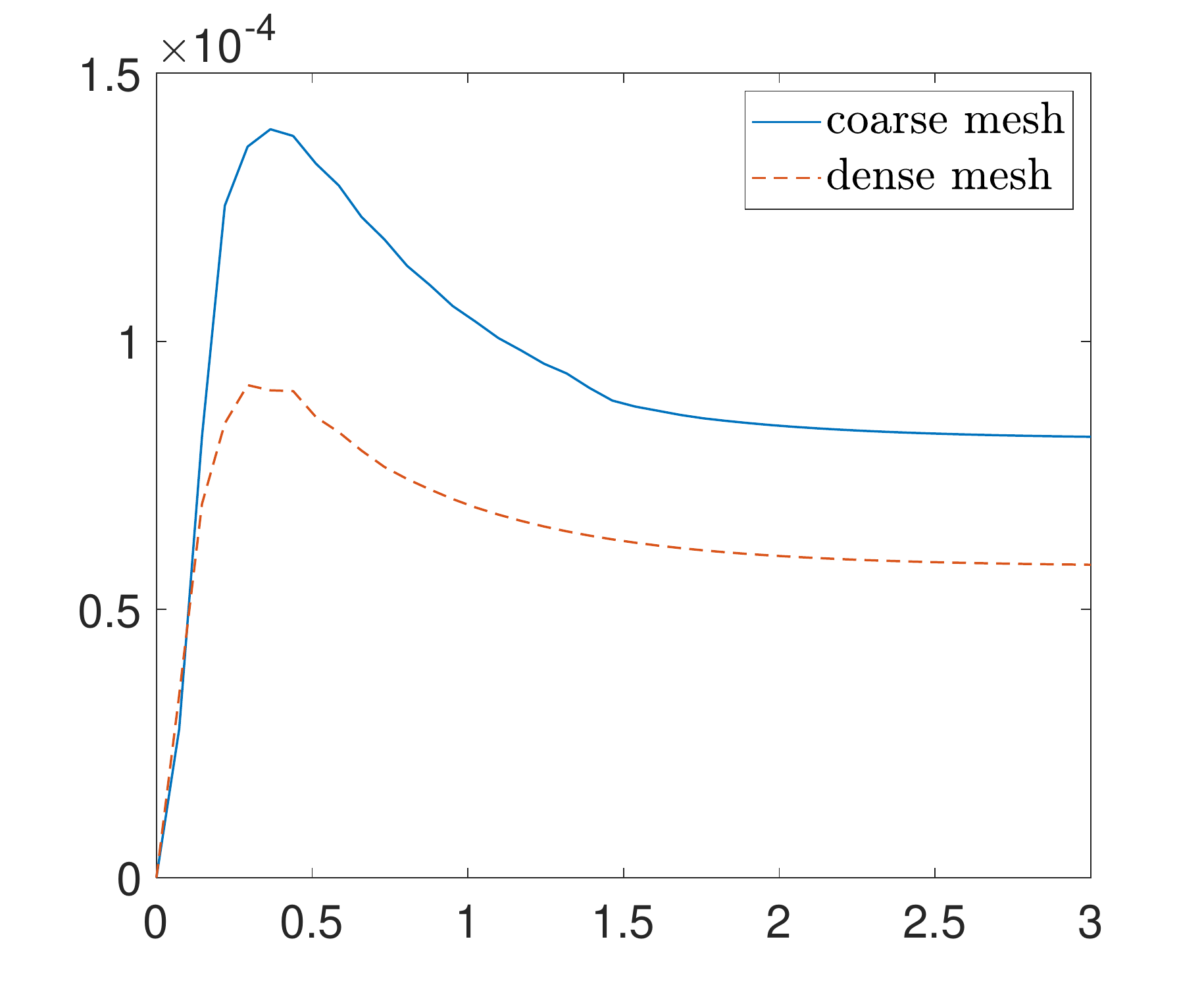}
\put(-215,92){$\delta_L$}
\put(-109,-5){$t$ [s]}
\hspace{0mm}
\includegraphics[scale=0.4]{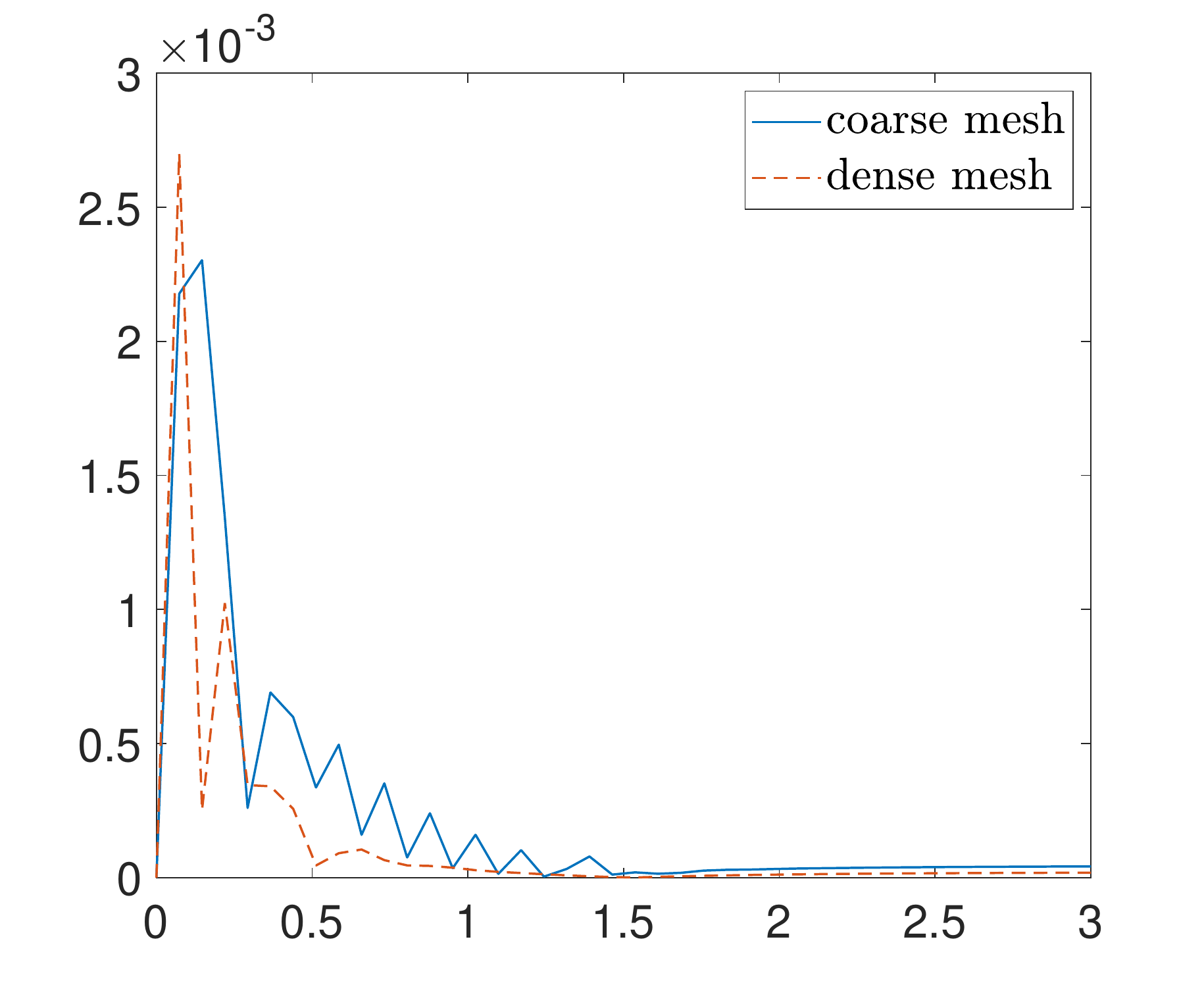}
\put(-109,-5){$t$ [s]}
\put(-215,92){$\delta_{v_0}$}
\put(-440,160){$\textbf{a)}$}
\put(-225,160){$\textbf{b)}$}
\caption{The error of solution for `variant 2' of the FEM problem configuration: a) the relative error of the crack length, $\delta_L$, b) the relative error of the crack propagation speed, $\delta_{v_0}$.}
\label{delta_L_v0_var_2}
\end{center}
\end{figure}

The corresponding distributions of computational errors for `variant 2'  of the FEM pattern geometry are depicted in Figures \ref{delta_w_v_var_2_rough} -- \ref{delta_L_v0_var_2}. The general trends and levels of accuracy very much resemble those obtained with `variant 1'. Slightly higher errors, however within the same orders of magnitude, are reported for the crack length and the crack propagation speed.

\begin{figure}[htb!]
\begin{center}
\includegraphics[scale=0.40]{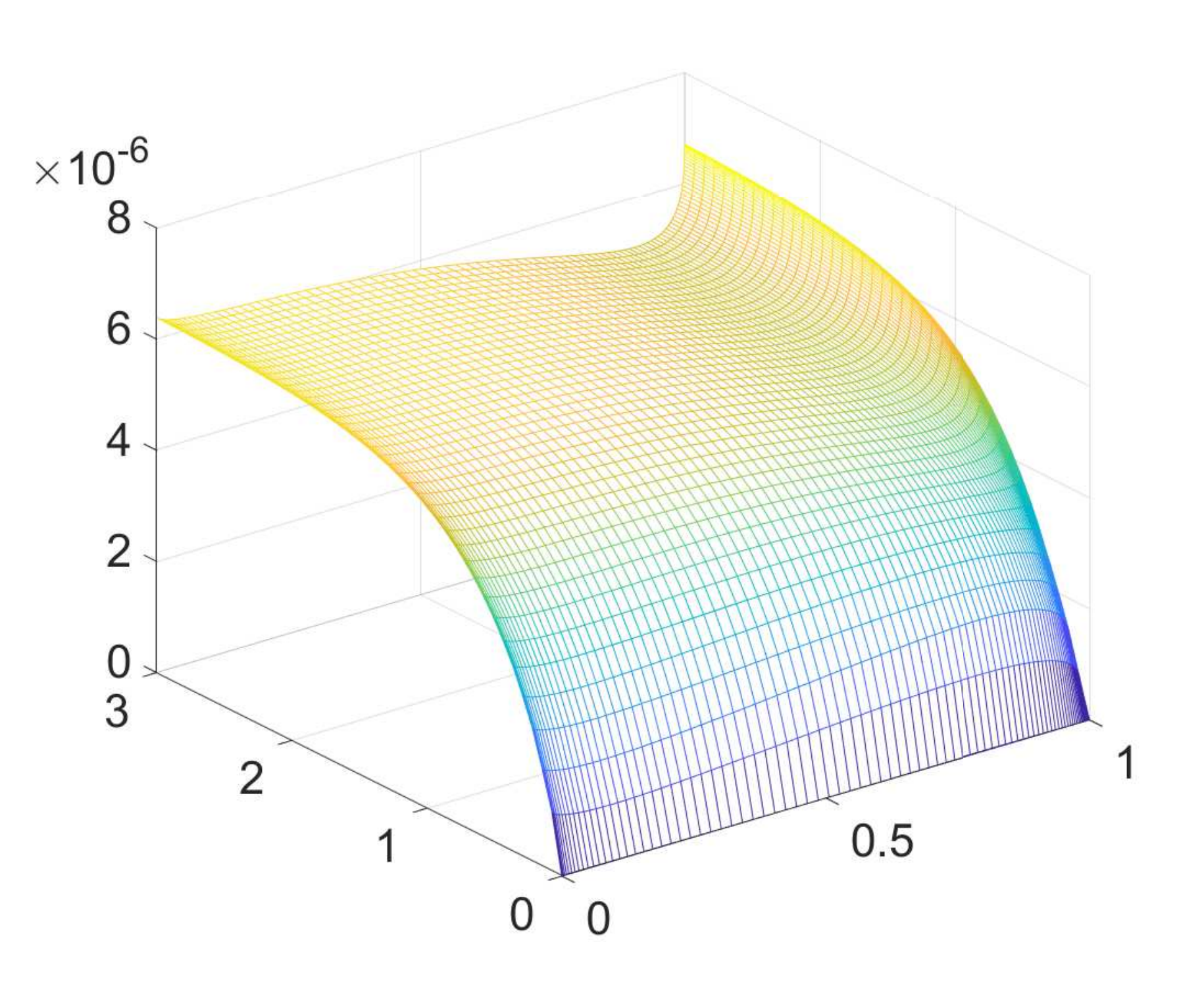}
\put(-215,92){$\delta_w$}
\put(-60,10){$\tilde x$}
\put(-170,15){$t$ [s]}
\hspace{0mm}
\includegraphics[scale=0.40]{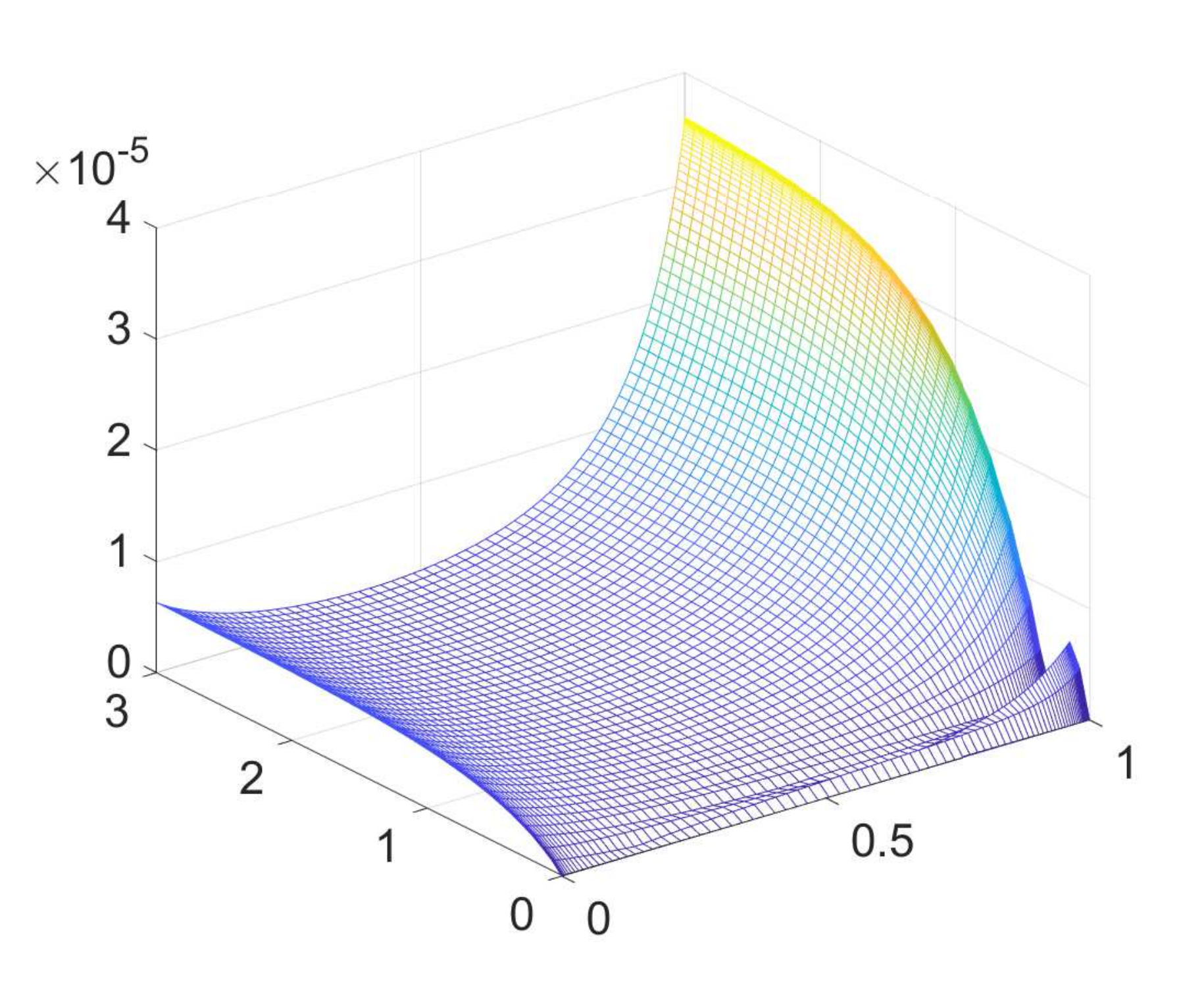}
\put(-60,10){$\tilde x$}
\put(-170,15){$t$ [s]}
\put(-215,92){$\delta_v$}
\put(-440,150){$\textbf{a)}$}
\put(-225,150){$\textbf{b)}$}
\caption{The error of solution produced by the algorithm from the paper by Wrobel $\&$ Mishuris\cite{Wrobel_2015}: a) the relative error of the crack opening, $\delta_w$, b) the relative error of the fluid velocity, $\delta_v$ . The scaled spatial variable $\tilde x=x/a$ is used.}
\label{delta_w_v_elast}
\end{center}
\end{figure}

\begin{figure}[htb!]
\begin{center}
\includegraphics[scale=0.4]{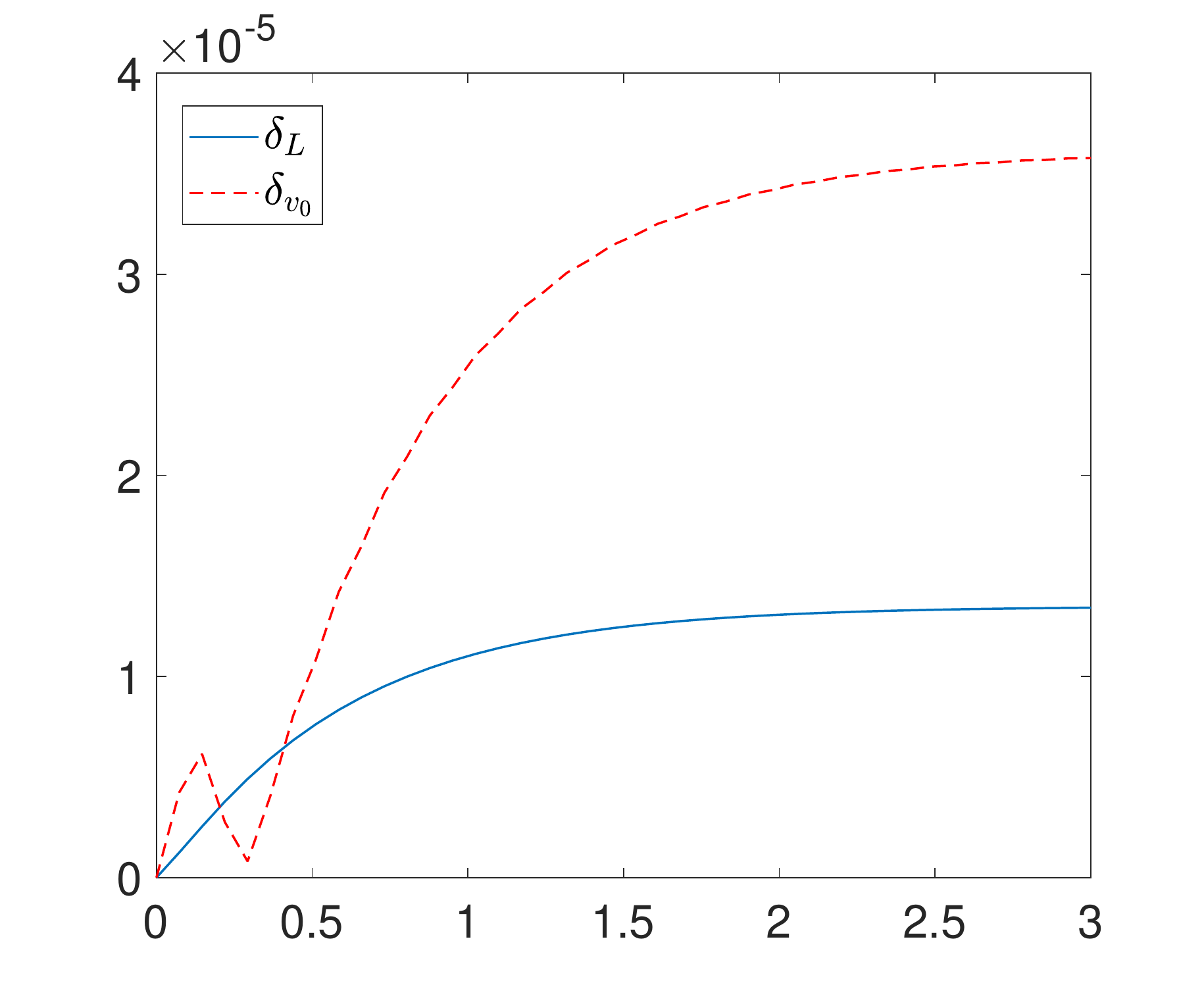}
\put(-109,-5){$t$ [s]}
\caption{The error of solution produced by the algorithm from the paper by Wrobel $\&$ Mishuris \cite{Wrobel_2015} for the crack length, $\delta_L$, and the crack propagation speed, $\delta_{v_0}$.}
\label{delta_L_v0_elast}
\end{center}
\end{figure}

In order to have a reference level for the results obtained with the proposed FEM-based algorithm we present in Figures \ref{delta_w_v_elast} -- \ref{delta_L_v0_elast} the computational errors produced by the original scheme from the paper by Wrobel $\&$ Mishuris \cite{Wrobel_2015}, where the crack opening was computed directly from the boundary integral equation of elasticity. The analyzed benchmark example and the time stepping strategy were exactly the same as those used in the above analysis.  The results show that the original algorithm yields up to three orders of magnitude better accuracy for the crack opening and up to two orders lower error for the fluid velocity. The crack length is computed with approximately 10 times better accuracy than with the FEM-based algorithm. On the other hand, after $\delta_{v_0}$ is stabilized with time, the accuracy of the crack propagation speed is similar in both compared algorithms. Nevertheless, even though the FEM-based version of the algorithm produces larger errors than the original scheme with the boundary integral equation of elasticity, the provided accuracy of solution is still relatively high and sufficient for any practical application.

Let us complement this subsection by discussing the efficiency of computations. To this end we compare the  total computational times observed for the respective variants of the FEM sub-problem and mesh densities. The normalized time of computations is defined as follows:
\[
t_\text{norm}=\frac{t_\text{total}}{t_\text{total}^\text{(el)}},
\]
where $t_\text{total}$ is the total computational time for the corresponding variant and $t_\text{total}^\text{(el)}$ is the total time of computations with the original algorithm introduced by Wrobel $\&$ Mishuris  \cite{Wrobel_2015} (i. e. the time recorded when producing results from Figures \ref{delta_w_v_elast} -- \ref{delta_L_v0_elast}). The obtained values of  $t_\text{norm}$ are:
\begin{itemize}
\item{for the `variant 1': 22.66 with the coarse mesh and 17.76 with the dense mesh;}
\item{for the `variant 2': 23.11 with the coarse mesh and 17.16 with the dense mesh;}
\end{itemize}
The above data reveals a surprising and counterintuitive trend. Namely, the computational cost is lower when using the refined FEM meshing. It stems from the fact the overall convergence rate of the algorithm is lower when utilizing the coarse mesh. Thus, employing finer FEM mesh is conducive not only to accuracy but also efficiency of computations.

\subsection{Numerical example}
\label{num_ex}

In order to demonstrate the potential of the developed algorithm we present a computational example of hydraulic fracture propagating through an interface between two neighbouring rock layers with differing material properties (another example of application can be found in the publication of Wrobel et al.\cite{Wrobel_plasticity} where the algorithm was used to analyze the problem of an elasto-plastic HF). The corresponding physical problem is relevant in those cases where HF propagates in the complex geological settings. Then, various scenarios of fracture containment (usually concerning the vertical fracture growth) need to be analyzed either to optimize the fracking treatments or to prevent the creation of pathways for fracturing fluids and hydrocarbons to pollute aquifers (for a detailed discussion on the topic see e.g.  the paper by Huang et al.\cite{Huang_2018}). Naturally, even for linear elasticity and the simplest configuration of the rock strata, such a problem is very difficult to be modelled with the boundary integral equation of elasticity as it requires obtaining a solution to the general 2D problem of solid deformation. Thus, the application of the FEM based version of the universal algorithm is justified here. Obviously, a complete analysis of the fracture containment problem goes far beyond the study presented below. Our intention is just to demonstrate the capability of the proposed algorithm to account for the respective physical features of a more complex phenomenon. 

For the sake of the computational example let us consider a hydraulic fracture propagating in a domain composed of two parts of essentially different stiffnesses. Respective subdomains are defined by the $x$ coordinate as: i) $x \in [0,1]$ m - the first subdomain, ii) $x>1$ m - the second subdomain. The material properties attributed to the subdomains are listed in Table \ref{mat_tab}. The fracture is oriented perpendicular to the interface between the two subdomains. Thus, before crossing the interface  the crack propagates solely in `material 1'. 
We assume that the fracturing fluid is a solution of partially hydrolyzed polyacrylamide (HPAM) with the concentartion of 150 weight parts per million. The rheological properties of this fluid can be described by the truncated power-law model:
\begin{equation}
\label{TP_bench}
\eta_\text{a}=
  \begin{cases}
		\eta_0       & \quad \text{for } \quad |\dot \gamma |<|\dot \gamma_1|,\\
    C |\dot \gamma|^{n-1}       & \quad \text{for} \quad |\dot \gamma_1|<|\dot \gamma|<|\dot \gamma_2|,\\
		
    \eta_\infty  & \quad \text{for } \quad |\dot \gamma|>|\dot \gamma_2|,
  \end{cases}
\end{equation}
where the corresponding parameters were provided by Wrobel \cite{Wrobel_2020}: $\eta_0=0.2668$ Pa$\cdot$s, $\eta_\infty=4.1\cdot 10^{-3}$ Pa$\cdot$s, $n=0.476$, $C=7.27\cdot 10^{-2}$ Pa$\cdot$s$^n$, $|\dot \gamma_1|=8.37\cdot 10^{-2}$ s$^{-1}$,  $|\dot \gamma_2|=241$ s$^{-1}$.

\begin{table}[]
\begin{center}
\begin{tabular}{c|c|c|c|}
\cline{2-4}
                                 & \begin{tabular}[c]{@{}c@{}}$E$\\ {[}GPa{]}\end{tabular} & $\nu$ & \begin{tabular}[c]{@{}c@{}}$K_I$\\ {[}MPa$\cdot \text{m}^{1/2}${]}\end{tabular} \\ \hline
\multicolumn{1}{|c|}{material 1} & 10                                                      & 0.24  & 1                                                         \\ \hline
\multicolumn{1}{|c|}{material 2} & 20                                                      & 0.3   & 1.5                                                       \\ \hline
\end{tabular}
\caption{Material properties attributed to the respective subdomains.}
\label{mat_tab}
\end{center}
\end{table}

The influx magnitude increases from zero for $t=0$ s to the maximum  $\bar q_0=5 \cdot 10^{-4}$ $ \frac{\text{m}^2}{\text{s}} $ at $t_1=0.1$ s and then is kept constant according to the following formula:
\begin{equation}
\label{q0_def}
 q_0(t)=
  \begin{cases}
	\left(\frac{3}{t_1^2}t^2-\frac{2}{t_1^3}t^3\right) \bar q_0      & \quad \text{for} \quad t<t_1,\\
    \bar q_0       & \quad \text{for} \quad t \geq t_1.
  \end{cases}
\end{equation}
Expression \eqref{q0_def} provides a smooth transition between the limiting values of the influx. The overall time of the process is set to $t_\text{end}=10$ s. The fluid leak-off to the rock formation is neglected. Initial fracture length and aperture are assumed zero. Consequently, the initial crack propagation speed is also zero. The computations are performed with the `variant 2' of the FEM problem configuration and the dense mesh of finite elements. The time stepping strategy is the same as the one accepted in the benchmark example from Subsection \ref{ver_alg}.

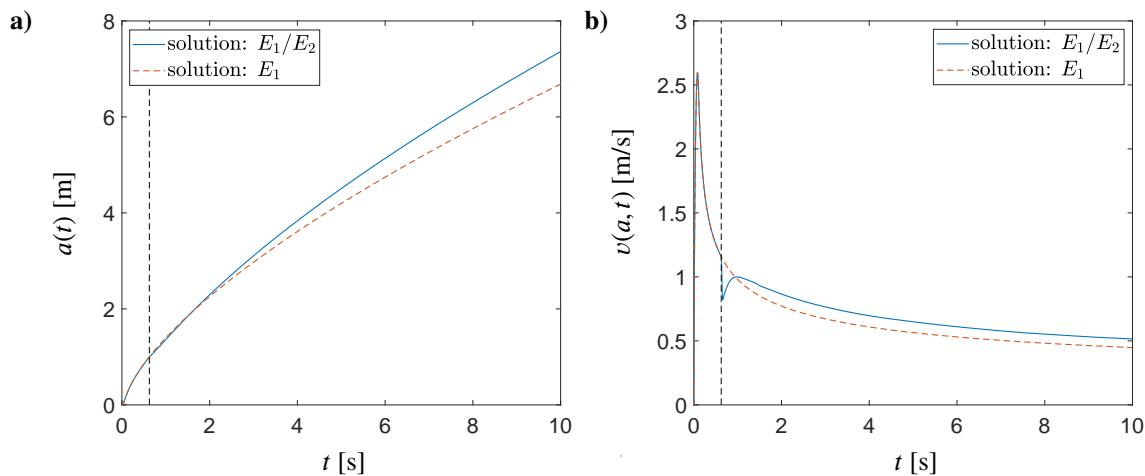
\begin{figure}[htb!]
\begin{center}
\includegraphics[scale=0.40]{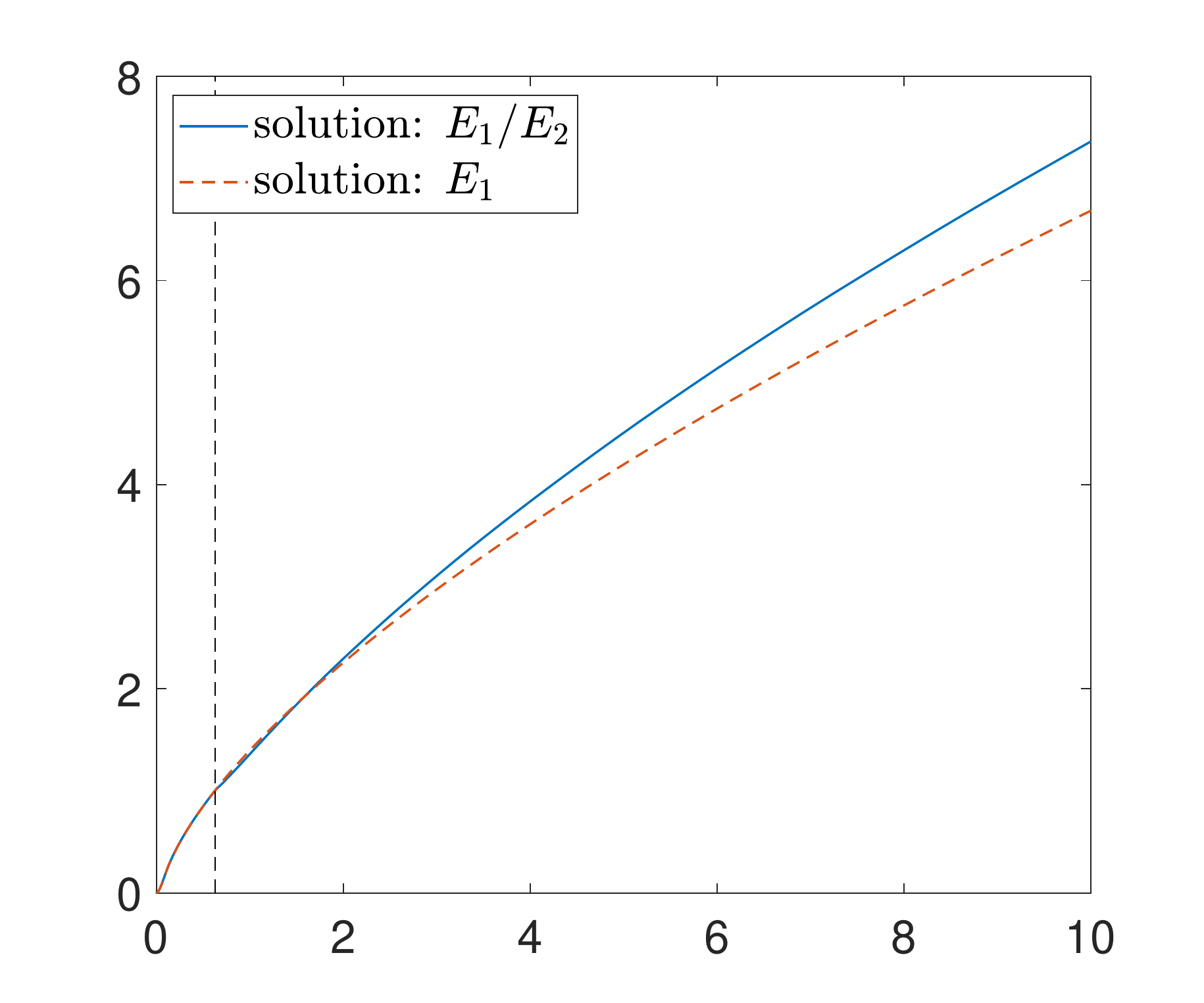}
\put(-210,75){\rotatebox{90}{$a(t)$ [m]} }
\put(-109,-5){$t$ [s]}
\hspace{0mm}
\includegraphics[scale=0.40]{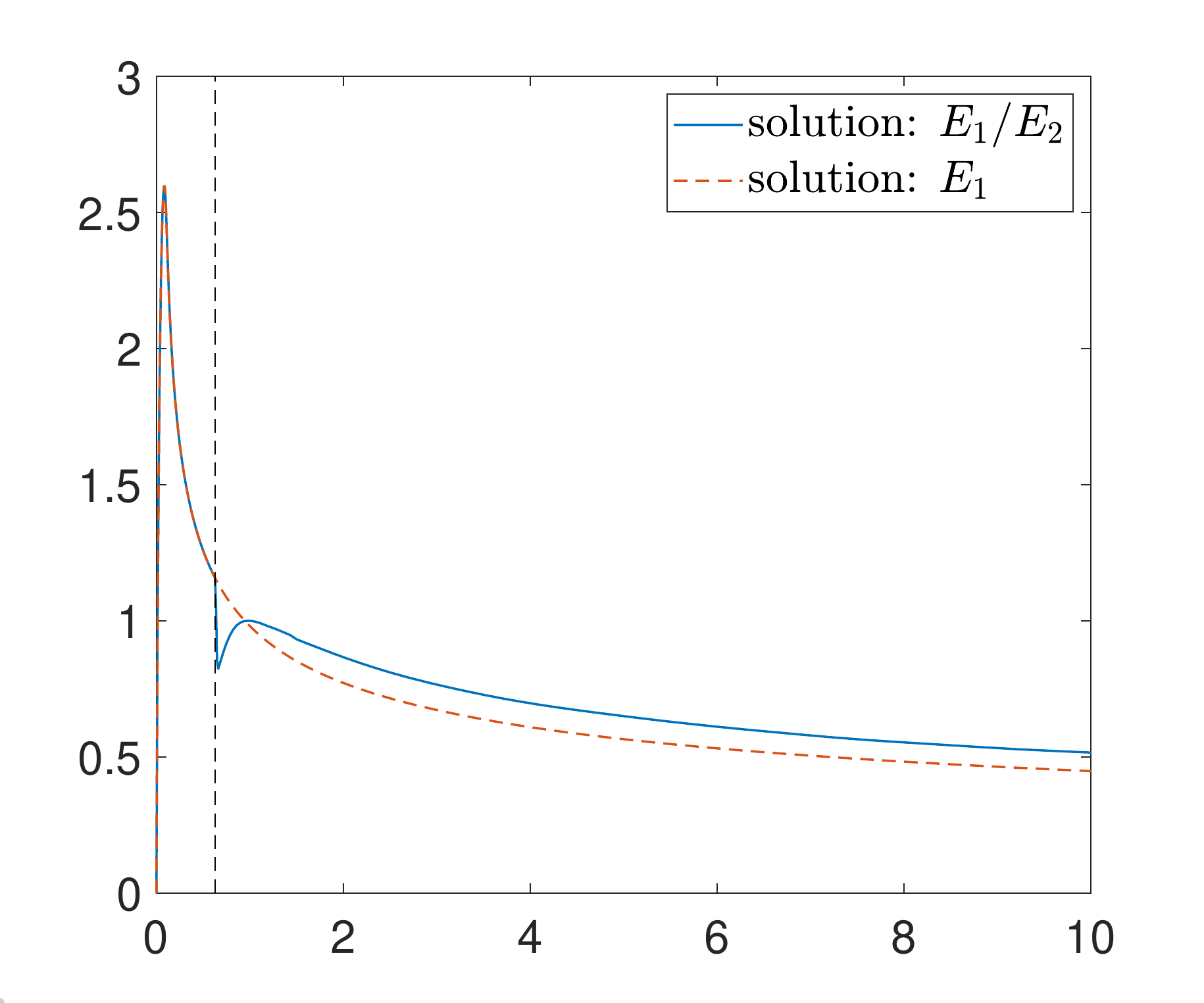}
\put(-109,-5){$t$ [s]}
\put(-215,75){\rotatebox{90}{$v(a,t)$ [m/s]}}
\put(-440,160){$\textbf{a)}$}
\put(-225,160){$\textbf{b)}$}
\caption{The evolution of: a) the crack length, $a(t)$ [m], b) the crack propagation speed, $v(a,t)$ [m/s]. The vertical dashed line marks the time instant ($t=0.63$ s) when the fracture crosses the interface between respective subdomains.}
\label{L_v0}
\end{center}
\end{figure}

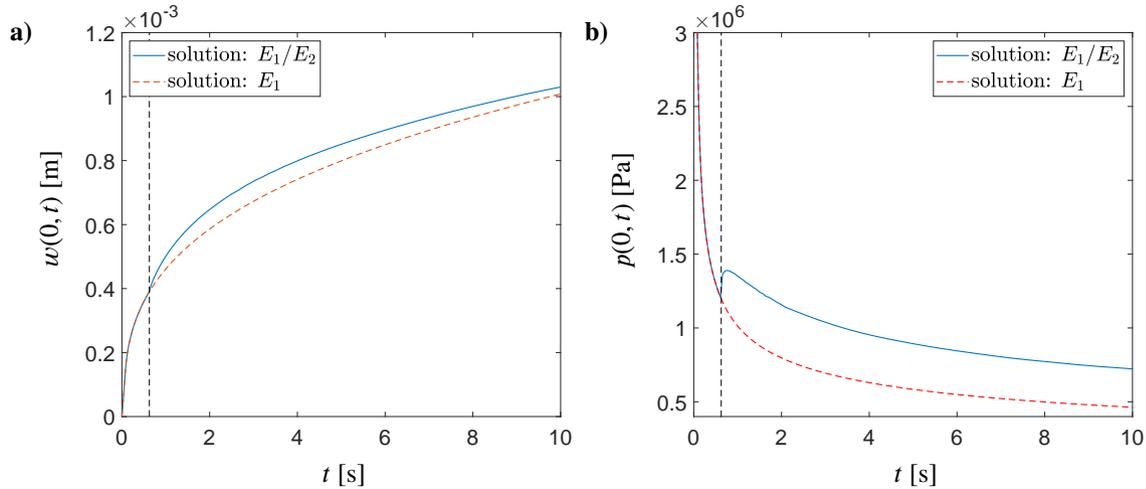
\begin{figure}[htb!]
\begin{center}
\includegraphics[scale=0.40]{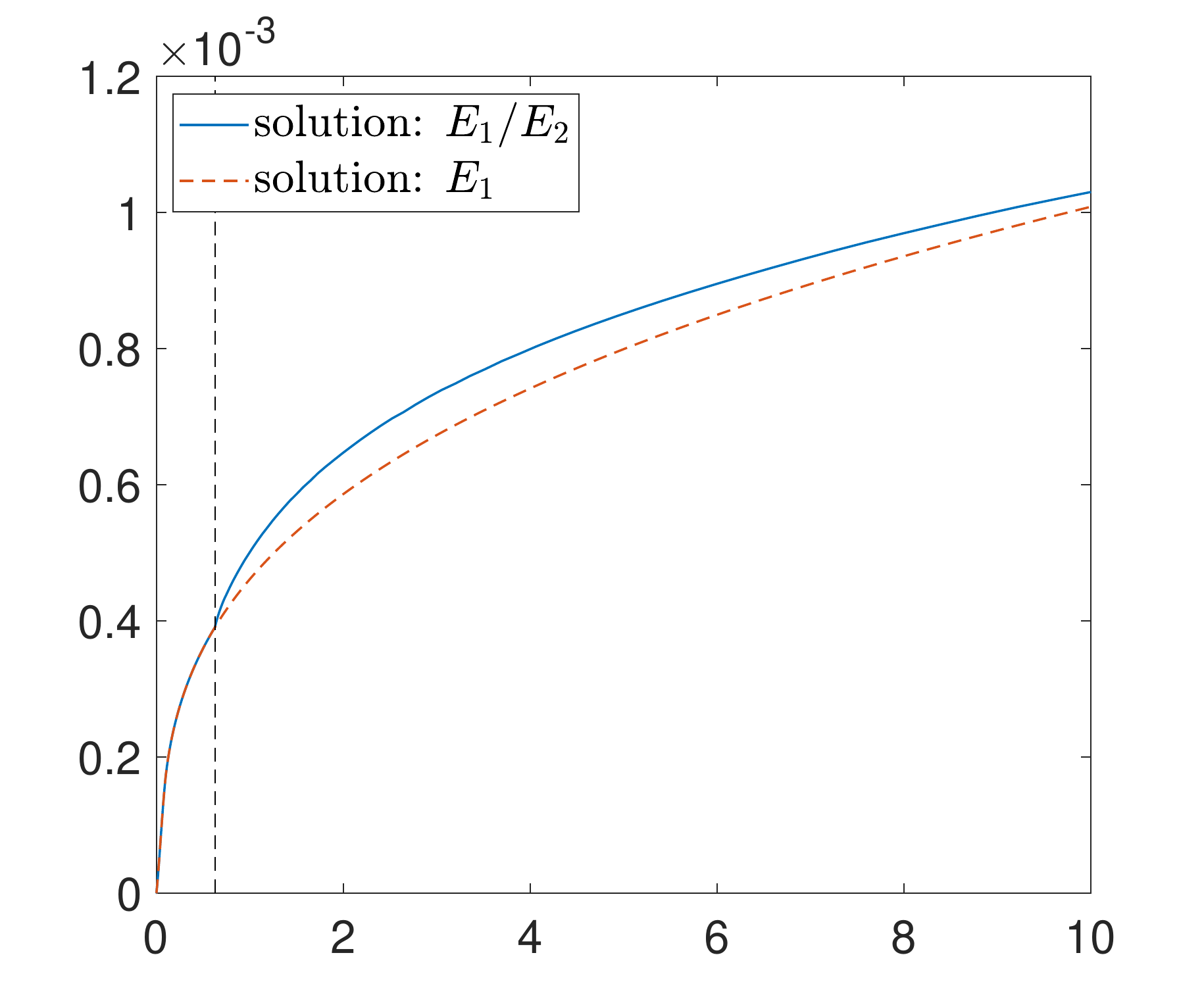}
\put(-215,75){\rotatebox{90}{$w(0,t)$ [m]}}
\put(-109,-5){$t$ [s]}
\hspace{0mm}
\includegraphics[scale=0.40]{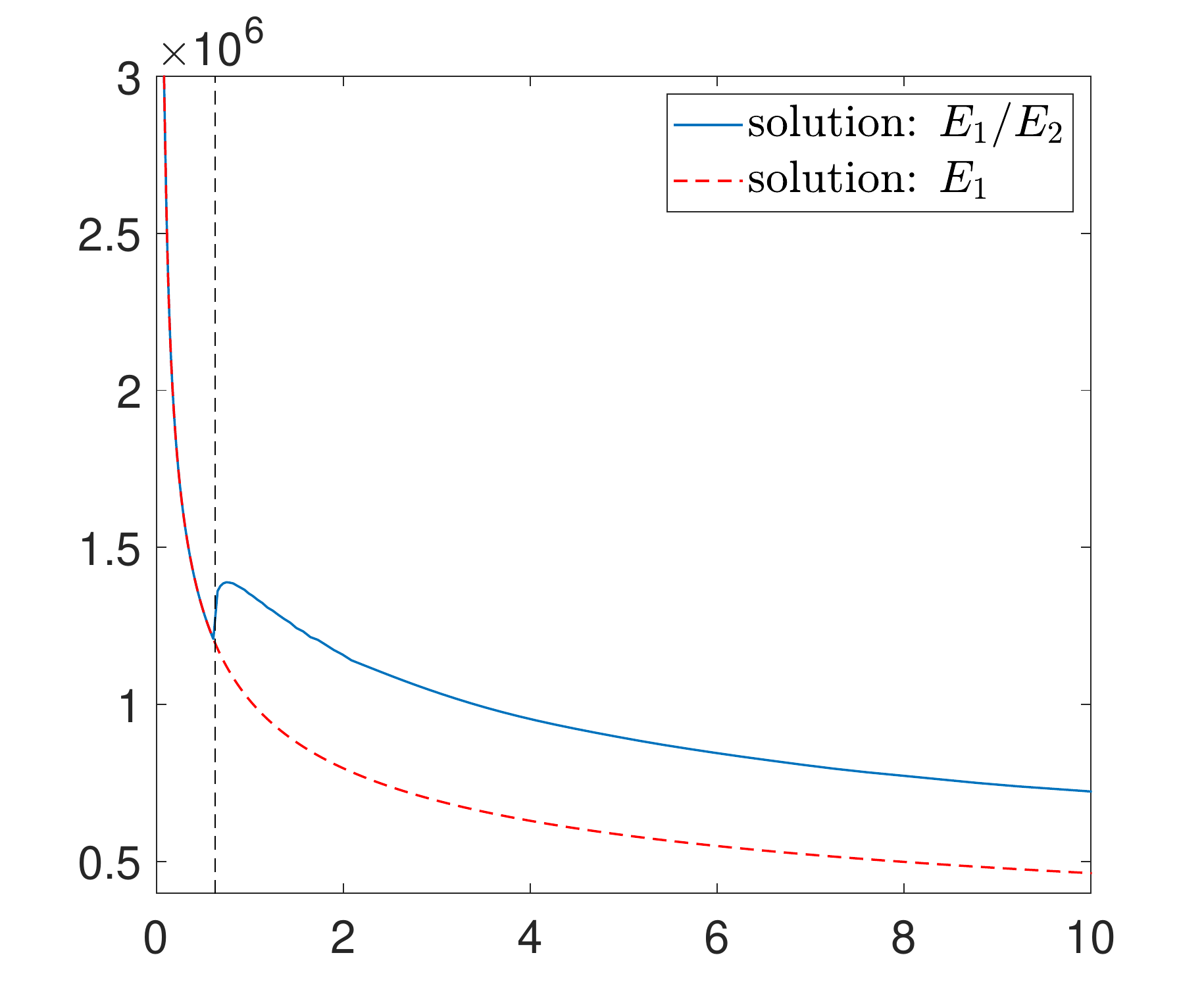}
\put(-109,-5){$t$ [s]}
\put(-215,75){\rotatebox{90}{$p(0,t)$ [Pa]}}
\put(-440,160){$\textbf{a)}$}
\put(-225,160){$\textbf{b)}$}
\caption{The evolution of: a) the crack opening at the crack mouth, $w(0,t)$ [m], b) the borehole pressure, $p(0,t)$ [Pa]. The vertical dashed line marks the time instant ($t=0.63$ s) when the fracture crosses the interface between respective subdomains.}
\label{w_p}
\end{center}
\end{figure}

\begin{figure}[htb!]
\begin{center}
\includegraphics[scale=0.40]{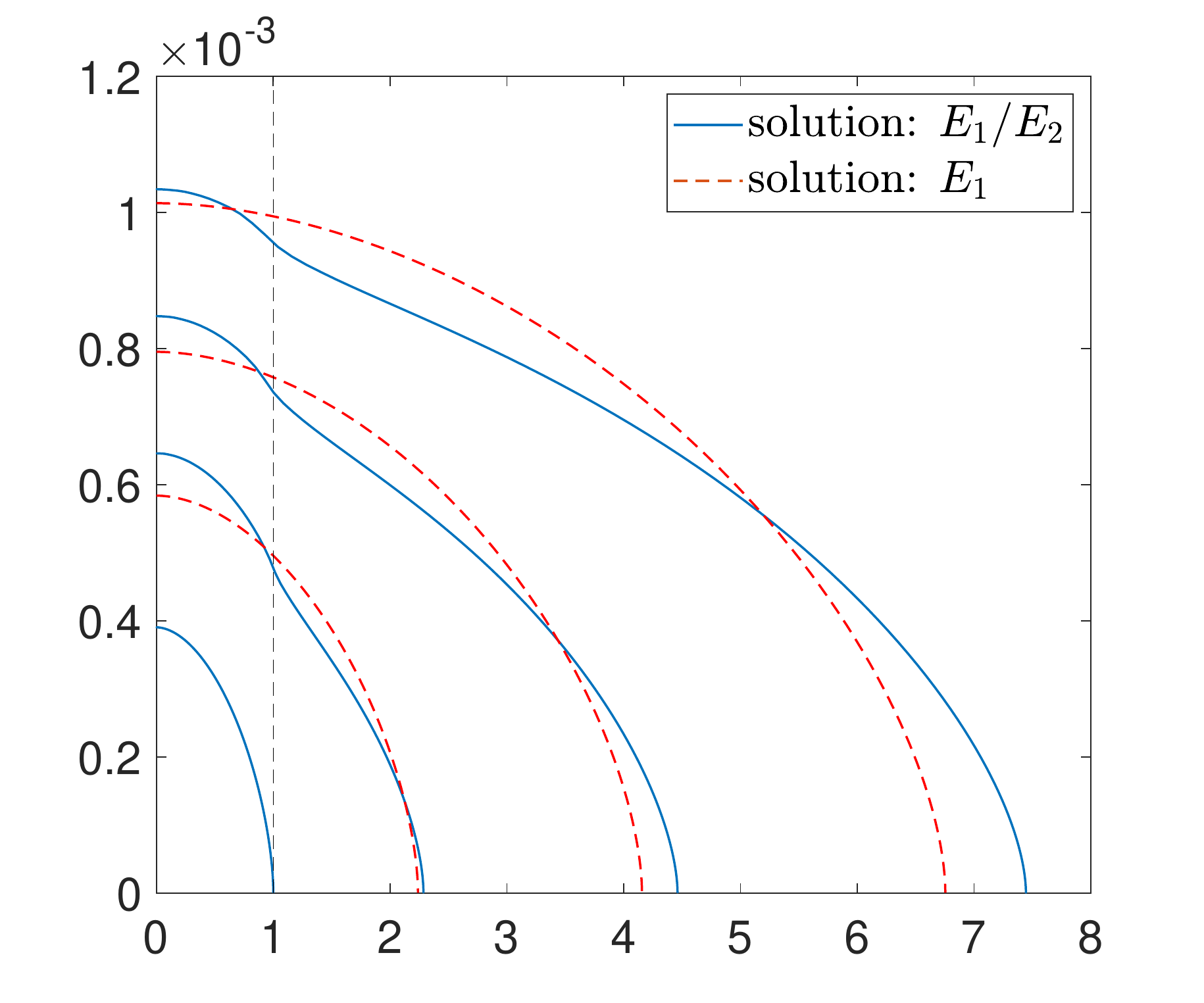}
\put(-215,75){\rotatebox{90}{$w(x,t)$ [m]}}
\put(-109,-5){$x$ [m]}
\put(-183,25){$t=0.63$ s}
\put(-138,35){$t=2$ s}
\put(-110,65){$t=5$ s}
\put(-70,85){$t=10$ s}
\hspace{0mm}
\includegraphics[scale=0.40]{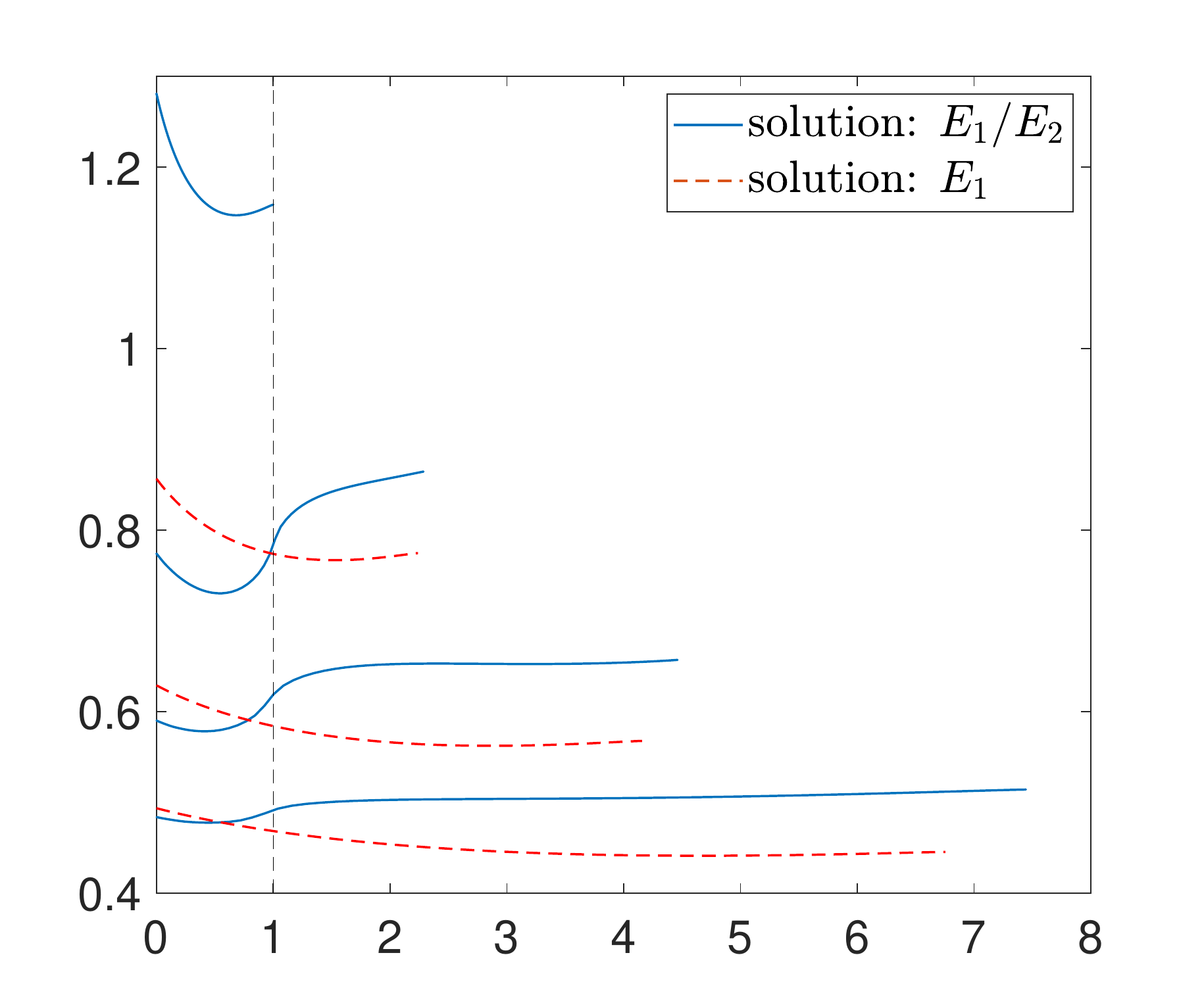}
\put(-109,-5){$x$ [m]}
\put(-215,70){\rotatebox{90}{$v(x,t)$ [m/s]}}
\put(-160,140){$t=0.63$ s}
\put(-138,83){$t=2$ s}
\put(-90,50){$t=5$ s}
\put(-57,28){$t=10$ s}
\put(-440,160){$\textbf{a)}$}
\put(-225,160){$\textbf{b)}$}
\caption{The distribution of: a) the crack opening, $w(x,t)$ [m], b) the fluid velocity, $v(x,t)$ [m/s], for four different instants of time. The vertical dashed line ($x=1$ m) marks  the interface between respective subdomains.}
\label{w_v_distribution}
\end{center}
\end{figure}
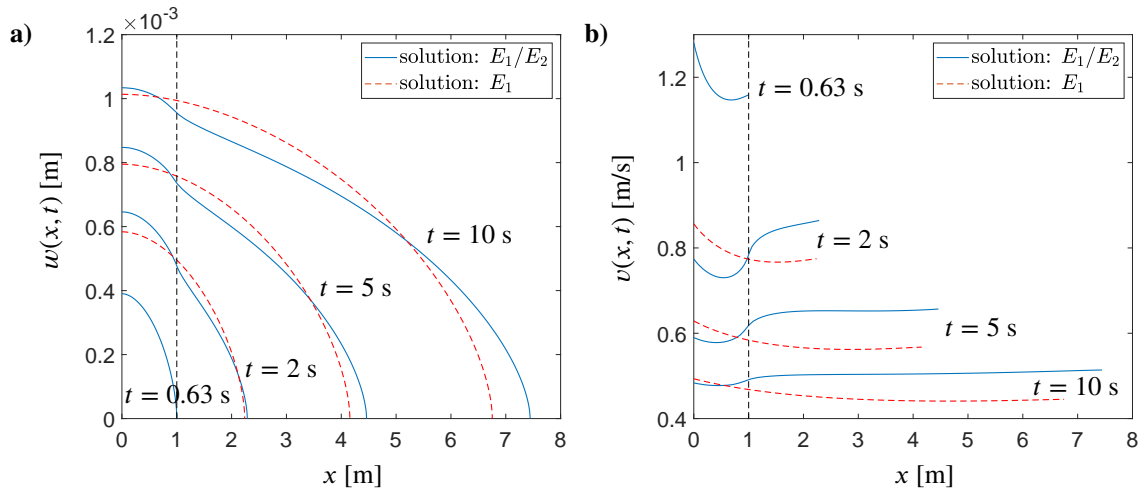

\begin{figure}[htb!]
\begin{center}
\includegraphics[scale=0.40]{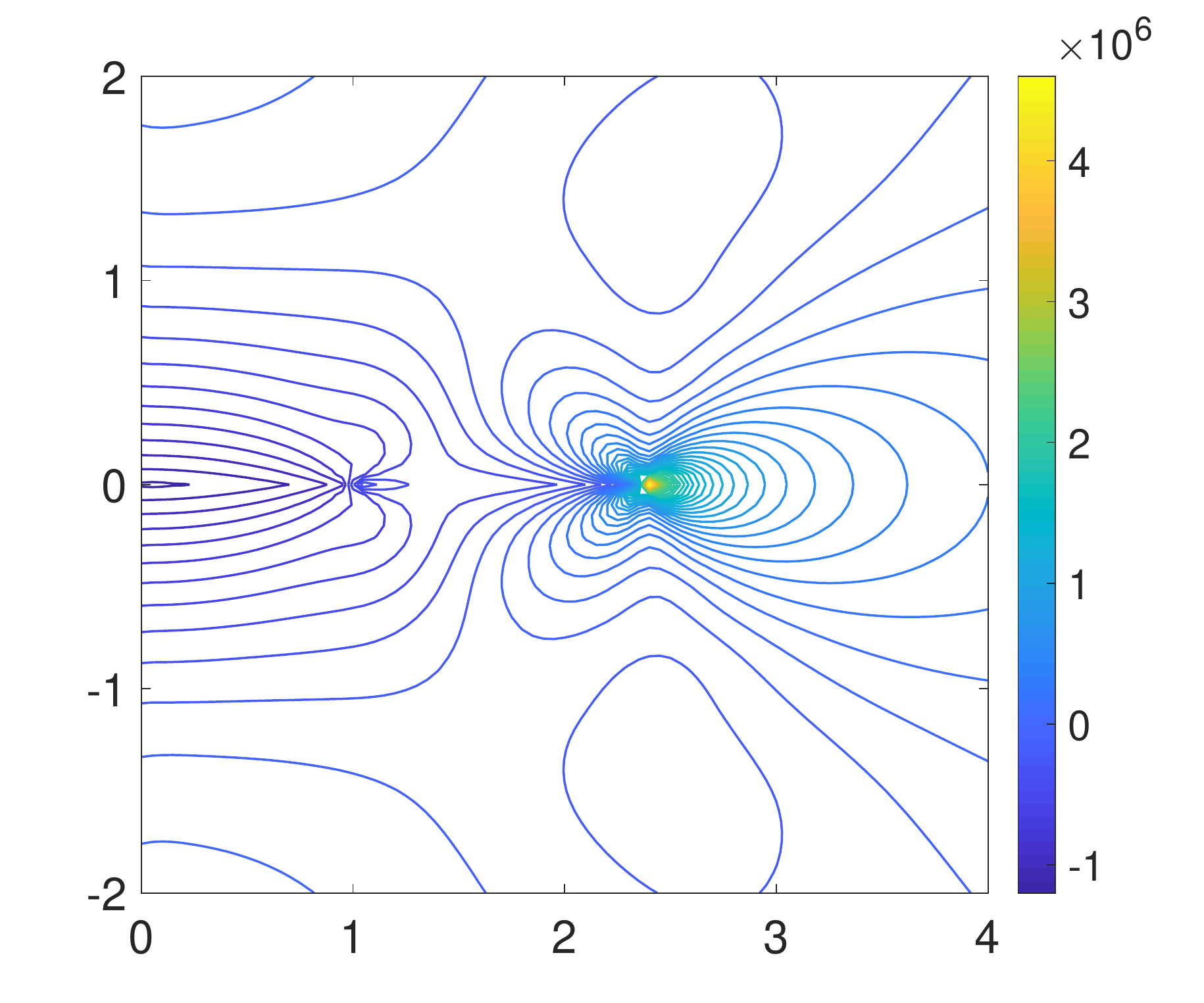}
\put(-205,83){\rotatebox{90}{$y$ [m]}}
\put(-121,0){$x$ [m]}
\put(-116,170){$\sigma_{11}$}
\hspace{0mm}
\includegraphics[scale=0.40]{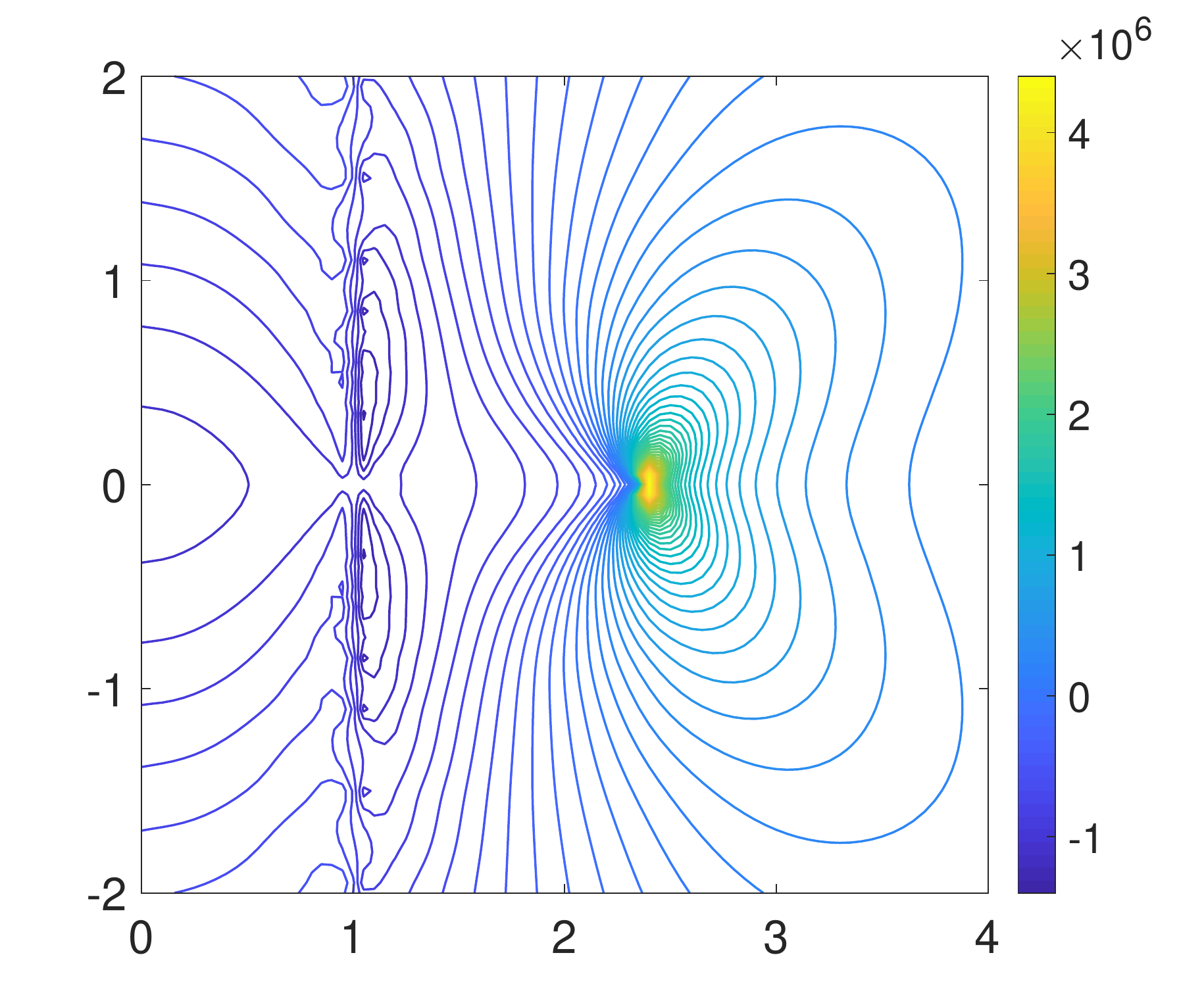}
\put(-205,83){\rotatebox{90}{$y$ [m]}}
\put(-121,0){$x$ [m]}
\put(-116,170){$\sigma_{22}$}
\put(-430,165){$\textbf{a)}$}
\put(-215,165){$\textbf{b)}$}

\includegraphics[scale=0.40]{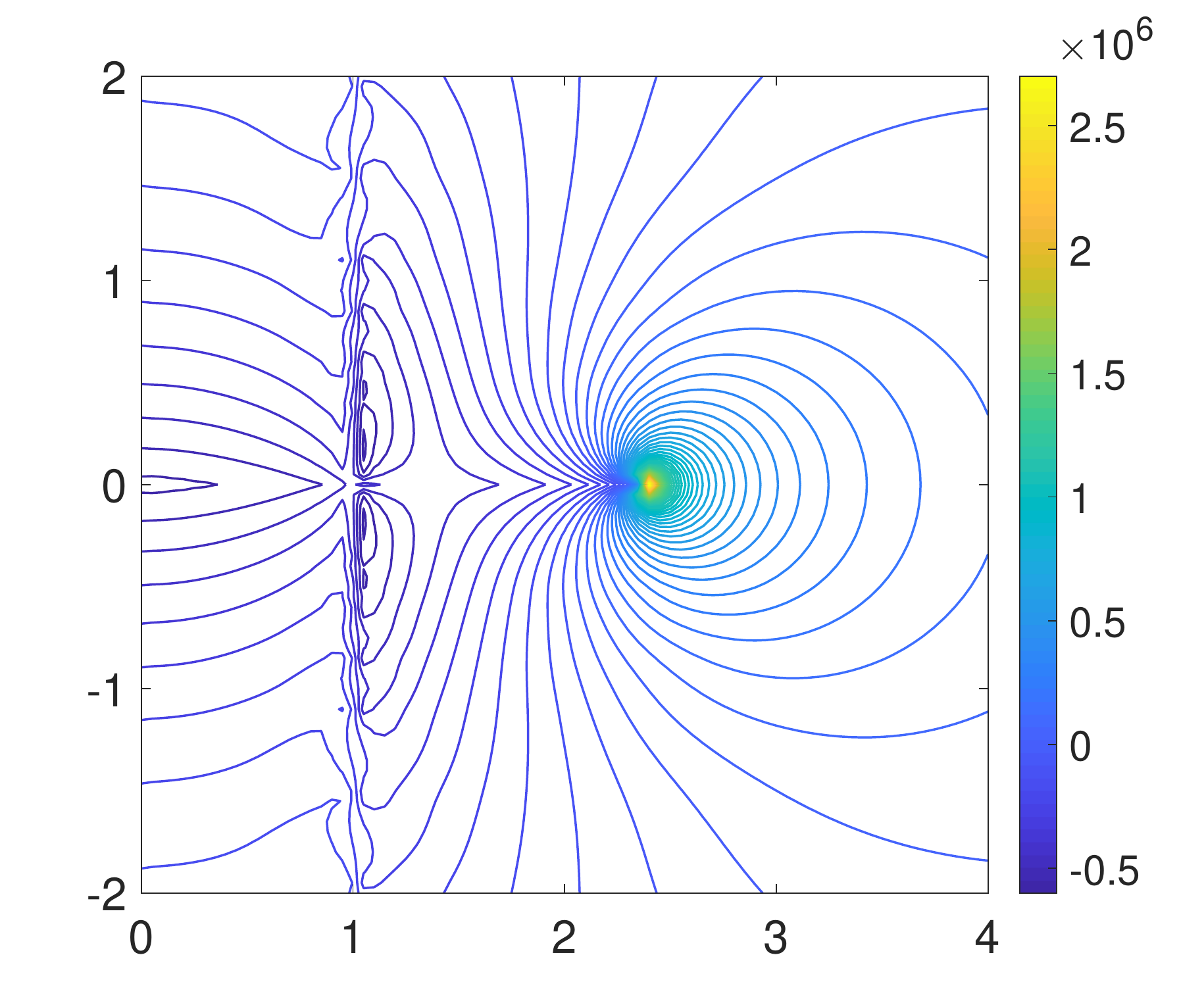}
\put(-210,160){$\textbf{c)}$}
\put(-205,83){\rotatebox{90}{$y$ [m]}}
\put(-121,0){$x$ [m]}
\put(-116,170){$\sigma_{33}$}
\caption{The principal stresses for $t=2$ s: a) $\sigma_{11}$ [Pa], b) $\sigma_{22}$ [Pa], c) $\sigma_{33}$ [Pa]. The interface between the two material strata occurs at $x=1$ m.}
\label{sigma_t2}
\end{center}
\end{figure}

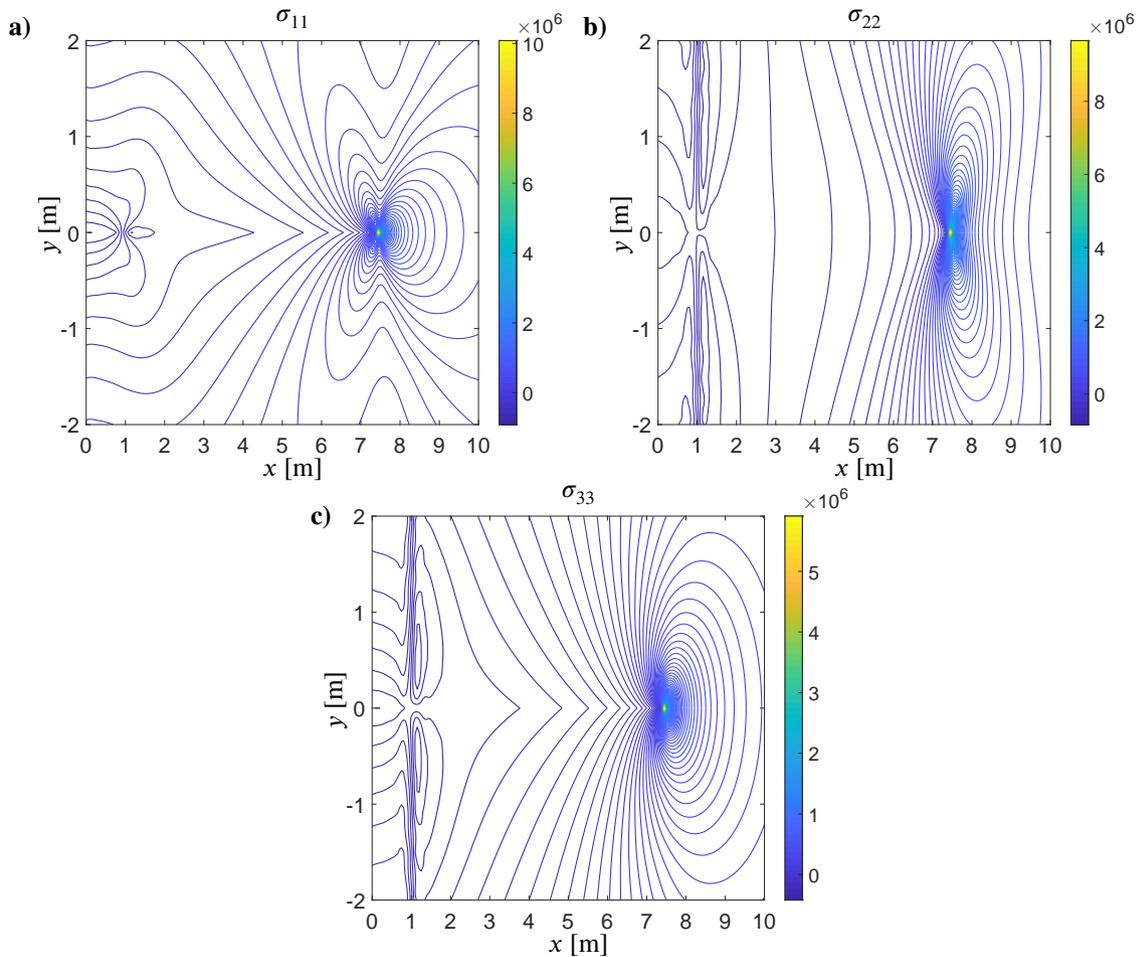
\begin{figure}[htb!]
\begin{center}
\includegraphics[scale=0.40]{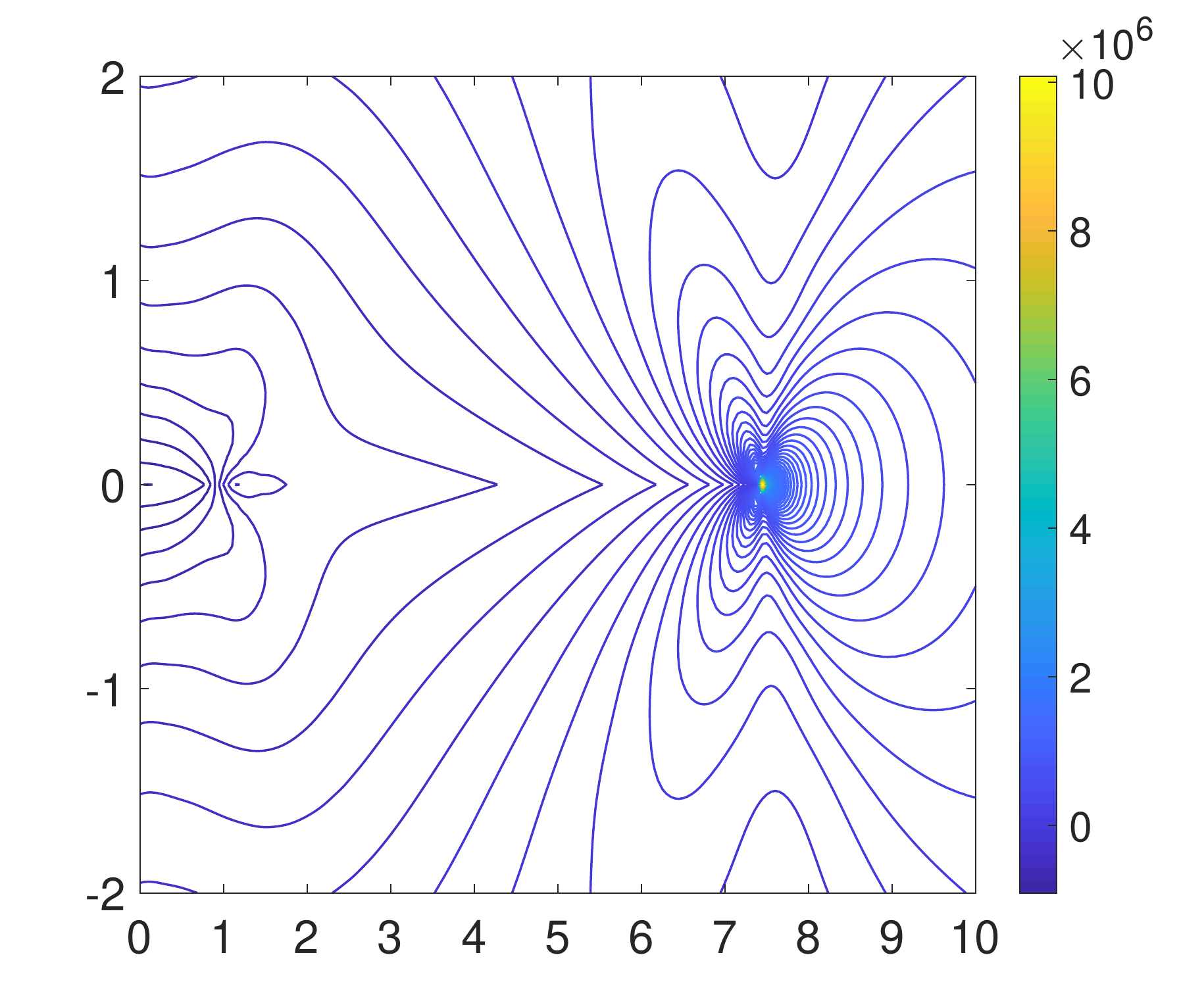}
\put(-205,83){\rotatebox{90}{$y$ [m]}}
\put(-121,0){$x$ [m]}
\put(-116,170){$\sigma_{11}$}
\hspace{0mm}
\includegraphics[scale=0.40]{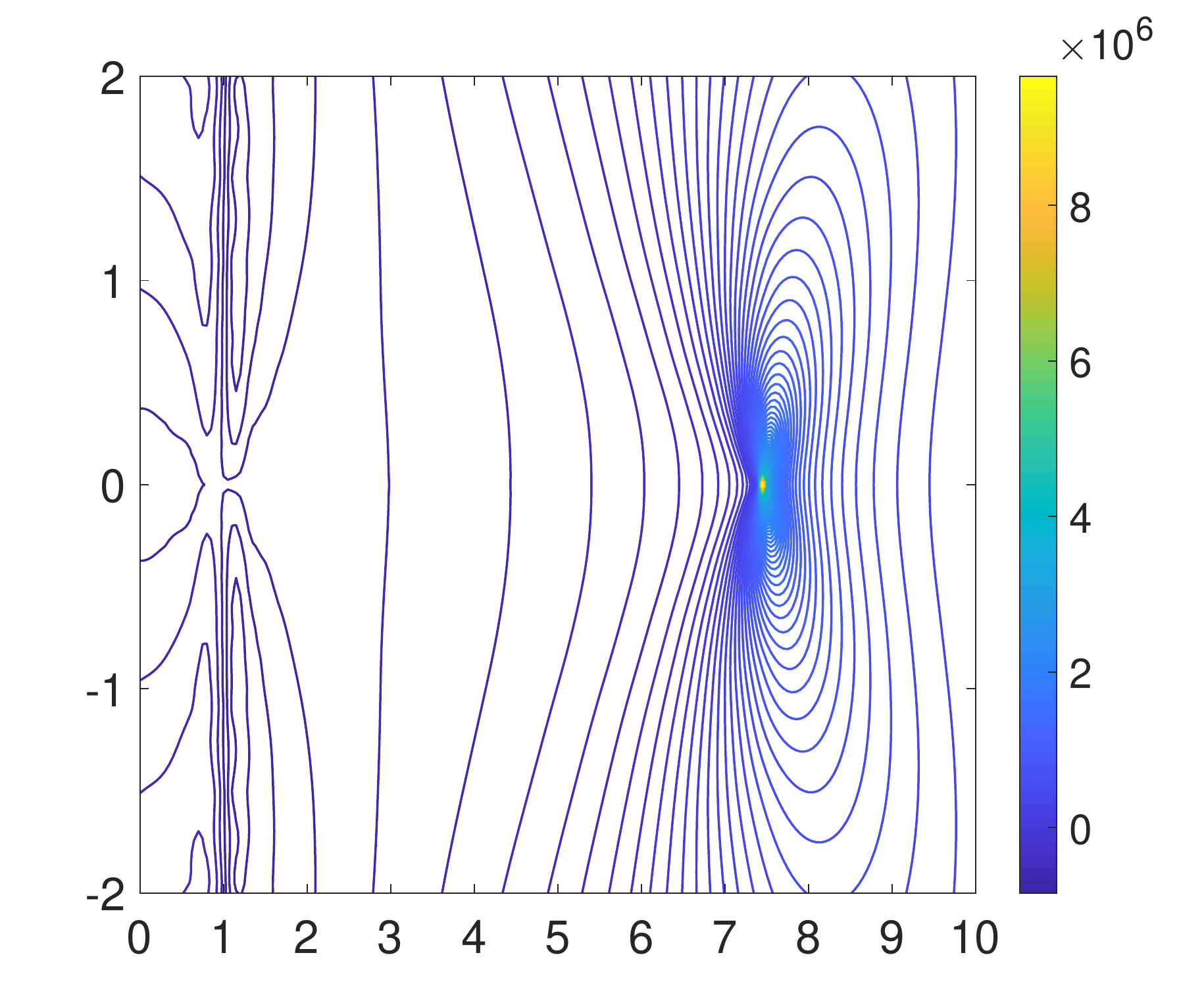}
\put(-205,83){\rotatebox{90}{$y$ [m]}}
\put(-121,0){$x$ [m]}
\put(-116,170){$\sigma_{22}$}
\put(-430,165){$\textbf{a)}$}
\put(-215,165){$\textbf{b)}$}

\includegraphics[scale=0.40]{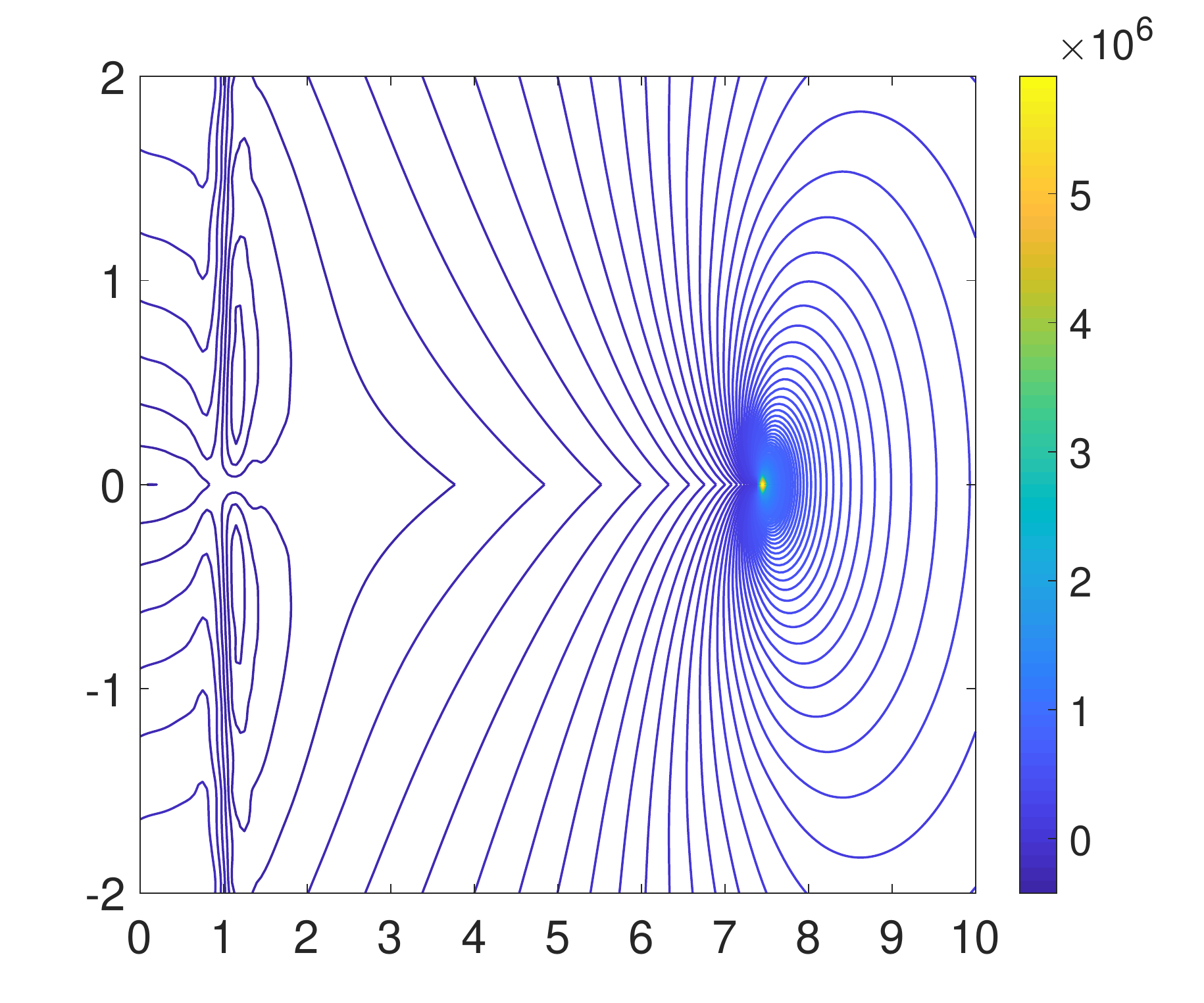}
\put(-210,160){$\textbf{c)}$}
\put(-205,83){\rotatebox{90}{$y$ [m]}}
\put(-121,0){$x$ [m]}
\put(-116,170){$\sigma_{33}$}
\caption{The principal stresses for $t=10$ s: a) $\sigma_{11}$ [Pa], b) $\sigma_{22}$ [Pa], c) $\sigma_{33}$ [Pa]. The interface between the two material strata occurs at $x=1$ m.}
\label{sigma_t10}
\end{center}
\end{figure}

The results of simulations are presented in  Figures \ref{L_v0} -- \ref{sigma_t10}. Results obtained assuming uniform properties of the fractured solid (identical to those of `material 1') are included in the figures for comparison. 
In the figures' legends the notation `solution $E_1/E_2$' is used to denote the solution obtained for the complex geological settings, whereas `solution $E_1$' refers to the case of uniform material properties.

To understand better the presented data let us recall that after crossing the interface between the subdomains the fracture is subjected to two counteracting mechanisms related to the material properties. As the fracture toughness of `material 2' is greater than the one of `material 1' the crack is expected to increase its aperture (with respect to the crack propagating in `material 1' only) at the expense of its length. On the other hand, the higher value of Young's modulus of `material 2' contributes to a reverse trend. Thus, the overall fracture geometry results from the interplay between these two processes.

In Figure \ref{L_v0}b) one can observe an instantaneous drop of the crack propagation speed  after the subdomain's interface is crossed. Clearly, this is a result of the increased fracture toughness. However, very shortly after crossing the threshold the crack begins to accelerate, with the velocity soon exceeding that obtained for the uniform material (`material 1'). Consequently, the fracture length becomes greater (Figure \ref{L_v0} a)) than than that for `material 1'. This time it is the Young's modulus dependent mechanism that causes the change. 

A surprising observation can be made from Figure \ref{w_p} a). Namely, it shows that the crack aperture at $x=0$ is always larger with `solution $E_1/E_2$'. It is a counterintuitive trend as for times greater than approximately $t=2$ s the crack length is also higher for this variant of solution. In order to explain this apparent contradiction let us have a look at Figure \ref{w_v_distribution}a) where the fracture profiles are shown for a few time instants. We can see there a shape peculiarity in the proximity of the subdomain's interface.   The observed deflection of the crack profile causes that the resulting fracture can be simultaneously longer and wider at the mouth than its uniform counterpart, even though both variants of fracture contain the same volume of fluid at every time instant. Anyway, this trend becomes less pronounced with time and is expected to vanish completely after sufficiently large time (i.e. as the distance from the crack opening, $x=0$, m to the interface location, $x=1$, m becomes negligible relative to the fracture length).
The fracture profile deflection at the interface also results in a local increase of the fluid velocity, as shown in Figure \ref{w_v_distribution}b). 

In Figure \ref{w_p}b) the borehole pressure is depicted as a function of time. A very steep pressure increase can be seen when the fracture crosses the interface between the subdomains. Note that in the implementation of fracking treatments this information is important not only to estimate the power input required for fluid injection but also  to detect the transition of the fracture between different rock strata.

Finally, in Figures \ref{sigma_t2}--\ref{sigma_t10} we show the isoline plots of the principal stresses in the vicinity of the fracture surface for two time instants: $t=2$ s and $t=10$ s. In both cases a very clear stress contrast can be identified along the line $x=1$ m for $\sigma_{22}$ and $\sigma_{33}$.

\section{Conclusions}
\label{conc}

In this paper the problem of numerical simulation of hydraulic fractures was considered. An efficient and versatile computational algorithm that employs the Finite Element Method was constructed. The algorithm is based on the scheme first introduced by Wrobel $\&$ Mishuris \cite{Wrobel_2015} for the HF propagating in an elastic material and modified later to account for non-Newtonian fluid rheologies and tangential traction exerted by the fluid on the crack faces. In the current version of the algorithm the subroutine for computation of the crack opening was built as a FEM-based module. The computational accuracy of the developed algorithm was investigated against analytical benchmark solutions, while its capabilities were demonstrated using the example of a hydraulic
fracture propagating across an interface between dissimilar rock strata.

The following conclusions can be drawn from the presented analysis:
\begin{itemize}
\item{The universal algorithm first introduced by Wrobel $\&$ Mishuris \cite{Wrobel_2015} can be effectively combined with  the FEM module to compute the deformation of the fractured material. This combination provides  simplicity and stability of computations of the original scheme, while simultaneously enabling one to analyze advanced cases of material properties (such as poro-elasto-plastic behaviour) and complex geological settings.  }
\item{Different configurations of the employed FEM module can be utilized within the scheme, depending on
the particular HF variant being analyzed.  A number of features, such as: spatial layout of the domain, type and number of finite elements, mesh pattern and density, can be adjusted/optimized to best fit the analyzed case.  It should be noted that the solver is still highly effective even without fine-tuning of the FEM module, due to the aforementioned stability introduced by the other components of the scheme.}
\item{The accuracy of FEM computations is important for the effectiveness and computational cost of the proposed scheme. It has been shown the reduction of the FEM mesh density (and corresponding decrease of the FEM solution accuracy) can deteriorate the overall convergence rate of the algorithm and extend the computational time. From this point of view it is beneficial to keep dense FEM meshing not only for the accuracy but also efficiency of computations. }
\item{The accuracy of computation of the basic HF parameters with the FEM based version of the algorithm is lower than that achievable by the original scheme proposed by Wrobel $\&$ Mishuris \cite{Wrobel_2015} (provided that the analyzed problem can be solved by the latter algorithm). However, the presented algorithm provides the accuracy still sufficient for any practical application.}
\item{In these cases where the simulations can be performed by the algorithm introduced by Wrobel $\&$ Mishuris\cite{Wrobel_2015},  the FEM module of the proposed scheme can be utilized for post-processing of the results. In this way the complete  2D fields of stress and displacement can be determined, with the overall efficiency and accuracy of computations of the original (more effective) scheme. }
\end{itemize}

\vspace{10mm}
\noindent
{\bf Funding:} 
This work was funded by European Regional Development Fund and the Republic of Cyprus
through the Research Promotion Foundation (RESTART 2016 - 2020 PROGRAMMES, Excellence Hubs,
Project EXCELLENCE/1216/0481). DP would like to thank the Welsh Government's S\^{e}r Cymru II Research Programme, supported by the European Regional Development Fund.

\section*{Acknowledgments}
The authors are thankful to Professor Gennady Mishuris for his useful comments and discussions.

\appendix
\section{Self-similar formulation for the KGD model}
\label{ap_A}

The analytical benchmark solution given  in the Appendix \ref{ap_B} is based on the self-similar formulation of the classical KGD problem. Note that the KGD model constitutes a simplified version of the more general HF problem considered in this paper. The simplifications include the following elements: i) the fluid is assumed to be Newtonian of a constant viscosity $\eta$, ii) the hydraulically induced tangential traction on the fracture walls is neglected. The first assumption implies the function $F(x,t)$ from equation \eqref{Poiseulle} turns identically to unity. Consequently, equation \eqref{Poiseulle} transforms to the standard Poiseulle equation:
\begin{equation}
\label{Poiseulle_stn}
q(x,t)=-\frac{1}{M}w^3\frac{\partial p}{\partial x},
\end{equation}
where $M=12\eta$.
As a result of the second assumption the relation between the fluid pressure and the crack opening converts to the classical form of the boundary integral equation of elasticity:
\begin{equation}
\label{elast_bound_stn}
p(x,t)=k_2\int_0^{a(t) }\frac{\partial w(s,t)}{\partial s}\frac{s}{x^2-s^2}ds. \quad \quad 0\le x<a(t),
\end{equation}
It is possible to derive a benchmark solution without imposing the second simplification, which was shown in the paper by Wrobel et al.  \cite{Wrobel_2017}. In such a case an additional predefined function needs to be included in the elasticity operator (or in the fluid pressure introduced in the FEM module). However, in this paper for the sake of clarity we employ the simpler variant.

Let us employ the following self-similar scaling of the problem (under simplifications \eqref{Poiseulle_stn}--\eqref{elast_bound_stn}) for the normalized spatial variable $\tilde x=x/a(t)$:

\[
a(t)=a_0^{3/2}e^{\beta t}, \quad w(x,t)=\sqrt{a_0}e^{\beta t}\hat w(\tilde x), \quad p(x,t)=\hat p(\tilde x), \quad q(x,t)=e^{2\beta t} \hat q(\tilde x), \quad q_0(t)=\hat q_0 e^{2\beta t},
\]
\begin{equation}
\label{sel_sim_scl}
\end{equation}
\[
q_\text{L}(x,t)=\beta e^{\beta t}\sqrt{a_0}\hat q_\text{L}(\tilde x), \quad v(x,t)=\sqrt{\frac{1}{a_0}}e^{\beta t}\hat v(\tilde x), \quad K_I=\left(a_0\right)^{1/4}e^{\beta t/2}\hat K_I,
\]
where $\beta>0$ is predefined, while $a_0>0$ is to be determined. Note that by using representation \eqref{sel_sim_scl} together with the self-similar solution from Appendix \ref{ap_B} one can easily recreate the benchmark example employed in Section \ref{num_an} to verify the accuracy of computations.

Under the above scaling the continuity equation \eqref{cont} is transformed to the following ODE:
\begin{equation}
\label{cont_ss}
\hat w- \tilde x \frac{\text{d}\hat w}{\text{d}\tilde x}+\frac{1}{\beta a_0^2} \frac{\text{d}\hat q}{\text{d}\tilde x}+\hat q_\text{L}=0, \quad \tilde x \in [0,1],
\end{equation}
where the self-similar fluid flow rate is defined as:
\begin{equation}
\label{q_ss}
\hat q=-\frac{1}{M}\hat w^3 \frac{\text{d}\hat p}{\text{d}\tilde x}.
\end{equation}
Consequently, the self-similar fluid velocity yields:
\begin{equation}
\label{v_ss}
\hat v=-\frac{1}{M}\hat w^2 \frac{\text{d}\hat p}{\text{d}\tilde x}.
\end{equation}
The elasticity operator \eqref{elast_bound_stn} is converted to:
\begin{equation}
\label{p_ss}
\hat p=\frac{k_2}{a_0}\int_0^1\frac{\text{d}\hat w}{\text{d}\eta}\frac{\eta}{\tilde x^2-\eta^2}\text{d}\eta.
\end{equation}
From the condition \eqref{SE} the interrelation between the self-similar crack length, $a_0$, and the self-similar crack propagation speed, $\hat v_0$, is derived:
\begin{equation}
\label{SE_ss}
\hat v_0=\beta a_0^2.
\end{equation}
The respective conditions for the self-similar solution include:
\begin{itemize}
\item{the crack tip boundary conditions
\begin{equation}
\label{ss_tip}
\hat w(1)=0, \quad \hat q(1)=0,
\end{equation}}
\item{the influx boundary condition
\begin{equation}
\label{q0_ss}
\hat q(0)=\hat q_0.
\end{equation}}
\end{itemize}
\section{Analytical benchmark solution}
\label{ap_B}

The methodology of constructing the analytical benchmark example is directly adopted from the paper of Wrobel $\&$ Mishuris \cite{Wrobel_2015}. It assumes that a predefined form of the benchmark crack opening function, $\hat w_\text{b}$, is selected provided that: i) it complies with the desired asymptotic behaviour of the solution, ii) the corresponding fluid pressure, $\hat p_\text{b}$, can be computed from the elasticity operator \eqref{p_ss} in a closed form. Then the resulting fluid flow rate, $\hat q_\text{b}$, is obtained in an analytical form from \eqref{q_ss}. Finally, the computed benchmark functions are substituted to the self-similar continuity equation \eqref{cont_ss} and the benchmark form of the leak-off function, $\hat q_\text{L(b)}$, is defined (note that, in this benchmark representation of the problem, the leak-off function is not an element of solution but a known predefined relation). In this way, the set of benchmark functions satisfies identically the continuity equation and the boundary integral equation of elasticity. When using the benchmark example to test the algorithm of computations the respective boundary conditions are taken in accordance with the selected benchmark functions.

In our benchmark example let us accept the following form of the crack opening function:
\begin{equation}
\label{w_bench}
\hat w_\text{b}(\tilde x)=\sum_{i=0}^3 \hat w_i h_i(\tilde x),
\end{equation}
where:
\[
h_0(\tilde x)=\sqrt{1-\tilde x^2}, \quad h_1(\tilde x)=1-\tilde x^2, \quad h_2(\tilde x)=(1-\tilde x^2)^{3/2}\ln (1-\tilde x^2),
\]
\begin{equation}
\label{h_def}
\end{equation}
\[
h_3(\tilde x)=2\sqrt{1-\tilde x^2}+\tilde x^2\ln \left|\frac{1-\sqrt{1-\tilde x^2}}{1+\sqrt{1-\tilde x^2}} \right|.
\]
Note that the representation \eqref{w_bench}--\eqref{h_def} complies with the asymptotics of the so-called toughness dominated regime of crack propagation \cite{Wrobel_2015}.

When employing \eqref{w_bench}--\eqref{h_def} in the integral operator \eqref{p_ss} one arrives at the following formulae for the fluid pressure:
\begin{equation}
\label{p_bench}
\hat p_\text{b}(\tilde x)=\sum_{i=0}^3 \hat w_i \Pi_i(\tilde x),
\end{equation}
where:
\[
\Pi_0(\tilde x)=\frac{\pi k_2}{2 a_0}, \quad \Pi_1(\tilde x)=\frac{2k_2}{a_0}\left[1-\tilde x \arctanh(\tilde x) \right],
\]
\begin{equation}
\label{Pi_def}
\end{equation}
\[
\Pi_2(\tilde x)=\frac{\pi k_2}{2 a_0}\left[1-2\tilde x^2+\frac{3}{2}\left(1-4\tilde x \sqrt{1-\tilde x^2}\arcsin(\tilde x) +4\ln(2)\tilde x^2-\ln(4)\right) \right], \quad \Pi_3(\tilde x)=\frac{k_2}{a_0}\left(2\pi-\pi^2\tilde x\right).
\]

Representations \eqref{w_bench} and \eqref{p_bench} are used in \eqref{q_ss} and \eqref{v_ss} to produce the benchmark values of the self-similar fluid flow rate, $\hat q_\text{b}$, and the self-similar fluid velocity, $\hat v_\text{b}$. For the sake of brevity we do not specify here the respective formuale as they can be easily obtained by analytical transformations. The benchmark magnitude of the influx is defined as:
\begin{equation}
\label{q0_bench}
\hat q_{0\text{(b)}}=\hat q_\text{b}(0).
\end{equation}
The benchmark value of the self-similar fracture half-length, $a_{0\text{(b)}}$, is obtained by substituting the self-similar crack propagation speed into equation \eqref{SE_ss}. In this way a complete analytical benchmark solution for the self-similar formulation of the hydraulic fracture problem is constructed. This benchmark can be easily extended to the time dependent form by using the scaling \eqref{sel_sim_scl} for a chosen value of $\beta$.

\end{document}